\PassOptionsToPackage{table,xcdraw}{xcolor}

\documentclass[sigconf]{acmart} 
\AtBeginDocument{%
  \providecommand\BibTeX{{%
    \normalfont B\kern-0.5em{\scshape i\kern-0.25em b}\kern-0.8em\TeX}}}

\copyrightyear{2023}
\acmYear{2023}
\setcopyright{rightsretained}
\acmConference[CHI '23]{Proceedings of the 2023 CHI Conference on Human Factors in Computing Systems}{April 23--28, 2023}{Hamburg, Germany}
\acmBooktitle{Proceedings of the 2023 CHI Conference on Human Factors in Computing Systems (CHI '23), April 23--28, 2023, Hamburg, Germany}\acmDOI{10.1145/3544548.3580846}
\acmISBN{978-1-4503-9421-5/23/04}
\usepackage{multirow}
\usepackage{float} 
\usepackage{caption}
\usepackage{subcaption}
\usepackage{soul}
\usepackage{url}
\usepackage{cleveref} %%%%
\usepackage[labelfont=bf]{caption}
\usepackage{array}
\usepackage{multirow}
\usepackage{enumitem}    %%%%
\usepackage{graphicx}
\usepackage{caption}
\usepackage{tabularx}
\usepackage{csquotes}

\usepackage{blindtext,graphicx}
\usepackage[absolute]{textpos}

\definecolor{labelr}{HTML}{FBE7E5}
\definecolor{labell}{HTML}{B4D4DF}
\definecolor{labelc}{HTML}{E9E9F8}

\begin{document}

% \title{Assessing enactment of content regulation policies: An audit study  investigating  YouTube's effectiveness in removing  election misinformation }

\title[A post hoc crowd-sourced audit of election misinformation on YouTube]{Assessing enactment of content regulation policies: A post hoc crowd-sourced audit of election misinformation on YouTube }

% \title{Assessing enactment of content regulation policies: A crowd-sourced audit of election misinformation on YouTube }

% \title{Auditing the enactment  of  YouTube’s anti-election misinformation policies}
% Auditing YouTube’s anti-election misinformation policy Enforcement

% \title{Investigating YouTube’s Effectiveness in Removing Election Misinformation: an audit study}
% \title{Investigating YouTube’s Effectiveness in Removing Election Misinformation: A crowd-sourced audit of election misinformation on YouTube}

% Auditing the enactment of  YouTube’s anti-election misinformation policies

% \title{Assessing algorithmic regulation: A case study  investigating the effectiveness of YouTube's anti-election misinformation policies }


\author{Prerna Juneja}
\affiliation{%
 \institution{University of Washington}
 \city{Seattle}
 \state{Washington}
 \country{USA}}
 \email{prerna79@uw.edu}

\author{Md Momen Bhuiyan}
\affiliation{%
 \institution{Virginia Tech}
 \city{Blacksburg}
 \state{Virginia}
 \country{USA}}
 \email{momen@vt.edu}

\author{Tanushree Mitra}
\affiliation{%
 \institution{University of Washington}
 \city{Seattle}
 \state{Washington}
 \country{USA}}
 \email{tmitra@uw.edu}


\begin{abstract}
With the 2022 US midterm elections approaching, conspiratorial claims about the 2020 presidential elections continue to threaten users' trust in the electoral process. To regulate election misinformation, YouTube introduced policies to remove such content from its searches and recommendations. In this paper, we conduct a 9-day crowd-sourced audit on YouTube to assess the extent of enactment of such policies. We recruited 99 users who installed a browser extension that enabled us to collect up-next recommendation trails and search results for 45 videos and 88 search queries about the 2020 elections. We find that YouTube's search results, irrespective of search query bias, contain more videos that oppose rather than support election misinformation. However, watching misinformative election videos still lead users to a small number of misinformative videos in the up-next trails. Our results imply that while  YouTube largely seems successful in regulating election misinformation, there is still room for improvement.
\end{abstract}



\maketitle

\section{Introduction}

\begin{small}
\begin{displayquote}
\enquote{Oregon GOP frontrunner for governor embraces claims of election fraud... said he doubted Oregon’s vote-by-mail system}---The Texas Tribune, Feb 11, 2022 \cite{OregonGO89:online} 
\end{displayquote}
\end{small}

\begin{small}
\begin{displayquote}
\enquote{Election Deniers Go Door-to-Door to Confront Voters After Losses (in US primaries)}---Bloomberg, Aug 23 2022 \cite{USPrimar71:online} 
\end{displayquote}
\end{small}

\begin{small}
\begin{displayquote}
\enquote{With 10 weeks until midterms, election deniers are hampering some election preparations
Some election deniers have ``weaponized'' against us, one election official says.}---ABC News, Aug 30, 2022 \cite{With10we65:online} 
\end{displayquote}
\end{small}

Skepticism around the legitimacy of the US electoral process, which primarily gained momentum during the 2020 US presidential election, had serious ramifications.  For example, endorsement of election conspiracy theories was found to be positively associated  with lower turnout in the 2021  US Senate election in Georgia \cite{doi:10.1073/pnas.2115900119}. In 2022, the false narratives around the 2020 elections still persist \cite{Studyfin13:online,HowtoFig58:online} and continue to threaten  democratic participation in the upcoming US midterm elections  \cite{Studyfin13:online,HowtoFig58:online}.
% Additionally, the Committee on Oversight and Reform\footnote{https://oversight.house.gov/about}  reported how election misinformation led to violent death threats against   election officials leading them to leave their positions \cite{Microsof58:online}. In 2022, the false narratives around the 2020 elections still persist \cite{Studyfin13:online,HowtoFig58:online} and continue to threaten  democratic participation in the upcoming US midterm elections  \cite{Studyfin13:online,HowtoFig58:online}.
% % \cite{Studyfin13:online,HowtoFig58:online}and ``increase the likelihood that bad-faith actors willsuccessfully subvert legitimate election results (in future) '' \cite{Microsof58:online,Identify56:online}.
In the last two years, 19 US states altered voting procedures and enacted laws to make voting more restrictive, creating information gaps and fresh opportunities for   election misinformation to emerge and proliferate in the real and online world \cite{HowtoFig58:online}. Thus, battling election misinformation has never been more important.

Studies show that social media platforms have % like Facebook, Twitter, and YouTube have
%  are playing an ever expanding   role in political events.
become important mediums for political discourse \cite{allcott2017social,vitak2011s}. In particular, YouTube---the most popular platform among US adults \cite{Socialme29:online}---has emerged
 as a  political battleground  as demonstrated by the fact that both political parties extensively used the platform for election campaigning \cite{Trumpdep92:online}.  However, the platform came under fire from technology critics for being a hub of electoral conspiracy theories \cite{YouTubeh93:online,Election91:online}. Given  the concern that search engines can play a significant role in shifting voting
decisions \cite{epstein2015search,epstein2017suppressing} and can confine users into a filter bubble of misinformation \cite{hussein2020measuring}, there has been a push for online platforms to enact policies that minimize election misinformation \cite{Inelecti49:online}. In response to this push, YouTube introduced content policies to remove videos spreading election-related falsehoods  and claimed that misinformative videos would not prominently surface or get recommended on the platform \cite{Supporti69:online,Election56:online,YouTubet38:online,Howwills6:online}. However, the formulation of policies does not equate to effective enactment. It's evident from the results of two misinformation audits conducted on the platform for the same conspiratorial topics (such as vaccine controversies, and 9/11 conspiracies), first in 2019 \cite{hussein2020measuring} and second in 2021 \cite{tomlein2021audit}, both of which found echo chambers of misinformation on the platform. Despite changes to YouTube's misinformation policies in 2020 \cite{Managing54:online}, the authors of the second audit study did not find improvements when compared to the results of the first audit, rather they
 found recommendations worsening for topics like vaccination. These findings iterate the need to continuously audit platforms  to investigate how a platform's algorithms fare with respect to  problematic content and how effectively a platform's content policies are implemented \cite{simko2021towards}. While multiple studies have audited YouTube for misinformation  \cite{hussein2020measuring,tomlein2021audit,papadamou2022just}, these were mostly conducted using sock-puppets 
 (bot accounts emulating real users) in conservative settings\footnote{For example, sock-puppet building account history by watching videos that only promote misinformation.} which often do not reflect true user behavior.  There is a dearth of crowd-sourced misinformation audits that test the algorithms' behavior with real-world users (\cite{bisbee2022election} is one of the few exceptions).
 In this paper, we fill this gap by conducting a large-scale crowd-sourced audit on YouTube to determine how effectively YouTube has  regulated its algorithms---search and recommendation---for election misinformation.

 To conduct the audit, we recruited 99 participants who filled out a survey and installed \textit{TubeCapture}, a browser extension  built to collect  users' YouTube search results,  and recommendations. The extension conducted searches for 88 search queries related to the 2020 US presidential elections. We also seeded \textit{TubeCapture} with 45 seed videos with three differing stances on election misinformation---supporting, neutral, and opposing. The extension collected up-next recommendation trails---five consecutive up-next recommendation videos---for each seed video. \textit{TubeCapture}  simultaneously collected YouTube components from both personalized standard and unpersonalized incognito windows allowing us to measure the extent of personalization. This leads us to our first research question:
 
  \begin{itemize}[leftmargin=*]
\item[] \indent \textbf{RQ1 Extent of personalization: } What is the extent of personalization in various YouTube components?

\begin{itemize}
\item[] \indent \textbf{RQ1a:} How much are search results personalized for search queries about the 2020 US presidential elections and the surrounding voter fraud claims?
\item[] \indent \textbf{RQ1b:} How much are YouTube's up-next recommendation trails personalized for seed videos with different stances on election misinformation---supporting, neutral and opposing?
\end{itemize}
\end{itemize}
 
We find that while search results have very little personalization, up-next trails are highly personalized.  We next venture into 
%Our study also enables us to determine the impact of personalization on the amount of misinformation presented to users. Through our experimental design we report on the behavior of YouTube's algorithms and determine 
determining the amount of election misinformation real users could be exposed to under different conditions, such as following up-next trails for videos supporting or opposing election misinformation. 

\begin{itemize}[leftmargin=*]
\item[] \indent \textbf{RQ2: Amount of election misinformation:}  %What is the impact of watching  a sequence of YouTube up-next recommendation videos starting with seed videos with different stances on election misinformation (supporting, neutral and opposing) on the  amount of election misinformation returned in various YouTube components?
% What is the impact on various YouTube components when watching a sequence of YouTube up-next recommendation videos starting with seed videos with different stances on election misinformation (supporting, neutral and opposing)?
What is the impact of watching a sequence of YouTube up-next recommendation videos starting with seed videos with different stances on election misinformation (supporting, neutral, and opposing)  on various YouTube components?
\begin{itemize}
\item[] \indent \textbf{RQ2a: }  How much do search results get contaminated with election misinformation?
\item[] \indent \textbf{RQ2b: } What is the amount of misinformation returned in users'  up-next recommendation trails? 
\item[] \indent \textbf{RQ2c: } {What is the amount of misinformation that appears in users'  homepage video recommendations?}
\end{itemize}
\end{itemize}

We find that YouTube presents debunking videos in search results for most of the queries. We also observe an  echo chamber effect in recommendations where trails with supporting seeds contain more misinformation than trails with neutral and opposing seeds. Since election misinformation is closely entangled with political beliefs  with several right-leaning news sources amplifying the claims of voter fraud \cite{Theuniqu22:online,Republic22:online}, we also study the diversity and composition of the content presented by YouTube in its various components. 
We ask,
\begin{itemize}[leftmargin=*]
\item[] \indent \textbf{RQ3: Impact on composition and diversity:} %What is the impact of watching  a sequence of YouTube up-next recommendation videos starting with seed videos with different stance on election misinformation (supporting, neutral and opposing) on the diversity of content that users are exposed to in YouTube?
What is the impact on content diversity when watching  a sequence of YouTube up-next recommendation videos starting with seed videos with different stances on election misinformation (supporting, neutral, and opposing)?

\begin{itemize}
\item[] \indent \textbf{RQ3a: } How diverse are the search results ?
\item[] \indent \textbf{RQ3b: } How diverse are the up-next recommendation trails?
% \item[] \indent \textbf{RQ2c: } How do the homepages get impacted?
\end{itemize}

\end{itemize}

We find that YouTube ensures source diversity in its search results. We also find a large number of  impressions for left-leaning late-night shows (e.g. Last Week Tonight with John Oliver)   and right-leaning Fox news in users' up-next trails.
Overall, our work makes the following contributions:
% We make four major contributions to the CSCW/HCI community. 
% \setlist{nolistsep}

\begin{itemize}

\item We conduct a  post hoc audit on YouTube to determine how its algorithms  fare with respect to election misinformation; post hoc auditing comprises investigating a platform for a past topic or event which could have a significant impact on citizenry in the present and future. In turn, we are able to test the effectiveness of YouTube's content policies enforced to curb election misinformation.

\item We extend prior work on misinformation audits by conducting an ethical crowd-sourced audit to see the impact of performing certain actions on the searches and recommendations of real-world people with complex platform histories instead of conservative settings of sock puppet audits.

% \item Through our experimental design we can estimate the imp real world users' recommenthe counterfactual scenarios. what would have happened if the same participant watched a video supporting video? What would have happened if participant views a debunking video? We can answer these questions without one condition affecting the other?

% \item 

% \item determining belief vs reality: matching survey responses to actual results

\item {Our audit reveals that  YouTube   search results contain more
videos that oppose election misinformation as compared to videos supporting election misinformation, especially for search queries about election fraud in presidential elections. However, a filter bubble effect still persists in the up-next recommendation trails, where a small number of misinformative videos are presented to users watching videos supporting election misinformation.} 

\end{itemize}

\section{Related Work} \label{rel}

% \subsection{Algorithmic pathways}
\subsection{Algorithmic audits}
Search engines and social media platforms act as information gatekeepers, with their algorithmically generated feed, timeline, and recommendations affecting the information exposure of people. Given the ubiquitousness of the algorithms and the influence they  hold over the citizenry, scholars have emphasized the need for auditing online platforms, i.e., conducting a systematic investigation to determine whether the algorithmic output is aligned with  ``laws and regulations, societal values, ethical desiderata, or
industry standards'' \cite{abs-2105-02980}. As a result, several research studies have audited  algorithmic systems  for
distortion (e.g. hyper-personalization \cite{10.1145/3449148}, ideological skew \cite{bandy2021more,trielli2019search,10.1145/2998181.2998321} ), discrimination (e.g. racial and gender discrimination \cite{buolamwini2018gender,kyriakou2019fairness,asplund2020auditing}),   exploitation (e.g. exploiting users' private and sensitive information \cite{DBLP:journals/corr/DattaTD14,cabanas2018unveiling}) and misjudgment (e.g. incorrect algorithmic predictions \cite{AllSouls64,duwe2019better}) \cite{10.1145/3449148}.  These scholarly studies have  used a myriad of audit research methods, including code audits,  scraping audits,  sock puppet audits, and  crowd-sourced audits (see \cite{sandvig2014auditing} for a review).   Among them, sock puppet auditing, where researchers create bots or fake user accounts to impersonate  real-life users is the most popular since this audit method gives researchers the greatest control over experimental variables \cite{wilsonpromise} and doesn't require high participant recruitment cost like in the case of crowd-sourced auditing \cite{sandvig2014auditing}. Thus, several past studies have employed this audit method \cite{asplund2020auditing,hussein2020measuring,bandy2020auditing,trielli2019search,bandy2021more,juneja2021auditing}.  However, in sock-puppet auditing, the bot histories are built in very conservative settings that do not emulate real-world users'  complex account histories \cite{juneja2021auditing}. Thus, as an alternative, scholars have collected and audited algorithmic outputs from real-world users to study and identify problematic algorithmic behaviors in users' naturalistic settings \cite{robertson2018auditing,3274417,bisbee2022election,bandy2020auditing,venkatadri2019auditing}.  We add to the existing crowd-sourced audit studies by conducting a  crowd-sourced audit of YouTube to measure the amount of election misinformation in the searches, and recommendations of real-world users.  In our study, we use a list of pre-selected videos and search queries to collect data from users' YouTube accounts to  test  whether users' existing account histories  could lead them to misinformative content on the platform. In the next section, we present   the audits conducted specifically on YouTube and discuss how our work adds to the growing literature on platform audits.

\subsection{Auditing YouTube for problematic content}
Given the popularity of YouTube and the criticism the platform has faced for not regulating problematic content, several scholarly studies  have audited YouTube's search and recommendation algorithms  for the prevalence of misinformation, extremism, and echo chambers of problematic content. Sock puppet audits on YouTube revealed that while the platform's channel recommendations radicalize users by recommending  extreme channels \cite{ribeiro2020auditing},  video recommendations  drive users away from radical content by recommending videos from mainstream news channels  \cite{huszar2022algorithmic}. A crowd-sourced audit further revealed  that real users with high prior levels of  racial
resentment get more exposure to extremist content since they typically subscribe to extremist channels \cite{chen2022subscriptions}.
% of YouTube channel recommendations for alt-right and extremist content
% showed that  YouTube radicalizes users by recommending  extreme channels \cite{ribeiro2020auditing}.
% Another, however, found that YouTube's video recommendation algorithm drives users away from radical content and the platform recommendations favor mainstream news channels  \cite{huszar2022algorithmic}. Both these prior studies used sock puppets and did not observe real users'  recommendations.
% % under naturalistic circumstances. 
% To overcome this limitation, a crowd-sourced audit focused on passively tracking real-world  users' YouTube usage. The study found that users with high prior levels of  racial
% resentment get more exposure to extremist content since they typically subscribe to extremist channels \cite{chen2022subscriptions}. Instead of passive data collection, in our study, we use a list of pre-selected videos and search queries to collect data from users' YouTube account. This allows us to specifically  test  whether users' existing account histories and past actions could lead them to misinformative content on the platform. 
% Our work also tracks and analyze the searches and recommendation of real users to determine 
% We add on to the existing crowd-sourced studies by investigating the YouTube's searches, video recommendation pathways and homepage-page recommendations of real-world users for the election misinformation topic and study the composition and diversity in the YouTube's search results  and recommendations.
In another line of inquiry, several studies audited YouTube for conspiracy theories  \cite{sanna2020yttrex,hussein2020measuring,faddoul2020longitudinal,papadamou2022just}. 
Notably, first such audit on YouTube  was conducted by Hussein et al \cite{hussein2020measuring}. 
This audit   
% investigated how personalization  impacts the amount of misinformation that is recommended to users in search results and top-5 video recommendations  \cite{hussein2020measuring}. The study 
revealed the prevalence of echo chambers of misinformation in  YouTube's top-5 video recommendations for topics such as the moon landing, 9/11 conspiracies, etc. \cite{hussein2020measuring}. 
Recently, Tomlein et. al re-conducted the audit performed by Hussein and Juneja et al \cite{hussein2020measuring}  and found that video recommendations for   topics like 9/11 conspiracies have worsened on the platform \cite{tomlein2021audit}.
Another study (conducted in the fall of 2020), closest to this work collected real-world YouTube recommendations for election fraud videos by asking users to manually click on recommendations following certain traversal rules \cite{bisbee2022election}. The study aimed at proving that users skeptical about the legitimacy of elections receive more voter fraud videos in their recommendations. On the other hand,  we  audit  YouTube's searches, homepages, and default algorithmic pathway  (up-next videos that are auto-played by the platform) of users with different political leanings and investigate how its algorithm fares under different conditions (watching videos of different stances) for the same individual. Additionally, we conduct the audit two years after the presidential election event. Post hoc auditing of the platform allows us to determine how well the platform has enacted its content policies and regulated harmful content.

\section{Methodology} \label{method}
% In this section, we present our audit methodology, including our approach to compile search queries (Section \ref{search_queries}) and YouTube videos associated with presidential elections (Section \ref{videos}). We also provide an overview of our crowd-sourced audit experiment (Section \ref{design}) and annotation process (Section \ref{anno}). We discuss steps taken to minimize the potential harm of our experiments (Section \ref{ethics}), and our study recruitment process (Sections \ref{survey_sec} and \ref{rec}).

% In this section, we present our audit methodology in detail. We start by describing our approach to compile  search queries (Section \ref{search_queries}) and YouTube videos (Section \ref{videos}) associated with election fraud 2020. Then, we present an overview of our crowd-sourced audit experiment (Section \ref{design}) and annotation process (Section \ref{anno} and \ref{classifier}).  We also discuss the steps we took to minimize the potential harm of our experiments (Section \ref{ethics}). Finally, we discuss our study recruitment process in Sections \ref{survey_sec} and \ref{rec}.

\subsection{Developing search queries to measure election fraud based misinformation} \label{search_queries}
\begin{figure*}
    \centering
    \includegraphics[width=0.9\textwidth]{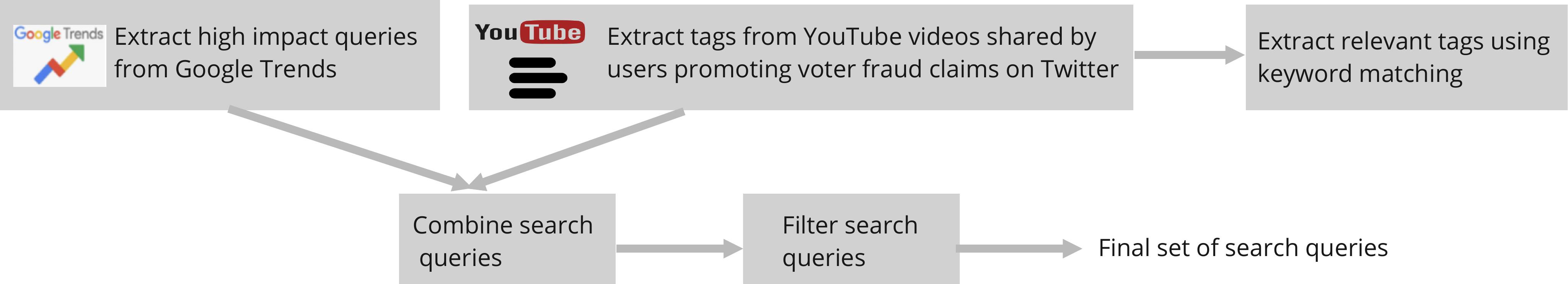}
    \caption{Figure illustrating our method to curate search queries for our audit experiment}
    \Description{The figure illustrates the two methods to curate search queries for the audit. First, we use high-impact queries from Google Trends. Second, use relevant YouTube video tags from YouTube videos that were shared by users promoting voter fraud claims on Twitter.}
    \label{fig:query}
\end{figure*}

The first methodological step in any algorithmic audit is to determine  a viable set of relevant search queries that would be used to probe the algorithmic system. For our study, we identified search queries that satisfy two properties. First, we select high-impact search queries that were used by people to search about Presidential Election as well as the voter fraud claims about the     2020 elections. Second,  we curate search queries that  have  a high probability of returning misinformative results which would result in meaningful measurements of algorithmically curated misinformation about the audit topic. To compile such queries, we used Google Trends and YouTube video tags (refer Figure \ref{fig:query}).

 \subsubsection{Curating high-impact queries via Google Trends} First, we  leveraged Google Trends which contain Google's daily and real-time search trends data. As the most popular search service, its trends are a good indicator for understanding the real-world search behavior of a large number of people. Using \textit{Election Fraud 2020} and \textit{Presidential Election} as  search topics, United States as location, April 2020 to Present as date range, and search service as YouTube search, we  extracted the top 15  most and least popular search queries that people used on  YouTube.  We choose April 7 as the start date since this was the day when Donald Trump  made one of his first fraudulent claims about the security of mail-in ballots \cite{Timeline1:online}. We  included  the most popular queries since they represent the ones that people mostly use to get information on elections. To explore the \textit{data-voids} \cite{golebiewski2019data} associated with our audit topic, we  also included the least popular search queries to determine if those terms have been hijacked by conspiracists  to surface misinformation. 
%  spreading false claims about the elections. 
%  Through this step, we curated a list of 40 search queries.

\begin{figure*}[!t]
\begin{minipage}{0.9\textwidth}
  \begin{minipage}[]{0.5\textwidth}
    \centering
    \includegraphics[width=0.8\textwidth,keepaspectratio]{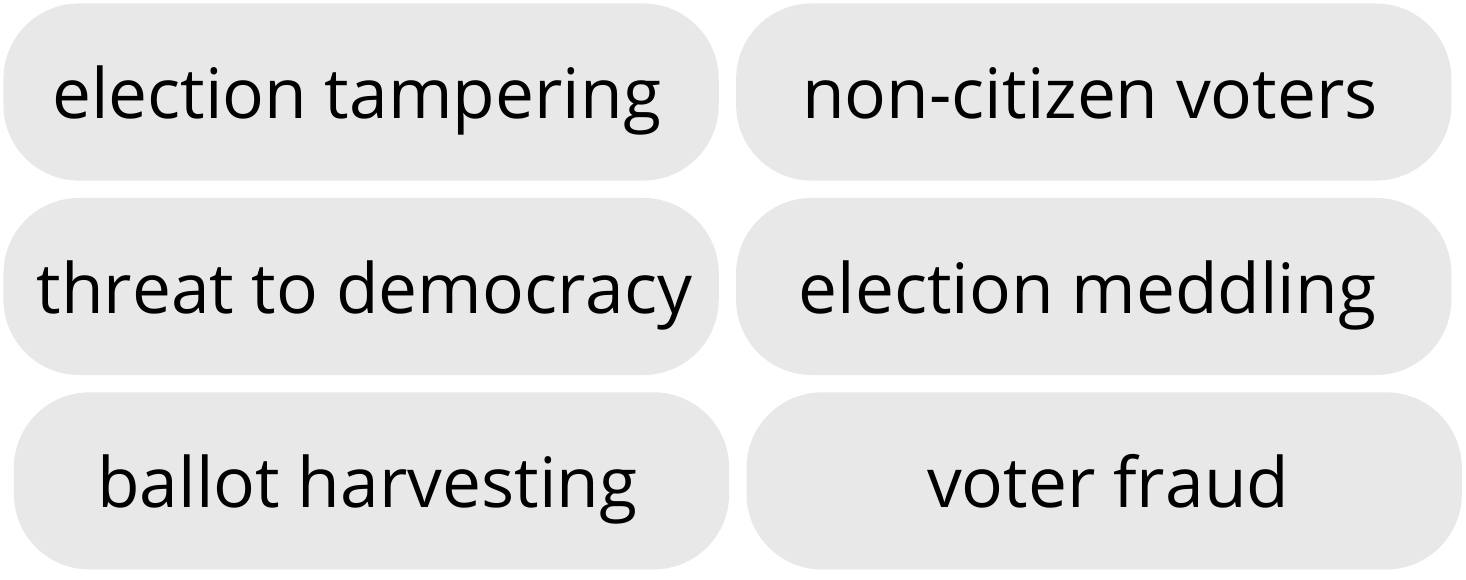}
    \captionof{figure}{List of video tags associated with YouTube video titled { \tt Is Voter Fraud Real?} (video id: { \tt RkLuXvIxFew}) that promotes voter fraud misinformation. Video tags are added by content creators while uploading YouTube videos on the platform. The tags can be extracted from videos via YouTube APIs or third-party tools. We use   tags associated with  videos shared by users promoting voter fraud claims on Twitter as search queries in our audit experiments.}
    \label{fig:tags}
    \Description{The figure shows a list of video tags associated with the YouTube video titled Is Voter Fraud Real?, such as election tampering, non-citizen voters, threat to democracy, ballot harvesting, etc. }
  \end{minipage}
%   \hfill
% \hspace{-1cm}
  \begin{minipage}[]{0.6\textwidth}
%   \begin{scriptsize}

    \centering
    \small
\begin{tabular}{l}

\hline

presidential election 2020 \\
us elections 2020 latest news \\
election fraud 2020 \\
rigged election \\
dominion voting exposed \\
mail in ballots 2020 \\
stop the steal \\
joe biden voter fraud \\
usps whistleblower \\
voter fraud evidence\\

trump biden general election\\

dominion voter fraud \\ \hline
\end{tabular}
      \captionof{table}{Sample search queries for our YouTube audit}
      \label{searchqueries}
    %   \end{scriptsize}
    \end{minipage}
  \end{minipage}
\end{figure*}

\subsubsection{Curating misinfo-queries queries using YouTube video tags}
Second, we  used YouTube video tags that content creators  associated with misinformative videos while uploading them on the YouTube platform (see Figure \ref{fig:tags} for an example). These tags could be 
thought of as search words representing how content creators would like their videos to be discovered. To extract video tags associated with election misinformation videos, we  leveraged a large-scale Voter Fraud 2020 dataset released by  Abilov et al \cite{Abilov}. The dataset contains over 12,002 YouTube video URLs that were shared on Twitter
% along with statistics indicating how many times they were shared 
by accounts that tend to refute   and  promote voter fraud claims. We extracted  YouTube video tags associated with videos shared by  accounts promoting voter fraud claims 
% and used them as \textit{misinfo-queries}
to probe YouTube (n=200K). 
% In total, we extracted 200K video tags.
To curate a viable number of search queries from the extracted video tags, we employed several steps. First, we manually curated a list of 10 keywords related to elections and  fraudulent claims surrounding the elections\footnote{  \textit{steal, fraud, ballot, elect, seal, dominion, sharpiegate, whistleblower, harvest, and sunrise zoom}} from the list of keywords provided by  Abilov et al  \cite{Abilov} as well election 2020 misinformation report produced by the Election Integrity Partnership \cite{eip}. Then for each of the keywords, we extracted  15 top and 15 least occurring video tags containing that term. For example, one of the most occurring tags containing keyword \textit{whistleblower} was
% `whistleblower', and 
`usps whistleblower'  while the least occurring tag was
% containing the term was
% `usps whistleblower did not recant', and
`whistleblower jesse morgan'. % and `was wisconsin stolen? whistleblower speaks up on 10000 backdated ballots'.
% After this step, we curated a list of 300 video tags to be used as search queries. 

\subsubsection{Filtering search queries to obtain the final set}
We combined  search queries obtained from both Google Trends and YouTube video tags in our final query set and employed several filtering steps to obtain a reasonable number of relevant search queries.
% to further to ensure that the  queries in the final set are relevant to the audit topic. 
First, we only kept queries related to the election  2020, for example, we 
kept `election fraud 2020' and removed `election fraud 2016'. We  removed duplicate and redundant search queries and replaced them with a single randomly selected query. For example, we replaced queries `voter fraud 2020', 'voter fraud',  and `vote fraud'  with `voter fraud 2020'. We removed  queries with lengths greater than five since they were overly specific  (e.g. `we've got pictures of the check stubs paid to people to ballot harvest'). We also removed  queries  containing names of news channels, news anchors,   and presidential candidates because they were too generic and not directly related to the audit topic. However, we kept the search queries where the names of the presidential candidates were together with the election or election fraud-related terms (e.g. 
% `trump election fraud',
`Joe Biden voter fraud').  We also removed search queries that were in languages other than English. Finally, we had 88 search queries in total. Table \ref{searchqueries} presents a sample.

% \begin{figure}
%     \centering
%     \includegraphics[width=0.95\linewidth]{figures/tags_new.pdf}
%     \caption{List of video tags associated with YouTube video titled {\small \tt Is Voter Fraud Real?} (video id: {\small \tt RkLuXvIxFew}) that promotes voter fraud misinformation. Video tags are added by content creators while uploading YouTube videos on the platform. The tags can be extracted from videos via YouTube APIs or third party tools. We use  video tags associated with misinformative videos as search queries in our audit experiments.}
%     \label{fig:tags}
% \end{figure}

% \begin{table}[]
% \begin{tabular}{p{7cm}}
% \hline
% election fraud 2020, rigged election, trump election fraud, voter fraud 2020, joe biden voter fraud, michigan voter fraud, biden voter fraud organization, presidential election 2020, trump whistleblower, usps whistleblower, whistleblower fraud, dominion investigation, dominion software, dominion voter fraud, dominion voter machine scandal, dominion voter systems, dominion voting exposed, sharpie ballots, sharpie gate, steal the election, stop the steal, mail in ballots 2020, ballot box fraud, ballot fraud, us elections 2020 state calls \\ \hline
% \end{tabular}
% \caption{Sample search queries for our YouTube audit}
% \label{searchqueries}
% \end{table}

\begin{figure*}[t]
    \centering
    \includegraphics[width=0.9\textwidth]{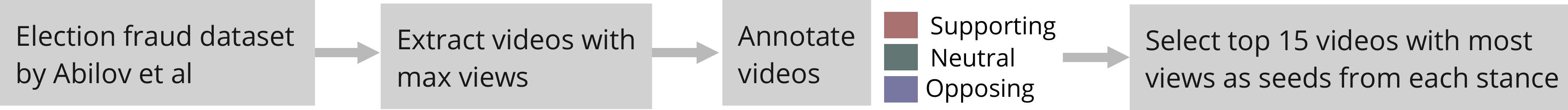}
    \caption{Figure illustrating our method to curate seed videos for our audit experiment}
    \label{fig:seed}
    \Description{The figure illustrates the method used to curate seed videos for the audit experiment that is described in detail in the Section titled ``determining popular seed videos to collect
up-next video trails''}
\end{figure*}

\begin{table*}[]
\small
\begin{tabular}{m{2.8cm}|m{8.5cm}|m{2cm}}
\hline
Annotation label & Video title & Video id \\ \hline
\multirow{2}{*}[-0.5em]{\begin{tabular}[c]{@{}l@{}}Supporting election\\ fraud misinformation\end{tabular}} & Poll worker gives his account of what happened when he tried to monitor the vote in Nevada & 4X2V5hPPp6w \\ \cline{2-3} 
 & Joe Biden says he's built most extensive "voter fraud" org in history & WGRnhBmHYN0 \\ \hline
\multirow{2}{*}[-0.5em]{Neutral} & Ex-Trump official shares his prediction if Trump loses 2020 & KuqhhrmhfCI \\ \cline{2-3} 
 & 'Don't be ridiculous': Rudy Giuliani learns about Biden win from reporters & Z0hEFa52Bdo \\ \hline
\multirow{2}{*}[-0.5em]{\begin{tabular}[c]{@{}l@{}}Opposing election\\ fraud misinformation\end{tabular}} & Voting by Mail: Last Week Tonight with John Oliver (HBO) & l-nEHkgm\_Gk \\\cline{2-3} 
 & Trump and the GOP Still Refuse to Accept Biden's Win: A Closer Look & QoPA3unjQgA
\end{tabular}
\caption{Sample seed videos curated for the audit experiment.}
\label{seedvideos}
\end{table*}

\subsection{Determining popular seed videos to  collect up-next video trails} \label{videos}
The second step of our audit experiment is to curate YouTube videos that would act as seed videos to  collect the up-next video recommendation  trails. We   again leveraged  Abilov et al's YouTube video dataset \cite{Abilov}.  
Recall, the authors identified clusters of Twitter users who either shared tweets promoting or detracting from voter fraud claims and released the YouTube videos related to election fraud 2020 shared by those users.  {At the the time of analysis, out of the $\sim$12K videos present in the dataset, ~8.9K were present on YouTube. The remaining videos were either removed or made private. Out of the videos that were still present, ~1K videos were shared by users in the detractor cluster, ~6.5K videos were shared by users in the promoting cluster, and the rest were shared by users who were suspended from Twitter. We sampled 445 videos that had accumulated the maximum number of views from both the promoting and detracting clusters (890 in total).}
% Recall, the authors identified clusters of Twitter users who  either shared tweets promoting or detracting voter fraud claims and released the YouTube videos related to election fraud 2020 shared by those users. {Out of the $\sim$12K videos present in the dataset,  1.1K YouTube videos  were shared by users in the promoting cluster, and 209 videos were shared by users in the detractor cluster that were still active on YouTube at the time of analysis.} We extracted 445 videos in total from the promoting and detracting clusters that had accumulated the maximum number of views. 
Since the videos were not annotated by the authors for misinformation, we could not assume that videos shared by users in the promoting cluster would contain misinformation. Therefore, we conducted an intensive and iterative process to determine the    labels and heuristics for annotating the YouTube videos for misinformation. We describe the process in detail in Section \ref{anno}. Through the annotation process, we labeled the videos as supporting, neutral, or opposing election  misinformation. Out of the 890 videos, 74 were opposing, 16 were neutral,  101  supported election misinformation while remaining were irrelevant. We selected the top 15 videos  that had accumulated maximum engagement, determined by the number of views, for each stance (except the irrelevant) as seeds. Figure \ref{fig:seed} illustrates the seed video curation method. Table \ref{seedvideos} presents a sample of seed videos.

% \usepackage{multirow}

% opposing: Opposes the misinformation narratives behind the 2020 election
% Irrelevant: Any video whose content is not related to the 2020 elections in any way will be marked as irrelevant.

% Annotation Process:
% While annotating the videos, use the following metadata in the priority order mentioned below to assign the annotation value:-
% Title and description
% Transcript/content of the video (Use the overall premise of the video)
% Channel bias 
\begin{figure*}[t]
    \centering
    \includegraphics[width=0.93\textwidth]{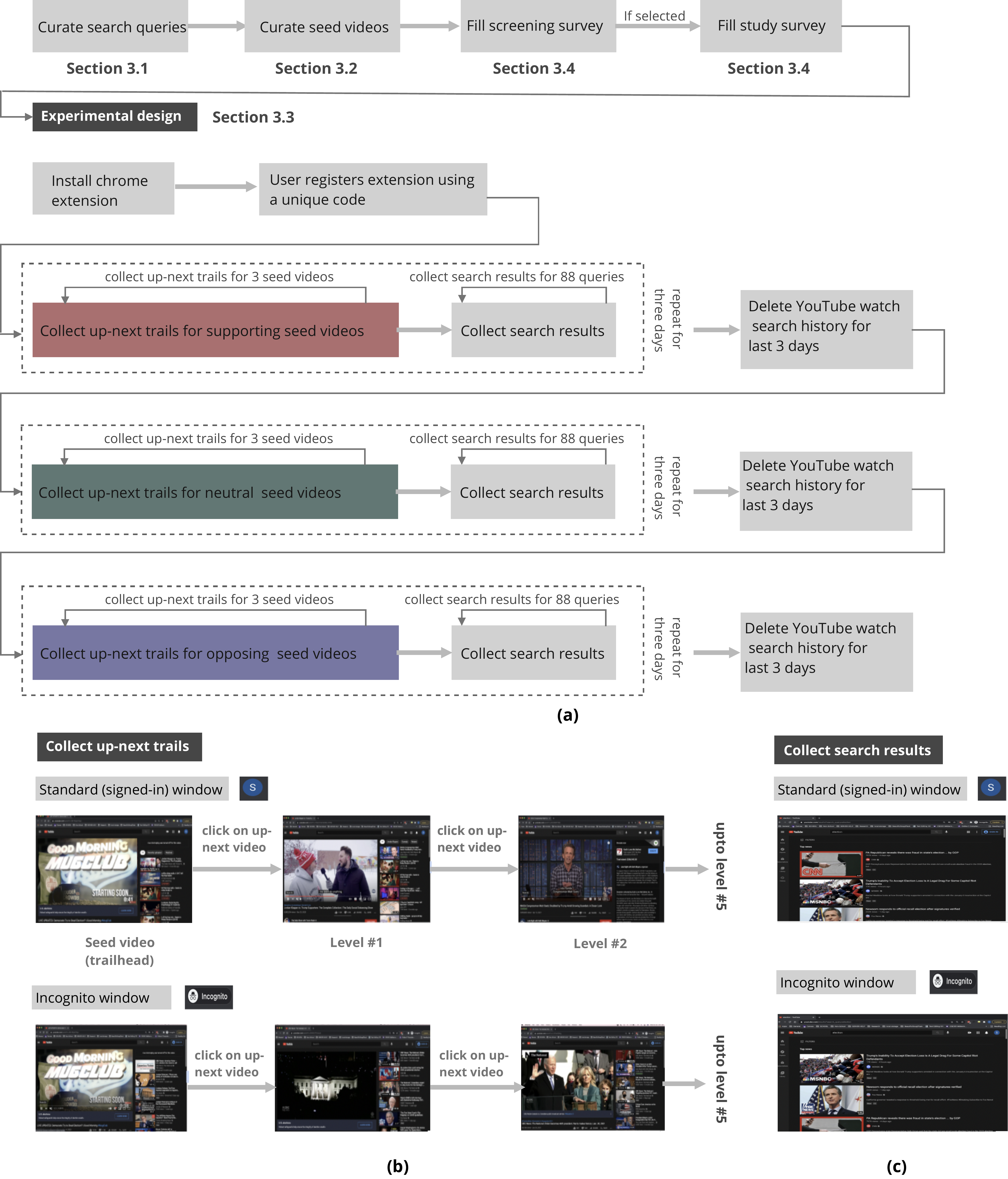}
    \caption{Figure (a) presents an overview of our crowd-sourced audit of YouTube for election misinformation, Figures (b) and (c) show how our extension \textit{TubeCapture} collected YouTube components from both standard and incognito windows simultaneously.}
    \label{fig:metafig}
    \Description{Figure (a) presents an overview of our crowd-sourced audit of YouTube for election misinformation, Figures (b) and (c) illustrate how our extension \textit{TubeCapture} collected YouTube components from both standard and incognito windows simultaneously.}
\end{figure*}
\subsection{Experimental design} \label{design}
% What kind of videos real users get exposed to when they search about  US 2020 presidential elections or the associated fraudulent claims? What happens when real users follow the up-next recommendation trails starting with seed videos having different stance on election misinformation? We answer these questions by conducting a crowd-sourced audit of the YouTube platform. 
To conduct the crowd-sourced audit,  we designed a chrome browser extension named \textit{TubeCapture} that enabled us to watch videos, conduct searches, and collect various YouTube components  from users' browsers. Figure \ref{fig:metafig} presents an overview of our  experimental design. 
% The extension also allowed us to explore counterfactual scenarios by deleting users'  account  histories built by our experiment after testing each condition of the experiment.
To select the study participants, we conducted a screening survey of a large sample of people (details in Section \ref{survey_sec}). Next, participants were instructed on how to use \textit{TubeCapture} and provided with a unique  code to activate the extension. Once activated, they used \textit{TubeCapture} for a period of 9 days. We  seeded our extension with 45 seed videos and 88 search queries. For each participant, each day the extension opened YouTube in two browser windows, one standard window and one incognito window. While the personalized results act as treatment for our experiments, results obtained from incognito act as control since YouTube does not personalize content in the incognito browsing window \cite{BrowseYo68:online}. By comparing the results from standard and incognito windows, we determine the role of YouTube’s personalization algorithms in exposing users to misinformative content. 
% leanings and investigate how its algorithm fares under different conditions (watching videos of different stances) for the same individual.
% And by comparing the standard personalized YouTube pages, we determine the effect of personalization on misinformation and partisan bias. 

\textit{TubeCapture} first collected and stored the user's YouTube  homepage from standard and incognito windows. The extension ensured that the user had signed in to their YouTube account in the standard window and remained logged in using the same YouTube account throughout the study period. We also ensured that the homepage from the standard window is stored without the user's email address to ensure the participant's anonymity. Next, the extension opened a seed video (previously selected) that supports election misinformation, watched it for 2 minutes, saved the video page, clicked on the up-next video, and again saved the video page of the up-next video. This process was repeated until we collected 5 levels of up-next  recommendations' video pages. We refer to the collection of 5 up-next video recommendations as up-next trails.   Each day we collected up-next trails for five seed videos.  Then, the extension again collected the user's homepage followed by personalized (via standard window) and unpersonalized (via incognito window) search results for the curated search queries. {The extension collected the search results for queries in the same order for every participant to control for carry-over effects of the search queries \mbox{\cite{hannak2013measuring}}}. For days 1-3, the extension collected up-next trails for seed videos supporting election misinformation. At the beginning of the fourth day, the extension deleted the  search and watch history created by the browser extension. According to YouTube, removing an item from search or watch history removes the impact of consuming that content on future searches and recommendations. This essential step helped us in two ways- 1) it ensured that the history created by our extension in the first three days does not impact the rest of the experiment,
% \footnote{{To test whether the Youtube algorithm discards the deleted history while making recommendations, the first author ran two test runs for two topics (including presidential elections, the audit topic)). They built  the account history of a brand new YouTube account for 3 days using  a few videos and search queries related to the topic after which they deleted the search and watch history. Then the author manually inspected the homepage recommendations and the video recommendations of the top five videos present on the homepage and found that the effect of history deletion is almost immediate.}},
% thereby helping us to estimate the counterfactual scenarios 
and 2) it also ensured that the user histories built by our extension did not pollute users' future recommendations and search results after the study period is over. For days 4-6, the extension collected up-next trails for seed videos that were neutral in stance. At beginning of the seventh day, again search and watch history developed by the extension was deleted. For days 7-9, the extension collected up-next trails for opposing seed videos. Towards the end of the 9th day, we again deleted the YouTube history developed by the extension. All the data collected by the extension was sent to a back-end server. The participants were instructed on how to remove the extension after the study period was over. 
% and we know that algorithm is impacted by order of things such as queries and watching content, it could compound and result could diverge significantly. then it becomes difficult to test differences.  and it would be really hard to account for all compound effects, we would need to consider large number of cases
{Our current mixed  design allows us to  test how YouTube's algorithm fares under different conditions---watching videos of different stances---for  individuals with different political beliefs. Note that we did not opt for a randomized assignment in a between-subject design since it would require a large number of participants to test all the conditions (3 political affiliations X 3 misinformation stances).  }

We built the YouTube capture extension  using 
JavaScript libraries. The back-end server was set up using
 Flask and Nginx. We load-tested the server using Jmeter and ensured that the server could simultaneously handle 500 GET and 200 POST requests and added mechanisms to handle errors and server timeouts. We used  a MySQL database for storing the data collected using the extension. The communication between the extension and our back-end  server was  encrypted
using SSL. Note that to collect data, TubeCapture opened windows in the background of the currently active browser window, thereby allowing participants to continue working on their device while the extension is running. In case, the participant accidentally closed any of the windows opened by our extension, we informed users via a pop-up window and instructed them on how to resume running the extension.

After building the TubeCapture extension, we tested it with our research
group and conducted three pilot studies. The aim of the pilot studies was to fix technical issues, examine the impact of running the extension on devices with different configurations, RAM, and operating systems as well as improve the usability of the extension. 
% The extension was  distributed to the study participants via  Chrome Web Store.

% is cleare

% builds search and watch history collects up-next trails, search results and video recommendations that YouTube returns to real users and sends it back to a backend server. 

% over a week by watching videos that either promote, debunk or are neutral in stance with respect to election fraud misinformation. The extension also collects up-next trails, search results and video recommendations that YouTube returns to real users and sends it back to a backend server. 

% The aim of our first study (corresponding to RQ1) is to determine what kind of videos real users are exposed to when they search for election fraud or watch videos that promote, debunk or discuss election fraud misinformation. 

% Through the study, we are also able to investigate whether the participants share any observable features collected via survey and plugin such as gender, political affiliation, types of channels subscribed, etc. that may illuminate why they are receiving biased results.  Figure 3 describes the experimental design of our crowdsourced audits.

\subsection{Screening and study survey} \label{survey_sec}
In order to select participants for our study, we screened users according to several criteria. To be eligible for the study, users should be 1) 18 years of age or older, 2) reside in the United States, 3) have a YouTube account, 4) consume content on YouTube primarily in the English language, 5) have a chrome browser installed, 6) willing to run a chrome browser extension for 9 days and 7) have at least 8GB  RAM on their device to ensure  smooth running of the extension\footnote{We warned  users against participating in the study if their device's RAM is less than 8GB and informed them that their device or browser might hang in such a situation}. The users who qualified for the screening survey were sent another study survey. The study survey contained questions about users' demographics, political affiliation, YouTube usage, trust in online information,  their opinion on personalization and bias in various components of YouTube, and their view on the results of the presidential elections 2020 as well as conspiracies surrounding the elections. We also included two attention-check questions.  The study survey was also used for screening participants. We disqualified users who 1) answered both attention check questions incorrectly, 2) did not frequently use YouTube, and 3) did not use YouTube to access news or information about the 2020 presidential elections.  We also used the survey responses to obtain a balanced number of participants across three political affiliations (Democrats, Republicans, and Independents). Later in the recruitment phase, we had enough democrats and independents as participants and thus, added being a republican as a qualifying criterion in the study survey.

\subsection{Recruitment and study deployment} \label{rec}
For our pilot studies, we recruited users from  a combination of platforms  such as Reddit\footnote{https://www.reddit.com/r/SampleSize/},  Facebook ads, Twitter, and Amazon Mechanical Turk (AMT). The retention rate was highest for participants recruited from Twitter and AMT. Thus, we used these two platforms to recruit participants for the main study. The pilots and the main study were approved by our university's Institutional Review Board. 
% Using TubeCapture, we ran the main study for 9 days.
Out of the 575 users who submitted the screening survey,  400 qualified, and 99 participated in the study. Out of the 99 participants, 94 ran the extension for  the entire study duration. Overall, our study sample of 99 users  constituted of 60.6\% males and 39.39\% females, was predominantly White/Caucasian (60.6\%) and the majority (53.53\%) of the participants had a bachelor’s degree. Politically, 39.39\% of our participants were Democrats, 34.34\%  independents, and 26.26\%  Republicans. Based on the results of 2020 presidential elections\footnote{https://www.politico.com/2020-election/results/president/},  66.67\% of our participants lived in the blue states, 32.32\% in red while one individual resided in Puerto Rico\footnote{Puerto Rico is not considered a state but is considered an unincorporated territory of the United States}. We report additional participants' characteristics in Appendix \ref{charac}.

\subsection{Developing data annotation scheme} \label{anno}
Developing the qualitative coding scheme to label YouTube videos for election misinformation was hard and time-consuming, requiring four rounds of discussions and 
consultation with an expert to reach a consensus on the annotation heuristics. In the first round,  the first author and an undergraduate research assistant   sampled 196 YouTube videos from  Abilov et al's YouTube dataset \cite{Abilov} and separately annotated the videos. They considered prior work on election misinformation narratives \cite{eip} and YouTube content policy \cite{Election56:online} as  references to identify election misinformation, and came up with an initial annotation scale and heuristics to classify videos. Then they came together to reach a consensus on the annotation values. However, even after multiple rounds of discussions,  annotations  diverged for 33.6\% of the videos. We then conducted additional rounds of annotation exercises with seven researchers, out of which five had  extensive work experience on online misinformation. 
In every round,  researchers independently annotated 15 videos 
and later discussed every video's annotation value and the researchers’ annotation process. 
We also reached out to a postdoctoral researcher who has extensive research experience on online multi-modal  election misinformation for feedback. Based on the insights provided by the external researchers and postdoc, we  refined the  annotation criteria and heuristics \footnote{It is important to note that all annotators and the post-doctoral researcher are left and center-left leaning individuals which may have affected how the content of YouTube videos was perceived and how the annotation heuristics were developed.}. Below we describe the annotation guidelines and  heuristics in detail.  
% For the second round, we  presented the refined annotation guidelines  developed using the insights of the post-doctoral researcher as a reference to the seven researchers and asked them to use the criteria to annotate a new set of  YouTube videos. 
% After three rounds, we were able to finalize the annotation scheme

\subsubsection{Annotation guidelines} In order to annotate a YouTube video, the annotators were required to go through several fields
present on the video  page in the following order: title and description, the overall premise of the video which could be determined by going through the video transcript or watching the video content, and considering channel bias. We encouraged participants to  perform an online search to gain more contextual information  about 
 events or individuals discussed in the video that they were unaware of. This strategy is grounded in the lateral reading technique that is often used by fact-checkers  for credibility assessments \cite{wineburg2017lateral}. Note that we did not ask participants to consider  video comments for the annotations because we found during our annotation exercises that comments could be misleading. For example,  video  \textit{Dominion Voting Systems representative demonstrates voting machines} (Q7kPSzYsR6Y) contains a demonstration of dominion voting machines, however, the comments indicate the video to be supporting misinformation.

\subsubsection{Annotation heuristics} In this section, we describe our annotation scale and heuristics.

\noindent\textbf{Supporting election misinformation (1)}: This category includes YouTube videos that  support or provide evidence for  misleading narratives around the presidential elections. 
% We also include videos that  exaggerate real incidents to support a misinformation narrative of widespread voter fraud. 
% For example, a real-life mail dumping incident could be presented in a way to indicate a widespread pattern of mail-dumping behaviour during 2020 presidential elections. 
We did not include videos showing incidents of mail dumping, destroyed ballots, etc. in isolation. However, if the videos use these incidents to push a specific narrative/agenda like undermining confidence in mail-in voting, 
% or advancing claims that the 2020 presidential elections were rigged, 
then we considered them as supporting misinformation. 
We also considered live YouTube videos (live press conferences, court hearings, etc.) that highlighted voter fraud claims without giving any additional context in the title, description, or beginning of the video  as supporting misinformation. A few examples of videos in this category include \textit{NO RETREAT! America Is About To \#StopTheSteal | Good Morning \#MugClub} (Xqcwzi8Onsk) where video's title, description, and content hint towards massive voter fraud incidents in the US 2020          presidential elections and  \textit{LIVE: Trump Legal Team Presents CLEAR Evidence of Fraud Before Georgia Senate Committee 12/3/20} (e35f4pUIYOg) which contains live footage capturing the testimony of individuals claiming occurrence of voter fraud in 2020 presidential elections. The video's description, title,         and beginning  do not contain any statements questioning or contradicting the claims of widespread voter fraud.

\noindent\textbf{Neutral (0)}: We consider videos as neutral when they are related to the 2020 elections  but do not support or oppose false narratives surrounding the elections.   % considered as neutral. 
For example,  video \textit{WATCH: The first 2020 presidential debate} (w3KxBME7DpM) 
is considered neutral since it 
covers the first presidential debate of the  elections. 

\noindent\textbf{Opposing (-1)}: We annotate videos as opposing when they oppose or debunk the misinformation narratives behind the 2020 US presidential elections. We also include
satire videos making fun of the misinformative claims  in this category.
For example, video \textit{Trump Has Yet To Show Real Evidence Of Fraud, But Getting Him Out Of Office May Be A Bumpy Ride } (7mJwuKhfvqY) whose title and description indicate that Donald Trump made false claims of massive voter fraud.

\noindent\textbf{Other annotations}: We mark a video as  \textit{Irrelevant} (2) if its  content is not related to the presidential elections,
% For example, the video titled \textit{France: Voter turnout concerns loom over election, Macron remains clear favorite| International News} (zBeDA\_rl-4c) is considered irrelevant because it covers news about French presidential elections. 
% We mark a video 
as \textit{URL not accessible} (3) if the YouTube video was not accessible at the time of annotation and as \textit{Other languages} (4) when the content, title, or description of the YouTube video was in a language other than English.

\subsection{Classifying YouTube videos for election misinformation} \label{classifier}
Our crowd-sourced audit experiments resulted in $\sim$47K unique YouTube videos and 35 unique YouTube shorts\footnote{YouTube shorts are short YouTube videos with lengths equal to or less than 60 seconds}. Given a large number of videos, we scaled the annotation process using a machine learning classifier. In this section, we present our method of creating the ground truth dataset, a description of features used in our classification model, model architecture, and the results of our classification.

\subsubsection{Creating a ground truth dataset} Two researchers manually annotated 1196 videos using the guidelines and heuristics mentioned in Section \ref{anno}.
% We also realized at the time of analysis that Y videos originally annotated by the authors were removed from the YouTube platform. 
We obtained annotations for  545 additional videos using AMT. We describe the process of obtaining video annotations from AMT workers in Figure \ref{fig:amt} and Appendix \ref{amt}. Overall, in our ground truth dataset, we had 1741 videos out of which 124 are supporting\footnote{Out of these 67 videos were removed from the platform at the time of analysis.}, 257 opposing, 228 neutral, and 1132 irrelevant videos.

\subsubsection{Feature description} 
% Below we describe the features that we considered for developing our machine learning classifier. We also present the pre-processing steps for these features.
We considered the following features for our classifier.

\noindent\textbf{Snippet (title+description)}: We concatenated the title of the YouTube video with its description together, as done by \cite{papadamou2022just}, and used the concatenated string as a feature. \\
\noindent\textbf{Transcript}: Transcript contains the textual content of the  video. We use  transcripts auto-generated by YouTube.  \\
\noindent\textbf{Tags}: Video tags are words that a content creator associates with their video  while uploading it on the platform.\\
\noindent\textbf{Video  Statistics}: Video  statistics include the number of views, likes, comments, and  date of publication.\\
\noindent\textbf{Channel Bias}: Since the election misinformation is closely entangled with the political beliefs \cite{Theuniqu22:online,Republic22:online}, we used partisan bias of YouTube channels as a feature. Using existing data sets on media bias and  manual annotations (described in Appendix \ref{partisanbias}), we annotated YouTube channels' partisan bias on a 5-point scale of far-left to far-right.
% We discuss this process next. \ref{partisanbias}

% Combining existing data set on media bias with our manual annotations (described in Section \ref{partisan}), we annotated channel bias for our dataset on a 5-point scale of far-left to far-right.\\
% We discuss this process next. \\

% In the classifier selection discussed next, we by all we mean these four features together.
% Before we converted the text to various type of vectors, words vectors or sentence vectors, we cleaned and lemmatized them, if needed.
Apart from the features listed above, we also tried several other features like LIWC dictionary~\cite{tausczik2010psychological}, Credibility Cues~\cite{mitra2017parsimonious}, 
and hashtag matching from the Voter Fraud dataset on the text features~\cite{Abilov} 
that didn't improve performance. Therefore, we do not discuss them in detail.  Recall, while manually annotating the videos, we discovered that   comments are not a good indicator of the veracity of the video. Therefore, we chose not to include those in our feature set. 

\begin{scriptsize}
\begin{table*}[]
    \centering
    \small
    \begin{tabular}{p{0.6\textwidth}p{0.15\textwidth}p{0.15\textwidth}}
         \textbf{Classifier[Feature + Vectorizer + Imbalance Handling + Data]} & Acc. & F1 \\
         \hline
         SVM[Video Engagement Statistics] & 0.38 & 0.14 \\
         SVM[Snippet + FastText] & 0.61 & 0.56 \\
         SVM[Transcript + FastText] & 0.58 & 0.51 \\
         SVM[Tags + FastText] & 0.59 & 0.53 \\
         SVM[Snippet,Transcript,Tag + FastText] & 0.63 & 0.57 \\
         SVM[Snippet,Transcript,Tag + Count] & 0.65 & 0.58 \\
         SVM[Snippet,Transcript,Tag + TFIDF] & \textbf{0.71} & \textbf{0.65} \\
         \hline
         SVM[Snippet,Transcript,Tag,Channel Bias + Sentence Transformer] & 0.73 & 0.69 \\
         SVM[Snippet,Transcript,Tag,Channel Bias + TFIDF] & 0.74 & 0.70 \\
         SGD[Snippet,Transcript,Tag,Channel Bias + TFIDF] & 0.64 & 0.57 \\
         KNN[Snippet,Transcript,Tag,Channel Bias + TFIDF] & 0.61 & 0.58 \\
         XGB[Snippet,Transcript,Tag,Channel Bias + TFIDF] & 0.74 & 0.68 \\
         Voting SVM+SGD+KNN+XGB [Snippet,Transcript,Tag,Channel Bias + TFIDF] & \textbf{0.75} & \textbf{0.71}\\
         \hline
         SVM[Snippet,Tag,Channel Bias + TFIDF + SMOTE + Additional Training Data] & \textbf{0.91} & 0.90 \\
         XGB[Snippet,Tag,Channel Bias + TFIDF + SMOTE + Additional Training Data] & \textbf{0.91} & \textbf{0.91} \\
         \hline
    \end{tabular}
    \caption{A sample of classifiers and feature set with the performance progression.}
    \label{tab:classifier}
\end{table*}
\end{scriptsize}

\subsubsection{Classifier Selection}
To find a classifier that performs well on our dataset, we applied a series of machine learning classifiers on several combinations of feature sets.
% we performed a series of steps to extract several sets of features and apply a set of machine learning classifiers on those feature sets.
% we applied on our dataset.
% We  cleaned  and lemmatized our textual features.  
To create feature vectors, we tested two types of word vectors (count  and tf-idf vectors) and two types of sentence vectors (FastText \footnote{\url{https://fasttext.cc/}} and BERT \cite{devlin2018bert}). For word vector generation, we cleaned the dataset by removing stop words and lemmatization, followed by up to 3-gram generation. To deal with data imbalance in our dataset, we used
 Synthetic Minority Over-sampling Technique \cite{chawla2002smote}
% First, we varied the feature set by including different subsets of the features.
% Second, for text data, we converted them to different types of text vectors including two types of word vectors (count vectors and tf-idf vectors) and two types of sentence vectors (FastText and SOTA sentence transformers).
% For word vector generation, we cleaned the dataset by removing stopwords and lemmatizing them, followed by upto 3-gram generation.
% Although we tried adding features such as LIWC, Credibility Cues and VoterFraud hashtags, those features did not improve the accuracy.
% Third, due to significant imbalance in our dataset with more than 50\% irrelevant data, we also used SMOTE to generate balanced dataset for model training.
% To partially offset this issue, we added more training data annotated both by one of the authors and AMT. 
We applied several 
% Fourth, among various 
classifier models on our feature set including
% , we applied models including linear, 
support vector machine, stochastic gradient descent, decision trees, nearest neighbor, and ensemble  models.
To find the best model, we performed a grid search on a five-fold cross-validation dataset by looking into standard parameter space for each classifier.
For the sake of brevity, we only show a sample of combinations tested in Table \ref{tab:classifier}.
% we are not including details about all the combinations tested.
% Table \ref{tab:classifier} shows a sample set of classifiers and features that we tested.
Out of all the combinations, both SVM and XGBoost performed the best (ACC=91\%) when trained with snippet, tags, and channel bias features and tf-idf text vectorizer~\footnote{If we merge irrelevant and neutral videos into one class resulting in a three-class classification problem,  SVM classifier performs with a 93\% accuracy.}.
Based on Occam's Razor principle \cite{Occamsra87:online}, we selected SVM as the final classifier, i.e., the simplest model with maximum accuracy. Using our final classifier, we determined the annotation labels for the remaining videos.
% Combining together the human annotation and the classifier annotation, 
In total, our dataset consisted of  431 supporting, 1868  opposing, 1658 neutral,  and 43041 irrelevant videos.

\section{Ethical considerations} \label{ethics}
% In our \textit{crowdsourced audit} experiment, we hired crowd workers to collect  YouTube data from the platform in order to test the algorithmic system.  
 Our browser extension \textit{TubeCapture} uses crowd workers' YouTube account to watch videos (including videos containing election  misinformation) and conduct searches on the platform. It was possible that participants would have seen more misinformation than they would have
otherwise during and also after the research study due to the watch and search history built during the audit. In order to eliminate the potential harm of our experiments, we  included two essential
steps in our experimental design. First, our extension always opened the browser window in the background so that participants don't actively see the videos being played. Second, the extension  deleted users' search and watch history built during the study period. 
Note that YouTube allows the deletion of  items from the search and watch history for a specific date range.
YouTube’s website \cite{Vieworde17:online,Learnabo12:online} clearly states that ``\textit{search entries you delete will no longer influence your
recommendations. At any time you can (also) remove videos (from watch history) to influence what YouTube recommends to you}''. We  explicitly informed users that their YouTube history during the study period would be deleted. We  ensured that the extension expires after the study period so that it does not perform any action. In addition, we ensured that the YouTube pages saved by our extension do not contain users' personally identifiable information such as email addresses.

\section{RQ1 Results: Extent of  Personalization} \label{rq1}
To measure the extent of  personalization in YouTube components, we compare the personalized list of video URLs present in the standard window with the baseline unpersonalized videos obtained from the incognito window.  
% Scholars have argued that meaningful comparisons of ranked lists to determine personalization is non-trivial \cite{bandy2020auditing}. 
Below we discuss the  metrics that we used to quantify personalization. 
% Multiple quantitative metrics have been proposed and used in prior audit studies to measure the extent of personalization. 

\textbf{Measuring personalization in web search:}
In our study, to determine personalization in search results, we employ two metrics: jaccard index and rank bias overlap (RBO). Jaccard index measures the similarity between two lists and has been used in several previous audit studies to measure personalization in web search \cite{kliman2015location,hannak2013measuring,juneja2021auditing}. However, Jaccard index does not take into account the rank of the lists being compared. Thus, we used the RBO metric introduced by Webber et al \cite{webber2010similarity} which takes into account the order of elements in the list. The RBO function includes a parameter \textit{p} which indicates the top-weightedness of the metric, i.e. how much will the metric penalize the difference in the top rankings. A previous  audit study used the click-through rate (CTR) of Google search results to estimate the value of \textit{p} \cite{robertson2018auditing}. Because of the lack of CTR statistics available for YouTube, we consider the default value of \textit{p} which is 1 (prior audit studies such as \cite{le2022crowdsourcing} opted for a similar approach), indicating that  differences in all rankings are equally penalized. Both jaccard and RBO scores range between 0 and 1, with 1 indicating that the two lists have similar elements while 0 indicating that the lists are completely different. 

\textbf{Measuring personalization in up-next trails:} 
To measure personalization in up-next trails, we employ jaccard index and Damerau-Levenshtein (DL) distance \cite{damerau1964technique}. DL distance is the enhanced version of edit distance that computes the number of transpositions in addition to insertions, deletions, and substitutions required to make the treatment list identical to the control list. %We use python's pypi implementation of 
DL distance has been used by prior audit work as a metric to estimate the ranking differences between two lists  \cite{c2020there}. It returns a score from 0  to 1 (identical lists) indicating how similar the two lists are.  We refrain from using the RBO metric to determine personalization in up-next trails because RBO is suitable for indefinite lists while the trails collected through our experiments have a known maximum length of five. We also refrain from using the Kendall tau metric since it requires the two ranked lists being compared to be conjoint\footnote{There are alternative versions of Kendall Tau that assume the dissimilar elements to be present at the end of the list. However, conceptually, the metric does not fit our collected trail data.}. Given, jaccard, RBO, and DL distance return similarity values, we define personalization as:-
\begin{equation}
1-similarity\_metric(URL_{incognito}, URL_{standard}).
\end{equation}

\subsection{RQ1a: Personalization in search results}

\begin{figure*}
  \begin{minipage}{\linewidth}
  \begin{subfigure}{0.32\textwidth}
    \centering
    \includegraphics[width=\textwidth]{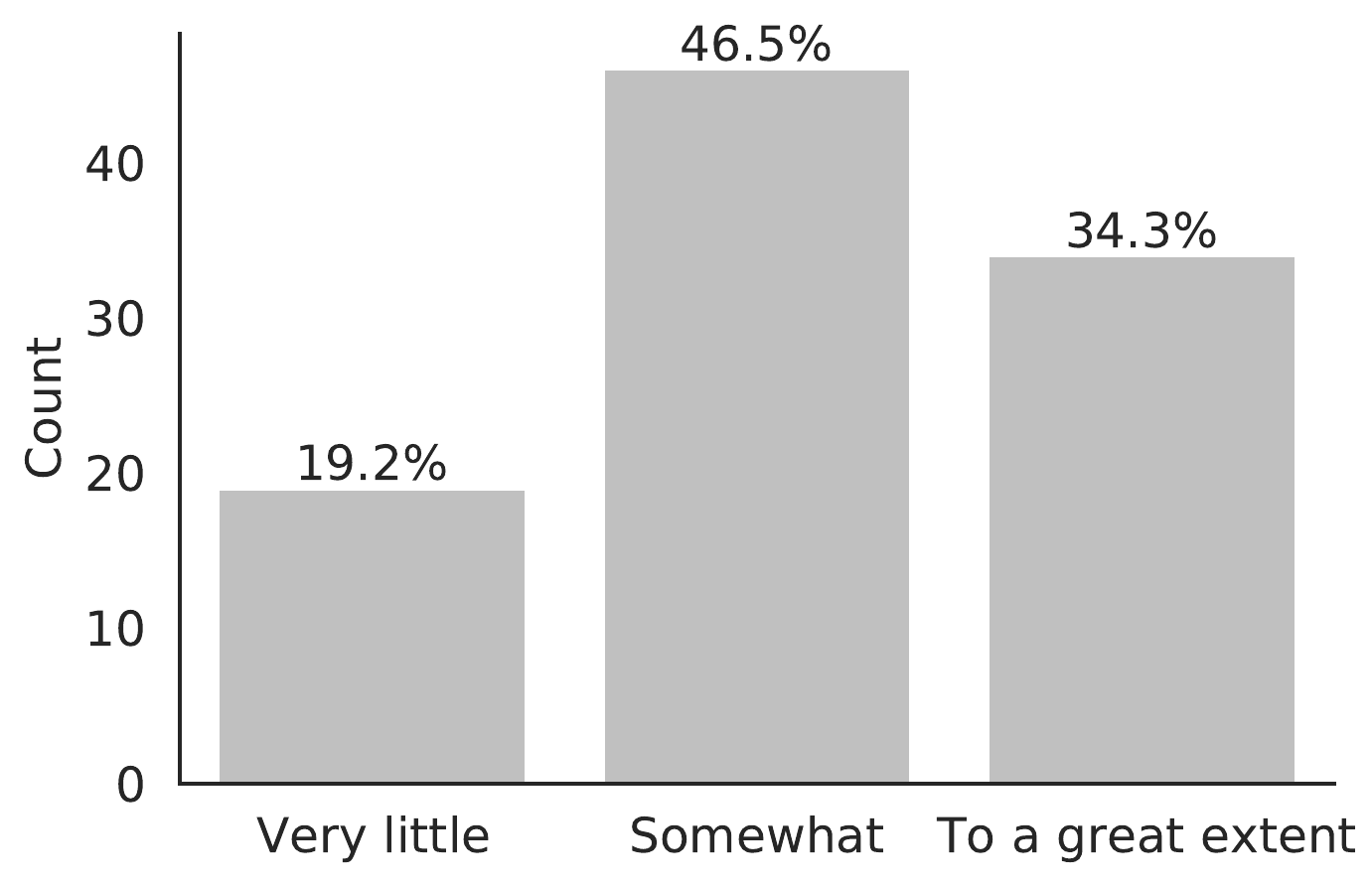}
    \caption{Participant's belief in extent of personalization in YouTube search results}\label{fig:persbelief_search}
  \end{subfigure}\hfill
  \begin{subfigure}{0.32\textwidth}
    \centering
    \includegraphics[width=\textwidth]{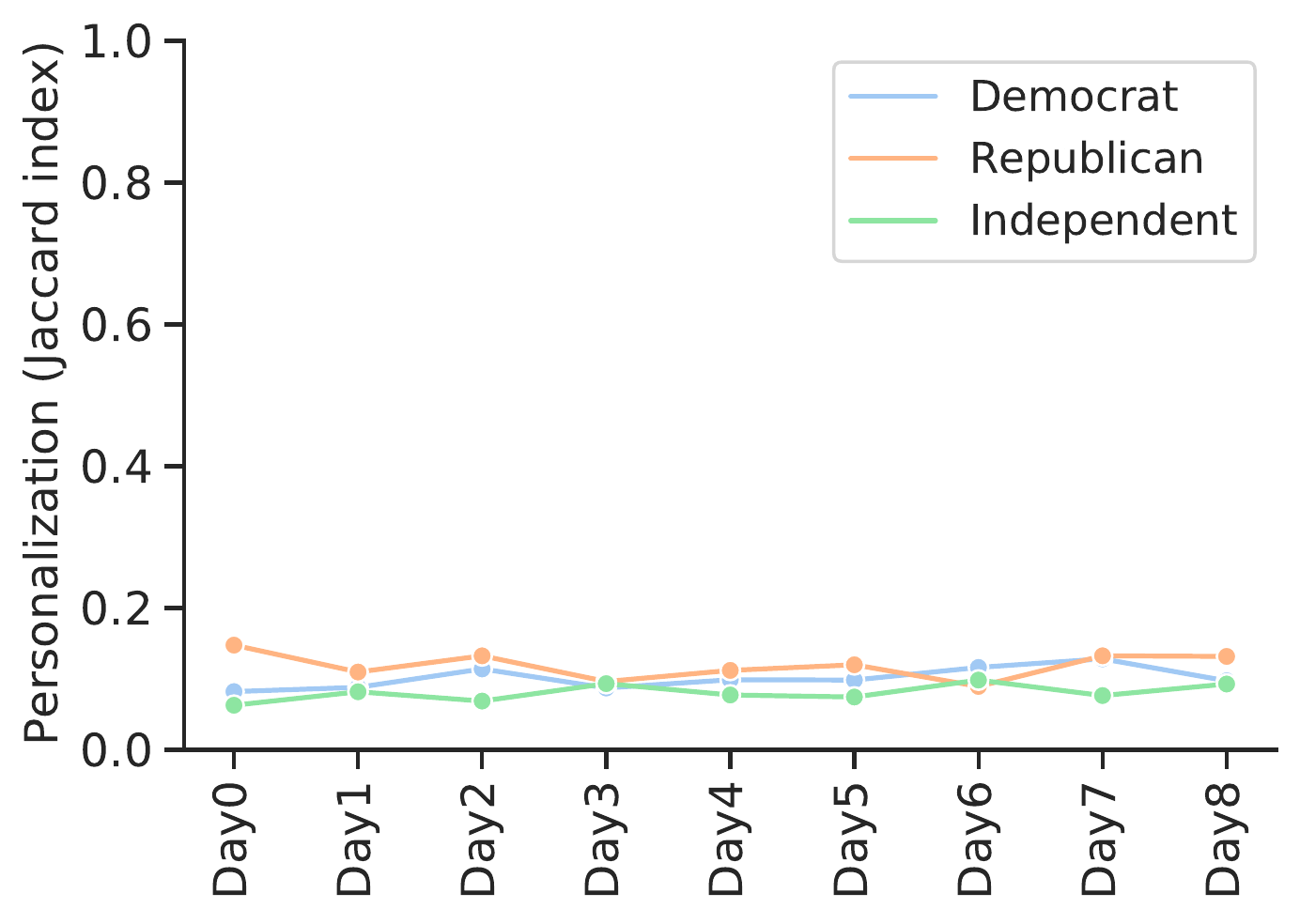}
    \caption{Measuring extent of personalization in SERPs using jaccard index}\label{fig:serpjac}
  \end{subfigure}\hfill
  \begin{subfigure}{0.32\textwidth}
    \centering
    \includegraphics[width=\textwidth]{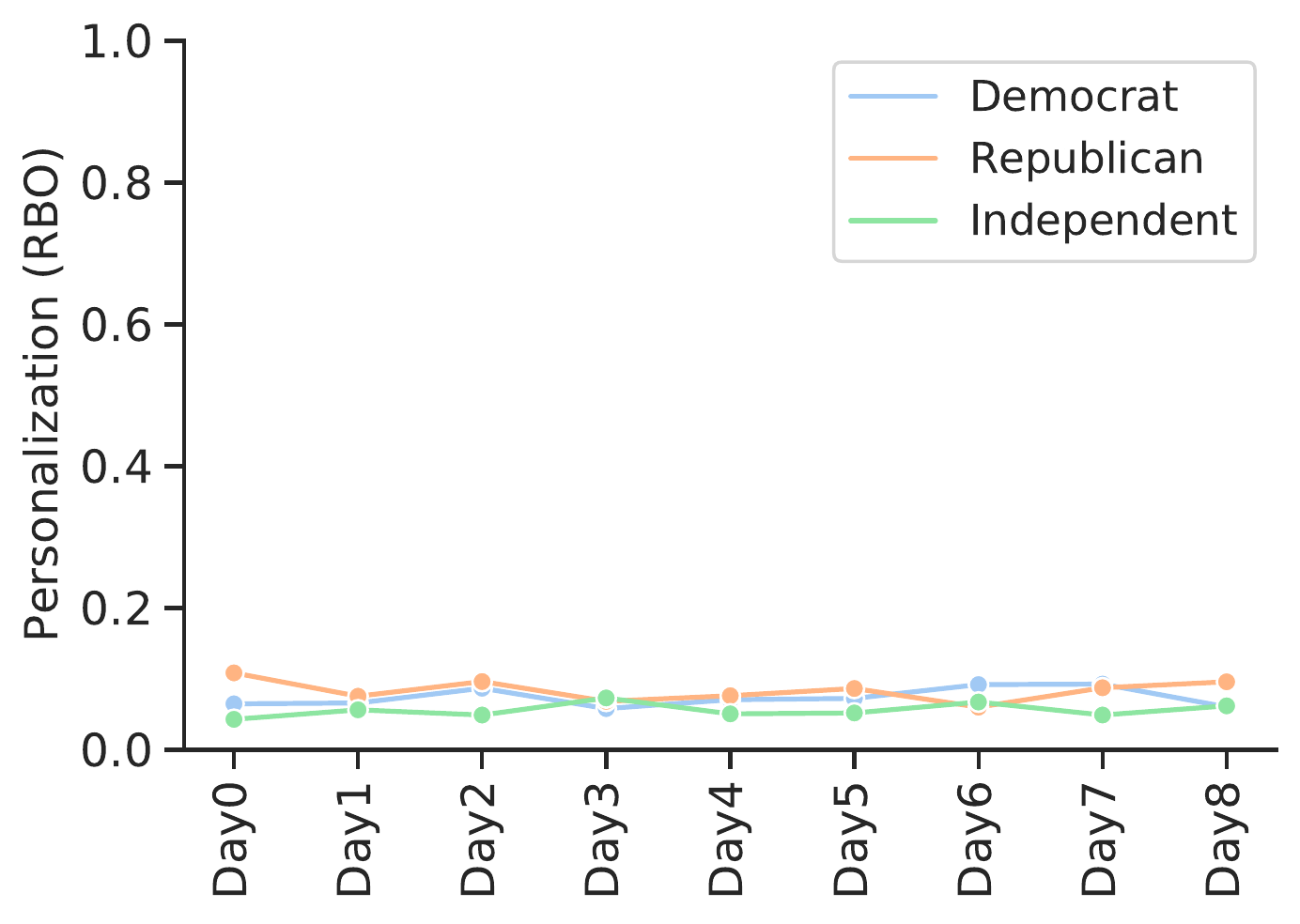}
    \caption{Measuring extent of personalization in SERPs using RBO}\label{fig:serprbo}
  \end{subfigure}
  \end{minipage}
 
  \caption{\textbf{RQ1a results:} Figure  (a) shows participants' response to the survey question: ``How much, if at all, do you think YouTube personalizes search results''. Figures  (b) and (c) show personalization calculated via  jaccard index values and RBO metric values respectively in YouTube's standard-incognito SERP pairs. We observe that search results are slightly personalized meaning  search results obtained from standard windows are very similar to the search results obtained from incognito windows.}
 \label{search}
 \Description{Figure  (a) shows participants' response to the survey question: ``How much, if at all, do you think YouTube personalizes search results''. 19.2\% believe there is little  personalization in YouTube search results, 46.5\% believe that search results are somewhat personalized, and 34.3\%  believe that YouTube personalizes
search results to a great extent. Figures (b) and (c) show personalization calculated via  jaccard index values and RBO metric values respectively in YouTube's standard-incognito SERP pairs for democrats, republicans, and democrats. The figures are line graphs with the x-axis showing the nine days of the experiment run (day0-day8). The y-axis shows the magnitude of personalization. The magnitude of personalization is very low (near 0) for all days for all users. }
\end{figure*}

% \subsubsection{User's belief in personalization of search results}
When asked in our study survey how much YouTube personalizes search results (Figure \ref{fig:persbelief_search}), 34.34\% believed YouTube personalizes search results to a great extent while 19.19\% believed  the extent of personalization to be very little.
% We found that the belief in extent of personalization does not have statistically significant association with participant's demographics, YouTube usage or political affiliation. 
On quantitatively measuring the extent of personalization in YouTube search results, we found little to no personalization indicating that search results present in standard and incognito windows are highly similar. Figures \ref{fig:serpjac} and \ref{fig:serprbo} show the extent of personalization in SERPs calculated using  jaccard index and RBO metric respectively for democrats, republicans, and independents for each day of the experiment run. 
% High values for both the metrics indicate that the search results present in standard and incognito windows are highly similar. 
We did not find any significant difference in the personalization values of SERPs for participants with respect to their political leaning.

% \subsubsection{Personalization by political preference}
% \subsubsection{Personalization by states}
% \subsubsection{Personalization by extent of watching election news on YouTube}
% \subsubsection{Personalization by search queries}
% \subsubsection{Personalization by belief in election misinformation}

% \begin{figure*}[t]
%     \centering
%     \includegraphics[width=0.3\textwidth,keepaspectratio]{RQ1/overlap.png}
%     \caption{Figure showing the distribution of percentage of YouTube videos recommended to our study participants from their subscribed channels.}
%     \label{fig:overlap}
% \end{figure*}

\begin{figure*}
%   \begin{minipage}{\linewidth}
  \begin{subfigure}{0.34\textwidth}
    \centering
    \includegraphics[width=\textwidth]{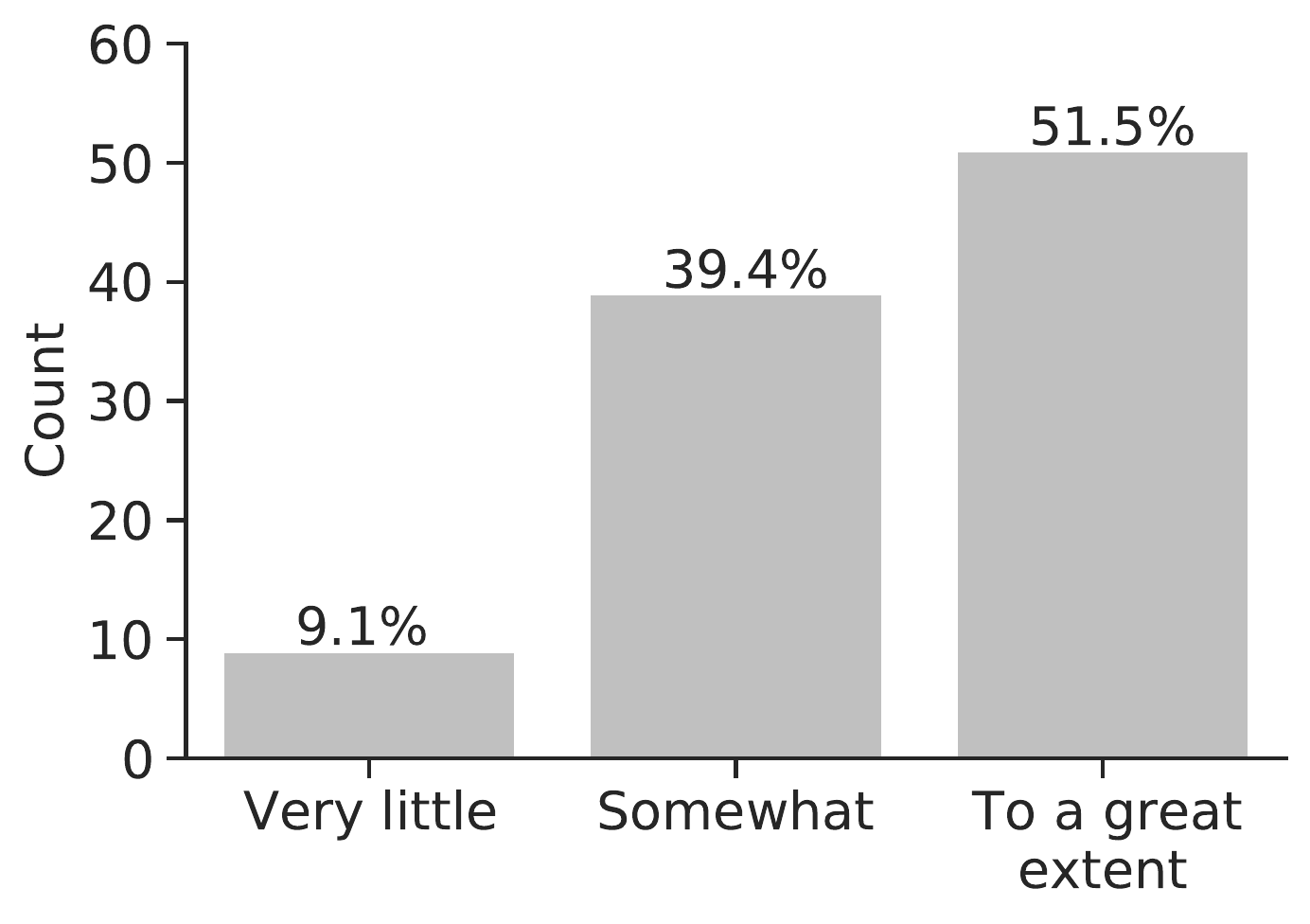}
    \caption{Participant's belief in extent of personalization in YouTube up-next recommendations}\label{fig:persbelief_trail}
  \end{subfigure} 
  \hspace{2cm}
  \begin{subfigure}{0.34\textwidth}
    \centering
    \includegraphics[width=\textwidth]{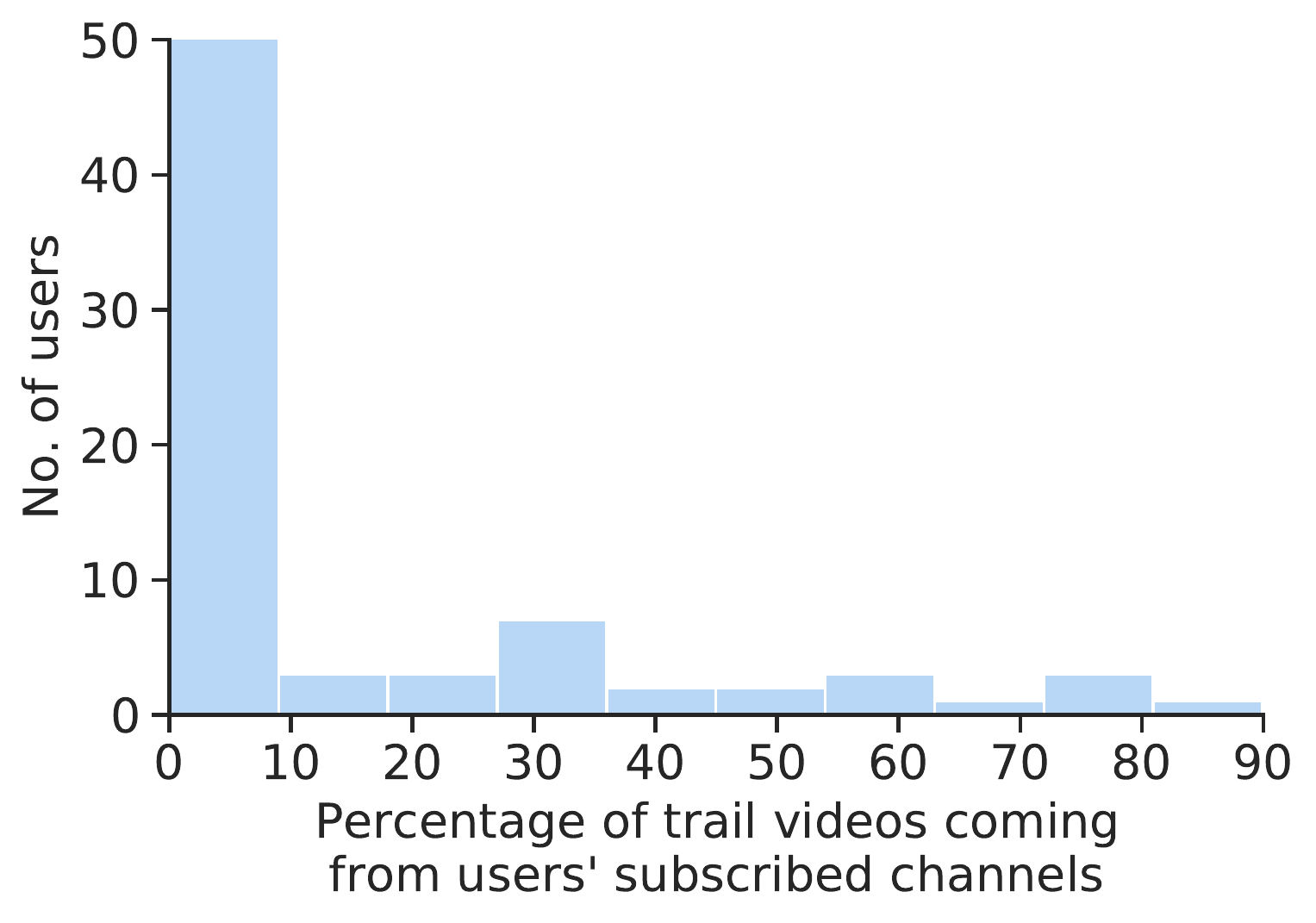}
    \caption{Distribution of percentage of up-next  video recommendations coming from users' subscribed channels. }\label{fig:overlap}
  \end{subfigure}\\
  \begin{subfigure}{0.34\textwidth}
    \centering
    \includegraphics[width=\textwidth]{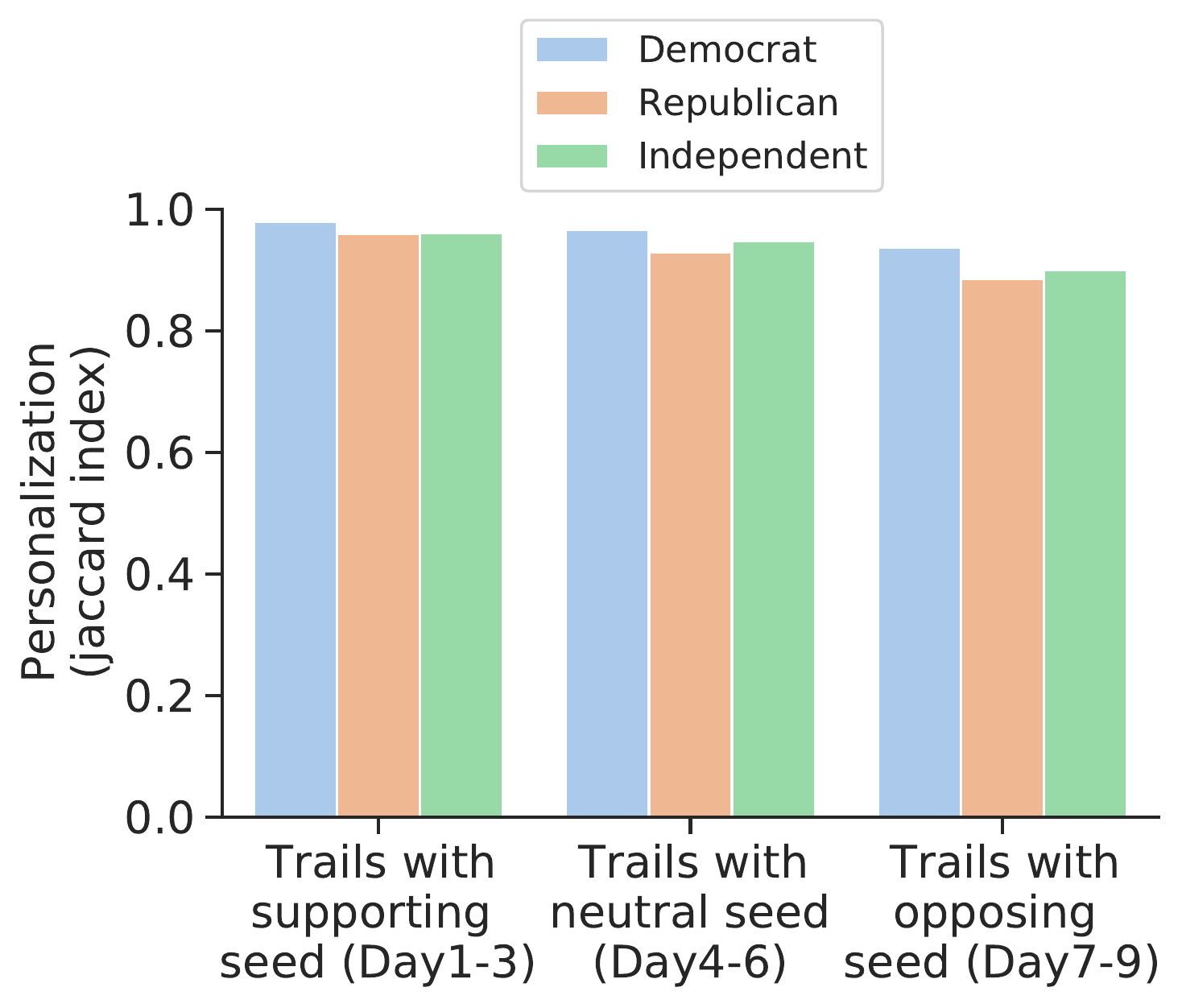}
    \caption{Measuring extent of personalization using jaccard index}\label{fig:pers_trail_jc}
  \end{subfigure}
  \hspace{2cm}
  \begin{subfigure}{0.34\textwidth}
    \centering
    \includegraphics[width=\textwidth]{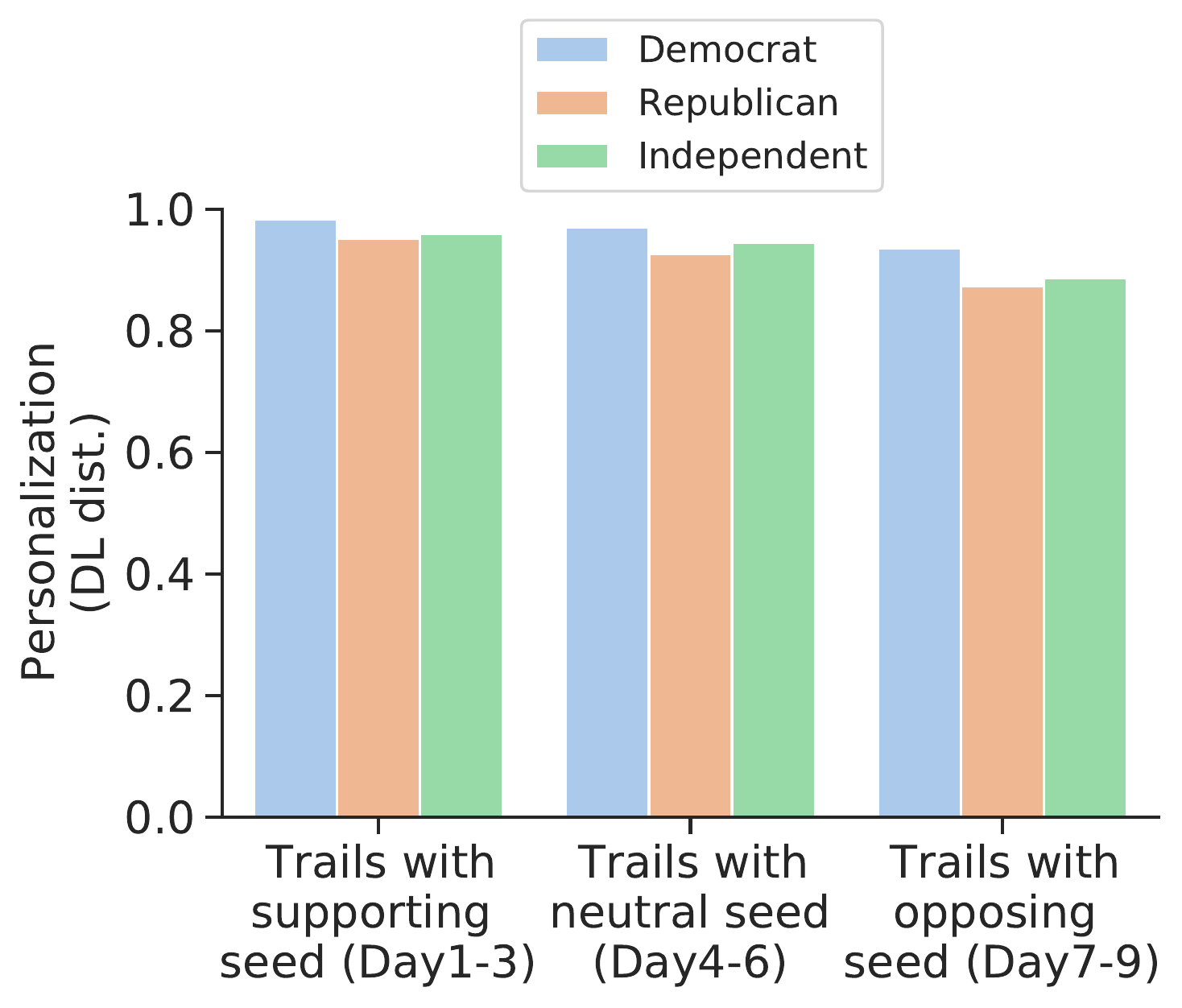}
    \caption{Measuring extent of personalization using DL index}
    \label{fig:pers_trail_edit}
  \end{subfigure}
%   \end{minipage}
 
  \caption{\textbf{RQ1b results:} Figure  (a) shows participants' response to the survey question: ``How much, if at all, do you think YouTube personalizes up-next recommendations''. Figure (b) shows the  distribution of the percentage of YouTube videos recommended to our study participants from their subscribed channels. Figures  (c) and (d) show personalization calculated via  jaccard index values and DL distance metric values respectively in YouTube's standard-incognito up-next trails pairs. We observe that up-next recommendation trails are highly personalized. }
 \label{trail}
 \Description{Figure  (a) shows participants' response to the survey question: ``How much, if at all, do you think YouTube personalizes up-next recommendations''. 9.1\% believe there is little  personalization in YouTube's up-next recommendations, 39.4\% believe that up-next recommendations are somewhat personalized, and 51.5\%  believe that YouTube personalizes
up-next recommendations to a great extent. Figure (b) shows the distribution
of the percentage of videos recommended to our participants in
up-next trails that are coming from their subscribed channels. For around 50\% of users,  10\% or fewer videos come from their subscribed channels. Figures (c) and (d) show personalization calculated
via jaccard index values and DL distance metric values respectively in YouTube’s standard-incognito up-next trails pairs. The personalization value is very high (between 0.8-1) for  the up-next trails collected from all the users.}
\end{figure*}

\subsection{RQ1b: Personalization in up-next trails}
When asked how much YouTube personalizes up-next recommendations, 51.5\%  of participants believed that YouTube personalizes up-next recommendations to a great extent (refer Figure \ref{fig:persbelief_trail}). 
% The belief in personalization is statistically independent of the participants' demographics (age, gender, education, etc.) and political affiliation. 
The quantitative measurements are in line with this belief showing that up-next trails are highly personalized. Figures \ref{fig:pers_trail_jc} and \ref{fig:pers_trail_edit} show the extent of personalization in up-next trails using jaccard index and DL distance. The graphs indicate that the  up-next trails obtained from the users' standard and incognito windows are highly dissimilar and thus, highly personalized. Statistical test revealed that the amount of personalization in trails with supporting, neutral, and opposing seeds is significantly different [F(2)=15.2, p<0.0001]. Post hoc test revealed that up-next trails with seed videos  opposing misinformation have lesser personalization (higher  jaccard index\footnote{The jaccard index values obtained were highly correlated with DL distance scores (pearson correlation coefficient = 0.96). Thus, we used jaccard index values to perform the statistical test.}) when compared with up-next trails with supporting and neutral   seed videos.

Next, we checked the influence of users' subscriptions on personalized trails. 81 (out of 99) participants had  subscribed to at least one YouTube channel (mean=109.4, median=31, SD=207.8).
The maximum number of subscriptions for a participant was 1073 and the minimum was 1. The participants had subscribed to 7670 unique channels out of which 79  either did not exist or were suspended due to violation of YouTube's moderation policy and thus, we did not consider these channels for analysis. To determine how many video recommendations in users' up-next trails were coming from their subscriptions, first, for each user we extracted the unique videos recommended in all the up-next trails collected for the user. Then we filtered and calculated the number of videos  coming from the users' subscribed channels. Figure \ref{fig:overlap} shows the distribution of the percentage of videos recommended to our participants in up-next trails that are coming from their subscribed channels. This percentage value is moderately correlated with the number of channels subscribed (r=0.61) and highly correlated with the number of news-related channels subscribed\footnote{To get a rough estimate of YouTube channels that broadcast news, we considered the news sources from \texttt{mediabiasfactcheck.com} and \texttt{allsides.com}. Additionally, we extracted the description of each channel and categorized it as a news channel if the description contained terms such as `breaking news', `politic*', `current affairs', `government', `national tv', `national news', `international news', `world news', `global news', `current affairs',  `wall street' etc. These terms were curated by the first author after manually going through the description of ~50 national and regional news channels on YouTube. We found that 44 users had subscribed to news and politics-related channels.}(r=0.71). 

\section{RQ2 Results: Amount of Misinformation} \label{rq2}
\begin{figure*}[]
  \begin{subfigure}[b]{0.4\textwidth}
    \includegraphics[width=\textwidth]{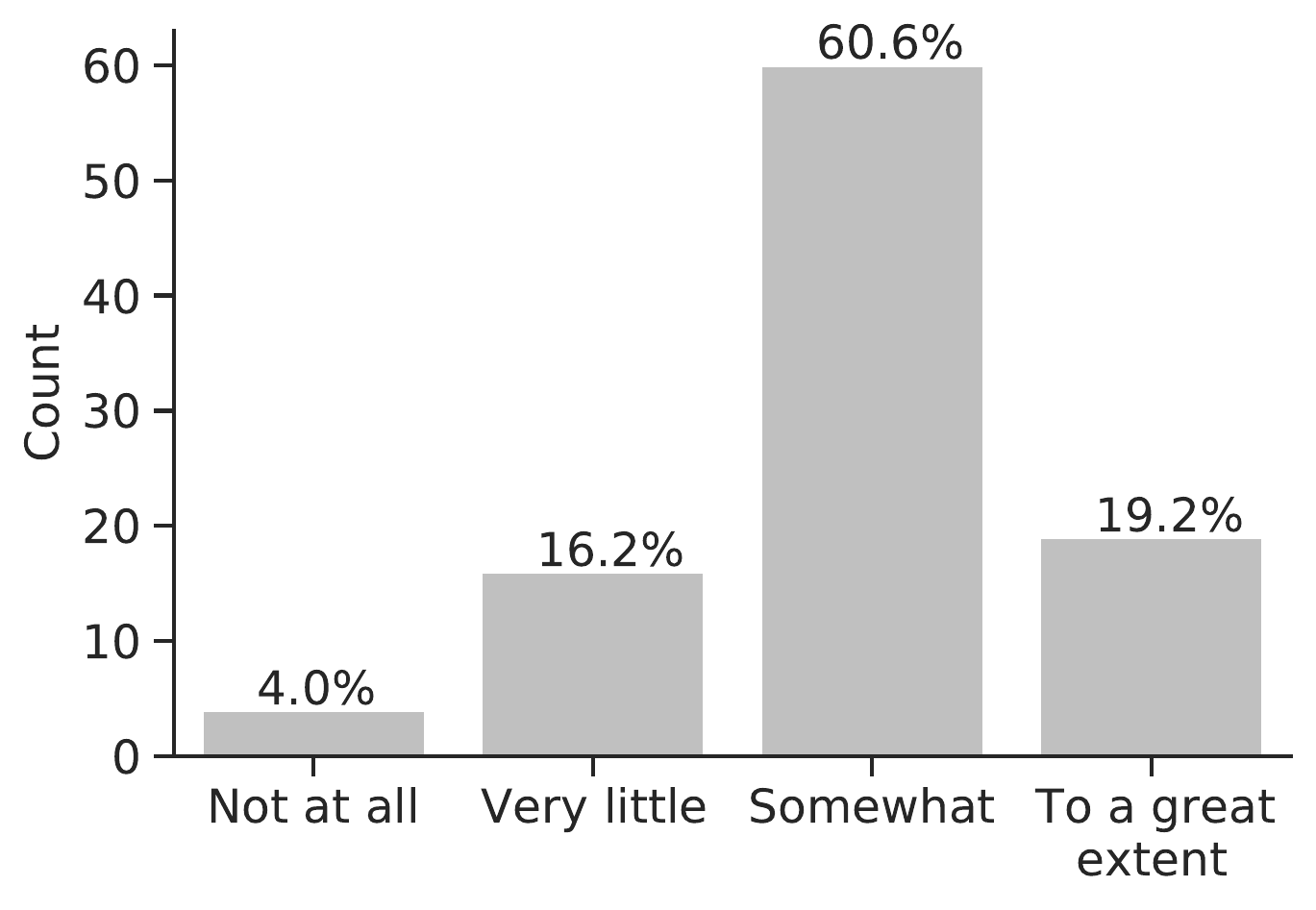}
    \caption{Participant's trust in the credibility of information presented  in search results}
    \label{cred_serach}
  \end{subfigure}
  \hfill
  \begin{subfigure}[b]{0.4\textwidth}
    \includegraphics[width=\textwidth]{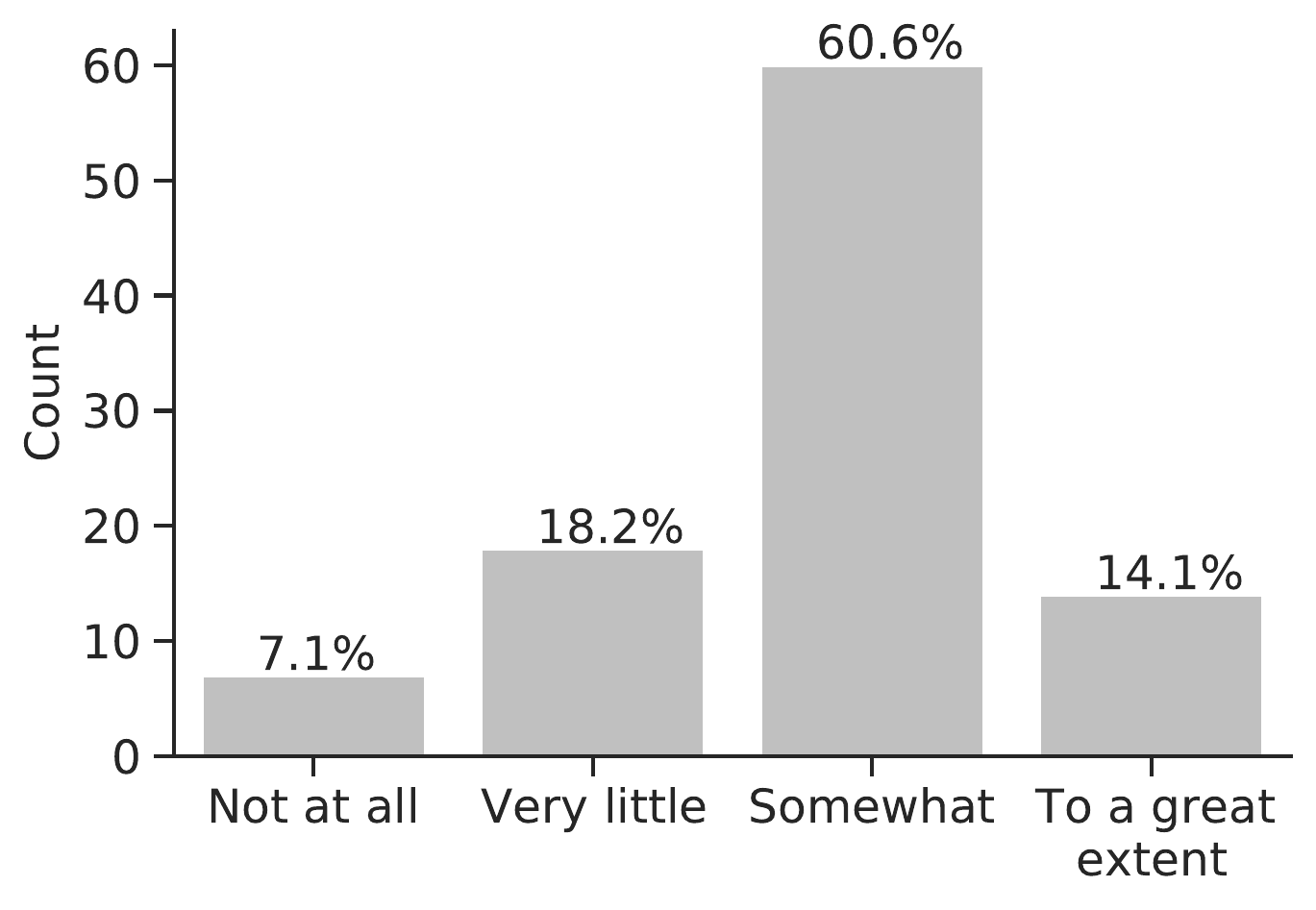}
    \caption{Participant's trust in the credibility of information presented  in up-next recommendations}
    \label{cred_recom}
  \end{subfigure}
  \caption{\textbf{RQ2:} Figure  showing participants' response to survey question: ``How much do you trust the credibility of information present in the '' a) search results and b) up-next videos recommended by YouTube.}
  \label{cred_bel}
  \Description{Figure (a) shows participants' responses to the survey question: How much do you trust the credibility of information present in the search results? 4\% say not al all, 16.2\% say very little. 60.6\% say somewhat, and 19.2\% say to a great extent. Figure (b) shows participants' responses to the survey question: How much do you trust the credibility of information present in the up-next videos recommended by YouTube? 7.1\% say not al all, 18.2\% say very little. 60.6\% say somewhat, and 14.1\% say to a great extent.}
  % \vspace{-2cm}
\end{figure*}

When asked how much do participants trust the credibility of videos in search results and recommendations, less than 20\% reported that they trust the credibility of content shown to them by YouTube to a great extent (Figure \ref{cred_bel}). To determine how much credible information is presented by YouTube to users in reality, we quantify the misinformation present in the YouTube components by adopting the misinformation bias score developed by Hussein and Juneja et al \cite{hussein2020measuring}. 
% which was originally developed to calculate  misinformation bias in SERPs. We adopt this metric 
% to quantify the amount of  misinformation  in both SERPs and up-next trails. 
The  score determines the misinformation in ranked lists and is calculated as $\frac{\sum_{r=1}^{n} {(x_r * (n - r+1))} }{ \frac{n * (n + 1)}{2}}$; where x is the video annotation,  $r$ is  rank of the video,  and $n$ is the total number of videos  present in the SERP/up-next trail. To conform to the video annotation scale in \cite{hussein2020measuring}, we map our annotation values to a normalized scale of -1, 0, and 1. We assign scores of -1 and 1 to videos opposing and supporting election misinformation respectively. Videos marked as irrelevant, neutral, belonging to a non-English language, or removed from the platform are assigned a 0 score. Thus, the misinformation bias score of a SERP/trail  is a continuous value ranging between -1 (all videos  are opposing election misinformation) to +1 (all videos  are supporting election misinformation). Note that a positive score indicates a lean towards misinformation, while a negative score indicates a lean towards content opposing misinformation. For analysis, we consider the top ten
search results  and  five consecutive videos in the up-next  trails.

\subsection{RQ2a: Misinformation in search results}
\begin{figure*}[]
\begin{minipage}{0.99\textwidth}
  \begin{minipage}[]{0.5\textwidth}
    \centering
    \includegraphics[width=0.7\textwidth,keepaspectratio]{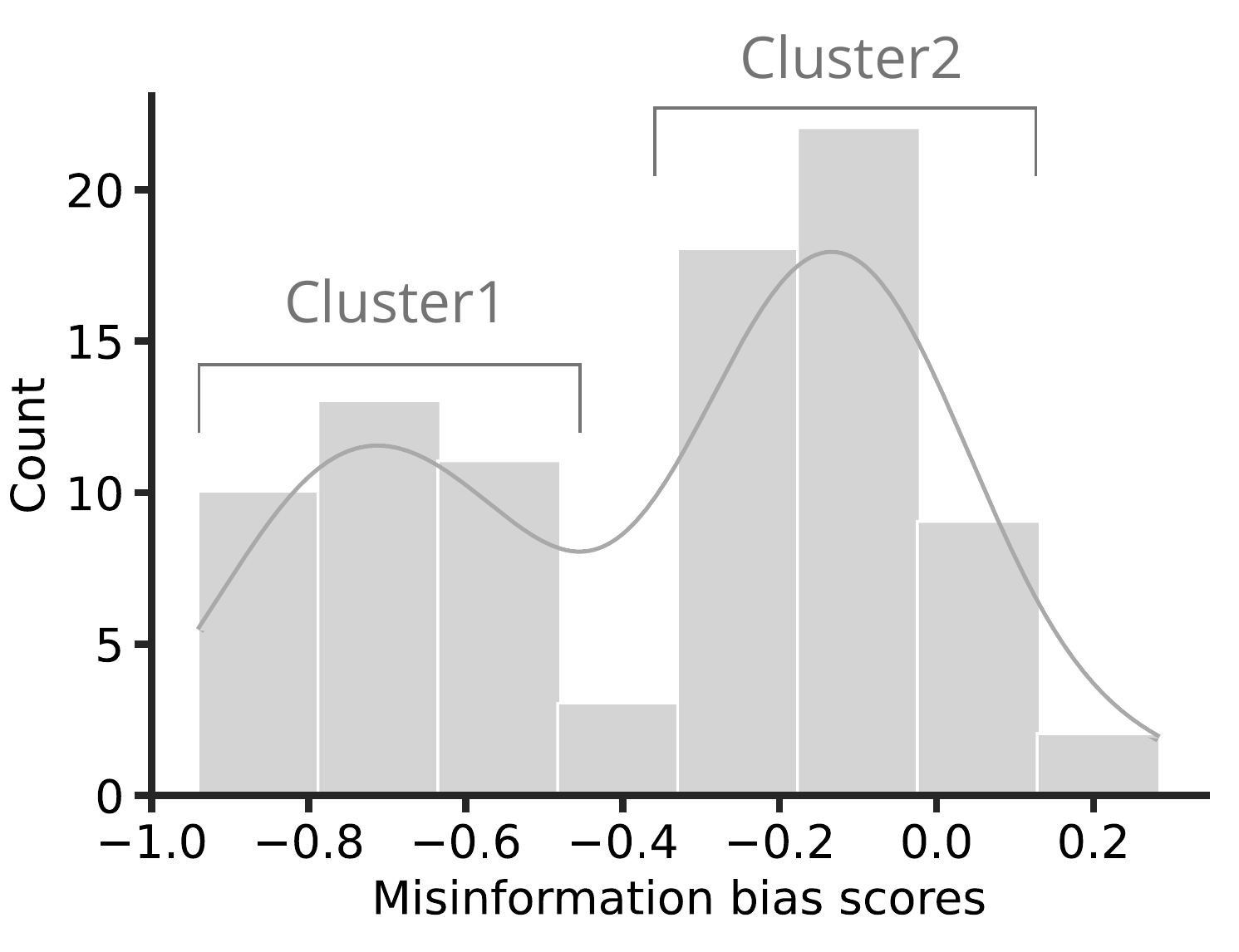}
    \captionof{figure}{\textbf{RQ2a results:} Mean misinformation bias scores for   88 search queries for all  participants. A  negative score indicates that SERPs contain more videos opposing election misinformation.}
    \label{dist}
    \Description{The figure shows the distribution of misinformation bias scores for   88 search queries for all  participants. The majority of search results have a negative score indicating that SERPs contain more videos opposing election misinformation.}
  \end{minipage}\hfill
    \begin{minipage}[]{0.45\textwidth}
    \centering
\small
\begin{tabular}{p{6.5cm}}

\hline
\rowcolor[HTML]{D5D3D3} 
\textbf{Cluster1: Search queries containing keyword fraud in conjunction with keywords voter, election, and dominion} \\ \hline
voter fraud evidence, dominion voter machine scandal, sharpie voter fraud, election fraud 2020, election fraud whistleblower \\ \hline
\rowcolor[HTML]{D5D3D3} 
\textbf{Cluster2: Search queries containing keywords election, and 2020} \\ \hline
trump biden general election, presidential election 2020, presidential election results 2020, mail in ballots 2020 \\ \hline
\end{tabular}
% \end{footnotesize}
% \end{table}
      \captionof{table}{The misinformation bias scores form a bimodal distribution, each constituting  a cluster of similar queries. This table describes the clusters and presents sample queries for each cluster.}
\label{tab:clusters}
    \end{minipage}
  \end{minipage}
\end{figure*}

The results of RQ1 showed that YouTube's SERPs are very slightly personalized  suggesting that search results present in the standard and incognito windows are mostly similar. Therefore, to quantify the misinformation bias in SERPs we only consider the  SERPs obtained from the standard YouTube windows of all the participants. We first calculated the average misinformation bias score for each of the 88 search queries for 9 days of the experiment run across all 99 participants. Figure \ref{dist} shows the distribution of misinformation bias scores for all the  search queries. We observe that the average misinformation bias scores of 84 (out of 88) search queries are negative indicating that  the search results contain more videos that oppose election misinformation as compared to videos supporting election misinformation\footnote{Only four search queries in our query set  (`stop the seal', `voting machine fraud', `ballots in garbage' and `ballots thrown out') have a positive misinformation bias. }. 
% Only four search queries (`stop the seal', `voting machine fraud', `ballots in garbage' and `ballots thrown out') have a positive misinformation bias while one search query (`us elections 2020 pennsylvania') has a bias value of zero. 

\begin{figure*}[]
  \begin{subfigure}[b]{0.45\textwidth}
    \includegraphics[width=\textwidth]{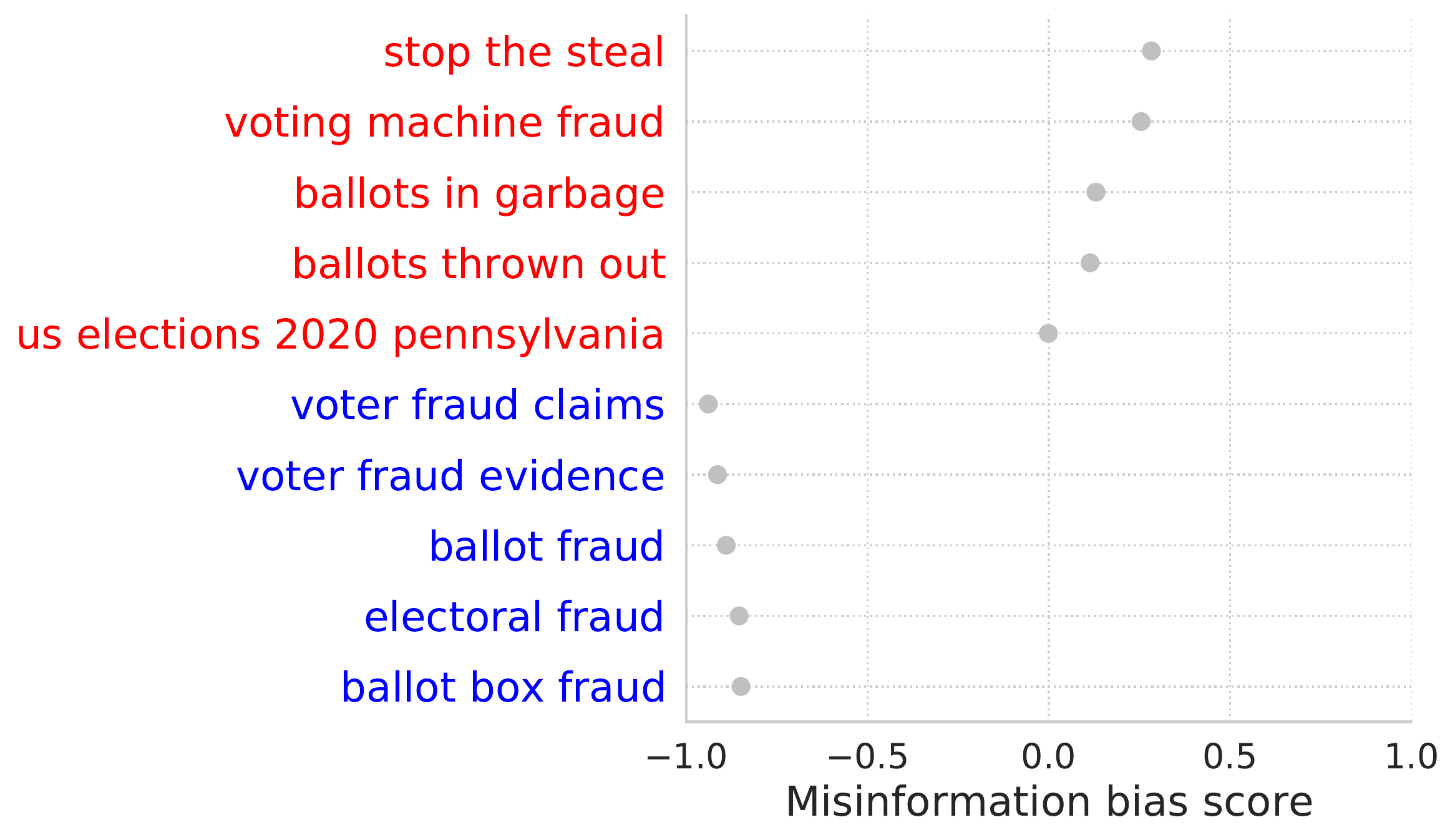}
    \caption{Search queries with highest and lowest  mean misinformation bias scores}
    \label{rank}
    \Description{The figure shows search queries with the highest (stop the seal, voting machine fraud, ballots in garbage, ballots thrown out, us elections 2020 pennsylvania) and lowest (voter fraud claims, voter fraud evidence, ballot fraud, electoral fraud, ballot box fraud) mean misinformation bias scores}
  \end{subfigure}
  % \hfill
  \hspace{1cm}
  \begin{subfigure}[b]{0.35\textwidth}
    \includegraphics[width=\textwidth]{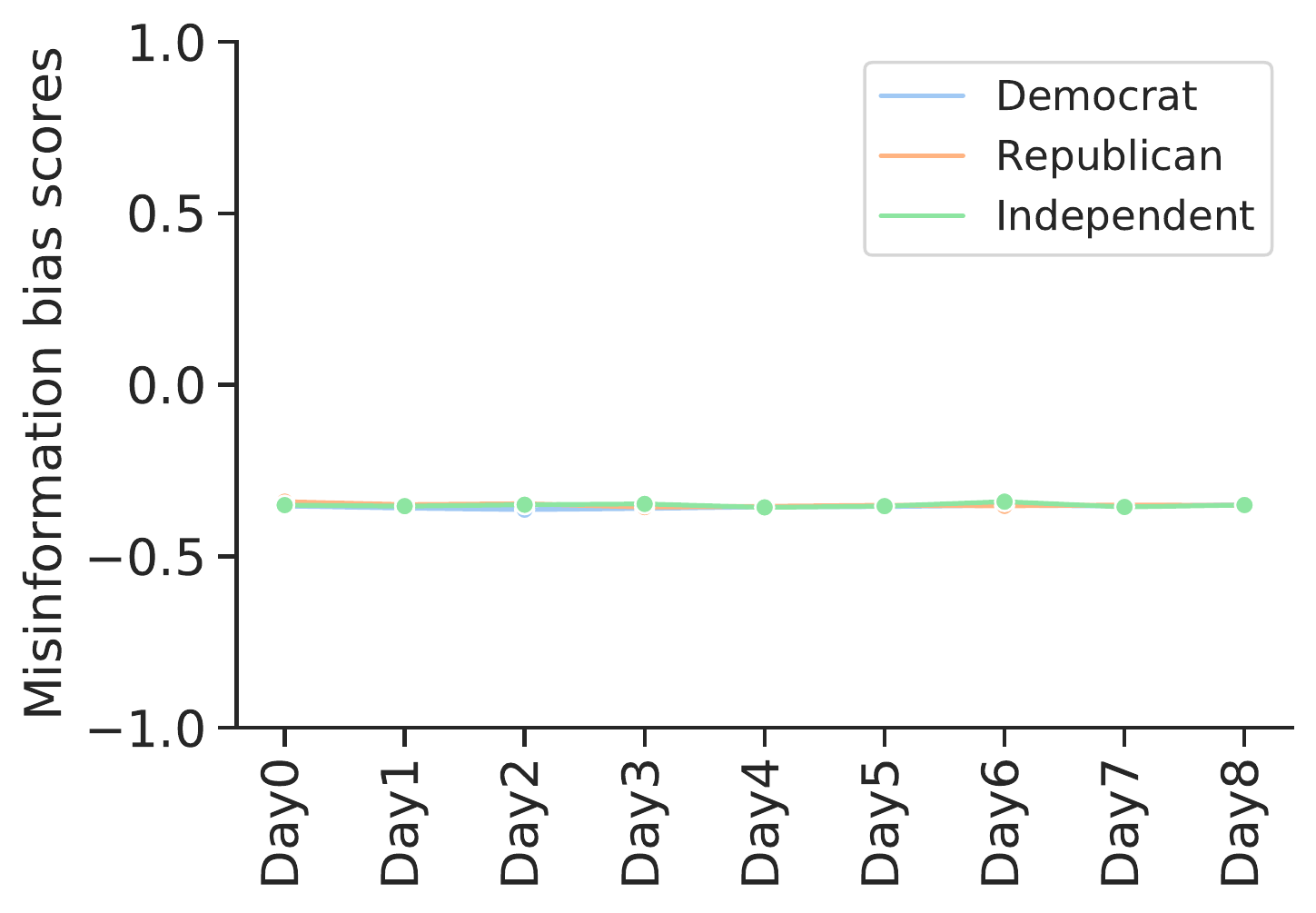}
    \caption{Misinformation bias scores of search queries for each day of experiment run}
    \label{fig:irdbias}
     \Description{ Figure shows  the misinformation bias scores (y-axis) for the participants belonging to the different
political leanings for the days of the experiment run (x-axis).  The bias scores coincide for all users on all days.}
  \end{subfigure}
  \caption{\textbf{RQ2a results:} a) Search queries with highest (labeled in red) and lowest (labeled in blue) mean misinformation bias scores. Positive misinformation bias scores indicate a lean toward misinformation where as negative bias scores indicate a lean toward information that opposes misinformation. b) Figure showing the distribution of misinformation bias scores of search queries for democrats, republicans, and independents. Note that the bias scores for the participants belonging to the different political leanings coincide indicating that misinformation bias in SERPs remains constant throughout for each participant.}
  \vspace{-0.4cm}
\end{figure*}
Furthermore, we observe
in Figure \ref{dist} 
that the misinformation bias scores of the SERPs form a bimodal distribution constituting two clusters of search queries (Table \ref{tab:clusters}). The cluster1 search queries have the most negative bias, i.e. they contain more opposing videos. This cluster mostly consists of search queries containing the keyword \textit{fraud} in conjunction with keywords \textit{voter}, \textit{election}, and \textit{dominion}. Cluster2 on the other hand consists of search queries with keywords \textit{election} and \textit{2020}. Overall, cluster1 consists of more search queries biased towards finding misinformation compared to search queries in cluster2. This indicates that YouTube pays more attention to search queries about election fraud and ensures that users are exposed to opposing videos when searching about  fraudulent claims surrounding the elections.
% that shows users' intent to find  information about election fraud.

Figure \ref{rank} shows five search queries with the highest and 5 search queries with the lowest misinformation bias. The search query `voter fraud claims' has the least amount of misinformation bias, indicating that most of the search results for this query oppose election misinformation. On the other hand, the search query `stop the seal' has the most amount of videos supporting election fraud claims. Next, we determine how do   misinformation bias scores in SERPs vary for democrats, independents, and republicans. Figure \ref{fig:irdbias} shows that the bias values for democrats, independents, and republicans for all days coincide indicating that the amount of misinformation bias is almost constant for all days for all participants irrespective of their partisanship. {Overall, our RQ2 results  indicate that YouTube  pushes debunking information in  search results, more for search queries about voter fraud claims as compared to generic queries about the presidential elections.} 

% for most of the queries about the presidential elections 2020 and the misinformative claims surrounding the elections.

\begin{figure*}
\hspace{-3cm}
  \begin{minipage}[]{0.55\linewidth}
    \centering
    \includegraphics[width=\linewidth]{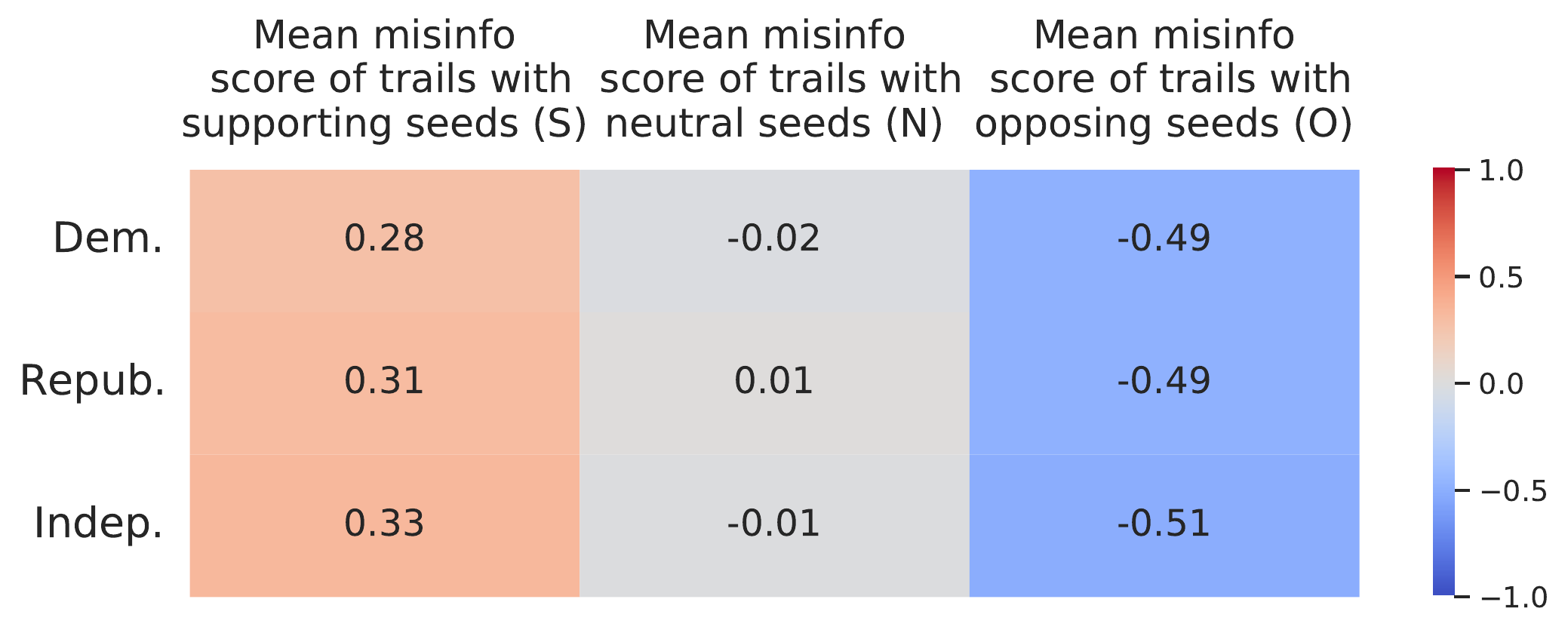}
    % \rule{6cm}{6cm} %to simulate an actual figure
  \end{minipage}
  \begin{minipage}[]{0.20\linewidth}
    \centering
    \begin{footnotesize}

\begin{tabular}{l|l|l}
\hline
Pol. aff. & Statistical tests & Mean diff. \\ \hline
Democrats & F(2,3407)=4035.1 , p=0 & S>N>O \\ \hline
Republicans & F(2,2265)=2981.4, p=0 & S>N>O \\ \hline
Independents & F(2,2941)=3593.8, p=0 & S>N>O
\end{tabular}
    \end{footnotesize}
% \caption{}
\label{tab:my-table}
\end{minipage}
\caption{\textbf{RQ2b results:} Mean misinformation scores of standard up-next trails with seed videos that are supporting (S), neutral (N), or opposing election misinformation (O) for Democrats, Independents, and Republicans. A positive  misinformation score indicates a lean toward misinformative content while a negative score indicates a lean toward content that opposes election misinformation.  Statistical tests reveal a significant difference in the amount of misinformation contained in up-next trails. We  find that    democrats, republicans, and independents  find more misinformation in supporting trails compared to neutral trails, and  more misinformation in neutral trails as compared to opposing trails.}
\label{tab:misinfo scores}
\Description{The mean misinformation score of trails with supporting seeds for democrats, republicans, and independents is 0.28, 0.31, and 0.33 respectively. The mean misinformation score of trails with neutral seeds for democrats, republicans, and independents is -0.02, 0.01, and -0.01 respectively. The mean misinformation score of trails with opposing seeds for democrats, republicans, and independents is -0.49, -0.49, and -0.51 respectively.}
\vspace{-13pt}
\end{figure*}

\subsection{RQ2b: Misinformation in up-next trails}

% \subsection{Misinformation due to personalization}
The results of RQ1 showed that participants' up-next trails are highly personalized. 
% In other words, the videos present in the up-next trails of  standard window are different from videos present in the up-next trails of incognito window. 
 In other words,  videos in up-next trails obtained from the standard window are different from videos in trails obtained from the incognito window.
Recall, that trails extracted from the incognito window  act as baseline unpersonalized trails while trails extracted from the standard window, where users had signed into their accounts, act as personalized treatment trails. Therefore, to determine the impact of personalization on the amount of misinformation in up-next trails, we compare the misinformation bias scores of trails collected in standard windows with the trails collected in incognito windows. We find that the difference in misinformation bias scores of  standard and incognito up-next trails  is not significant (t=-0.62, p=0.53). This means that although the standard up-next trails are very different from the incognito up-next trails, there is no difference in the amount of misinformation present in them. To avoid inflating our sample size, for further downstream analysis, we only consider up-next trails obtained from participants' standard windows. This similar strategy was adopted by Robertson et al  for analyzing bias in Google search results when they did not see any significant difference in the amount of partisan bias in incognito-standard SERP pairs \cite{robertson2018auditing}.

\subsubsection{Misinformation in standard up-next trails for different scenarios} In this section, we 
% estimate the counterfactual scenarios by determining
determine the amount of misinformation  encountered by our study participants in the standard up-next trails for seed videos with  different stances on election misinformation---supporting, neutral and opposing. 
% We estimate this by conducting  within group statistical tests on the misinformation scores of standard up-next trails for users. 
Figure \ref{tab:misinfo scores} shows the mean misinformation scores of different  up-next trails
% with seed videos that support election misinformation (supporting trails), are neutral in stance (neutral trails) and oppose election misinformation (opposing trails) 
collected from the standard windows of democrats, republicans, and independents. Recall that a positive misinformation score (>0) indicates a lean toward misinformation, while a negative misinformation score indicates a lean toward information that opposes election misinformation. We conduct within-group statistical tests to determine the difference in misinformation for the three scenarios (following trails for supporting, neutral, and opposing seed videos). The  tests indicate a filter bubble effect. If users watch supporting videos, they are led to supporting videos in the trails. But if they watch neutral videos, they are led to less misinformation  compared to when they watched supporting videos. However, if users watch opposing videos, they are led to more opposing videos in the up-next trails. The same trend is observed for democrats, republicans, and independents.

% the amount of content with that leaning. The mean misinformation scores of standard trails  is positive for democrats, independents and republicans indicating that misinformative videos about elections lead to a few more misinformative videos in participants' up-next trails. We also observe that the mean misinformation score of up-next trails starting with a seed video debunking election misinformation is negative for democrats, independents and republicans indicating an echo chamber effect where YouTube drives users watching videos opposing election misinformation  to more such videos in the up-next trails. Table \ref{tab:misinfo scores}  also shows that the magnitude of mean misinformation scores of up-next trails starting with a neutral video is close to zero for all participants indicating  that users watching neutral videos about presidential election are led to mostly more neutral content in the up-next trails.
Is the amount of misinformation in trails with different seeds different for democrats, republicans, and independents?
Between-group statistical tests reveal that the amount of misinformation in supporting trails (KW H(2)=11.9,p=0.002) and neutral trails (KW H(2)=8.69,p=0.01) for democrats, independents, and republicans is significantly different. We find that independents in our sample receive more misinformation in their supporting trails as compared to democrats. Additionally, republicans receive more misinformation in their neutral trails compared to democrats.

Overall, by observing Figure \ref{tab:misinfo scores}, we realize misinformation scores of supporting trails are positive and opposing trails are negative. However, the magnitude of misinformation scores of opposing trails is much more than the supporting trails indicating that the strength of the filter bubble effect was more when our study participants watched videos opposing election misinformation.

\begin{figure*}
\centering
\hspace{-0.8cm}
\begin{subfigure}{.32\linewidth}
    \centering
    \includegraphics[width=.8\linewidth]{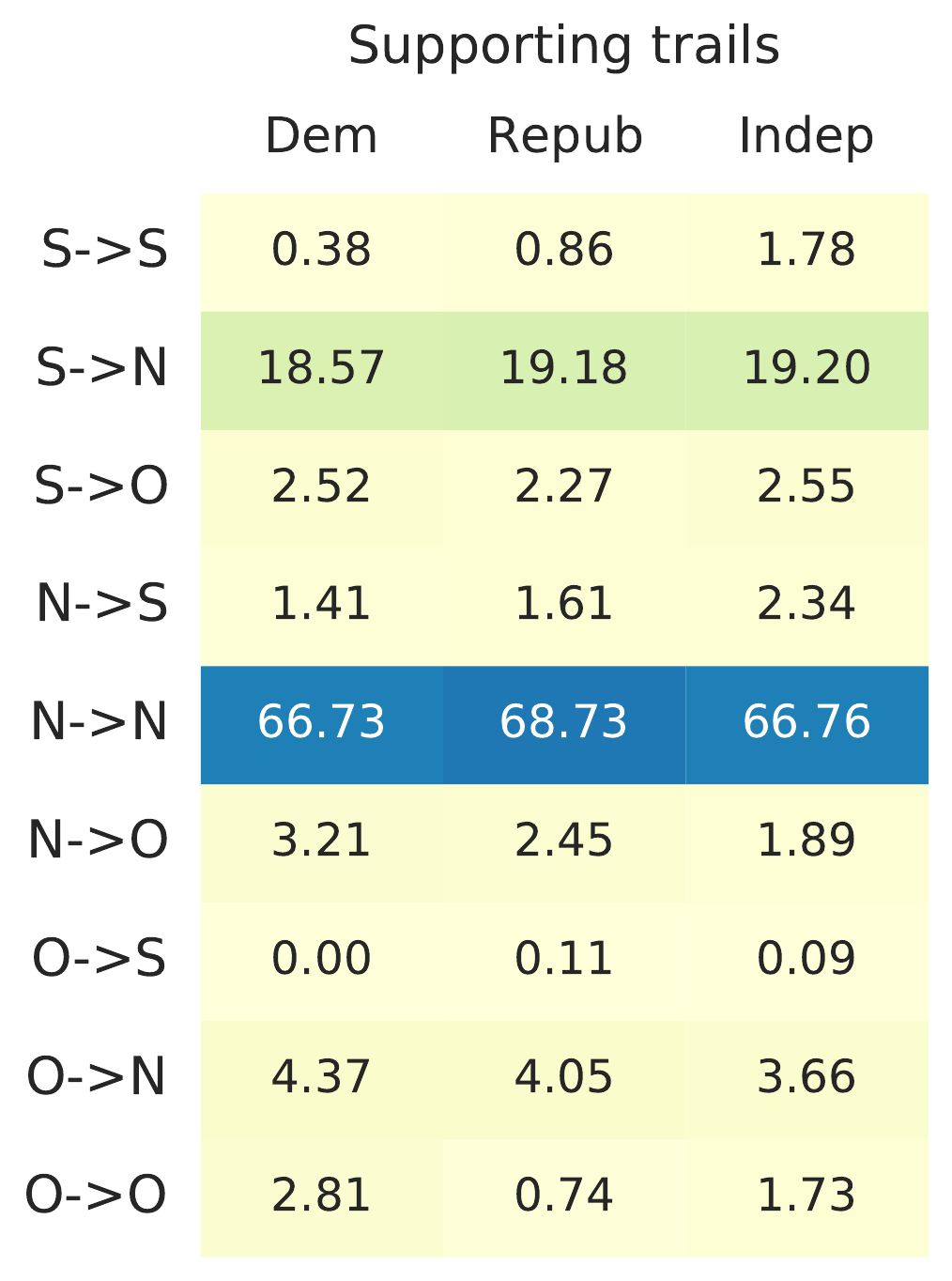}  
    \caption{Mean \% of transitions in trails with seed  videos supporting elec. misinfo.}
    \label{sm}
\end{subfigure}\hfill
\begin{subfigure}{.32\linewidth}
    \centering
    \includegraphics[width=.8\linewidth]{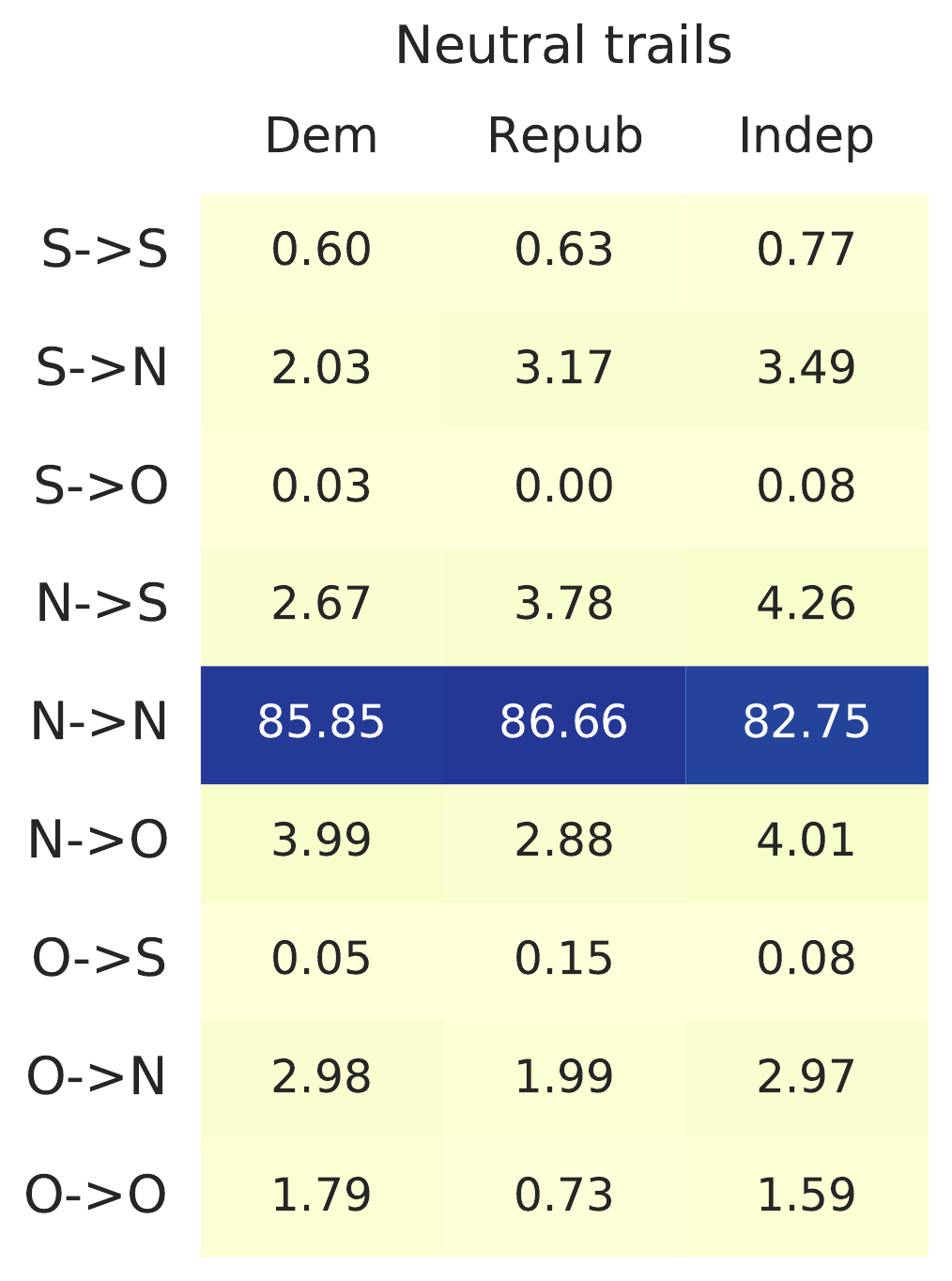}  
    \caption{Mean \% of transitions in trails with neutral seed videos}
    \label{nm}
\end{subfigure}\hfill
\begin{subfigure}{.389\linewidth}
    \centering
    \includegraphics[width=.8\linewidth]{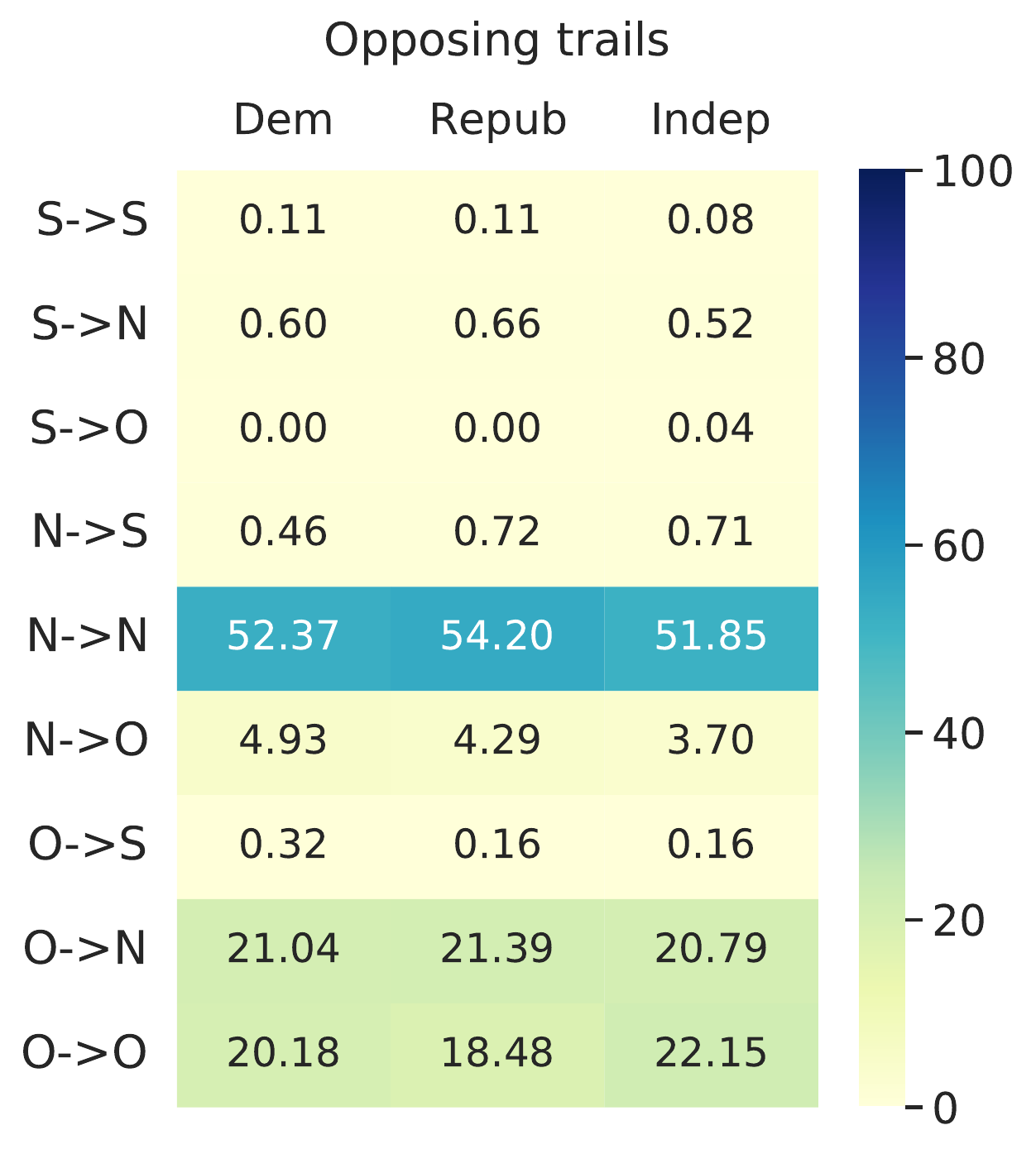}  
    \caption{Mean \% of transitions in trails with seed  videos opposing elec. misinfo.}
    \label{om}
\end{subfigure}
\caption{\textbf{RQ2b results:} Mean percentage of various transitions present in the standard up-next trails of democrats, independents, and republicans. S represents a video supporting election misinformation, N represents a neutral video and O represents a video opposing election misinformation. Transition S->S denotes that a YouTube video supporting election misinformation leads to an up-next video recommendation supporting election misinformation.}
\label{tab:transitions}
\Description{The figure shows the mean percentage of transitions (S->S, S->N, S->O, N->S, N->N, N->O, O->S, O->N, and O->O) in trails with seed videos that are promoting, neutral, and opposing. The percentage of N->N transitions are highest in all trails collected from all users. For the opposing trails, O->N and O->O transitions are around 20\% for all users. Problematic transition N->S is 2.67\%, 3.78\%, and 4.26\% in neutral trails of democrats, republicans, and independents.}
\end{figure*}

\subsubsection{Transitions in standard up-next trails}
In this section, we gain more insights into the anatomy of YouTube's up-next trails by studying the various transitions present in them. This allows us to determine how users get pushed towards misinformative or debunking videos in the  trails. Since our annotation scale consists of three values, supporting (S), neutral (N), and opposing (O), there  are 9 transitions possible in the trails (S->S, S->N, S->O, N->S, N->N, N->O, O->S, O->N, N->O). For each participant, we first individually determine the percentage of each of these transitions present in the three types of standard up-next trails collected (ones starting with a supporting seed video, neutral seed videos, and opposing seed video). Then we calculated the mean percentage of all of these transitions for democrats, independents, and republicans.  From Figure \ref{tab:transitions}, we see that the maximum number of transitions across all participants and all types of up-next trails is N->N. Problematic transitions like  S->S and O->S are less than 2\% in trails of all users. However, comparatively S->S transitions are still more in the supporting up-next trails of independents (1.78\%) compared to democrats (0.38\%) and republicans (0.86\%). In the neutral up-next trails of republicans and independents, N->S transitions dominate (after N->N transitions) indicating that independents and republicans are sometimes led to supporting videos in their up-next recommendations even when they are viewing neutral YouTube videos. We also observe that the opposing up-next trails  majorly consist of transitions O->N and N->O (after N->N transitions) indicating that once a user watches a video that opposes election misinformation, YouTube pushes more videos that are either neutral or opposing in stance in the up-next trails of all the participants.
We also observe that S->O transitions are less than S->N transitions in the supporting trails of democrats, republicans,  and independents. Previous work has shown that watching YouTube videos that debunk misinformation helps in bursting  filter bubbles of misinformation \cite{tomlein2021audit}. Our work also shows that opposing videos could lead to more opposing videos (O->O transitions in opposing trails). Thus, increasing the number of S->O transitions can lead users to trustworthy information on the platform.

\begin{figure}
    \centering
    \includegraphics[width=0.4\textwidth]{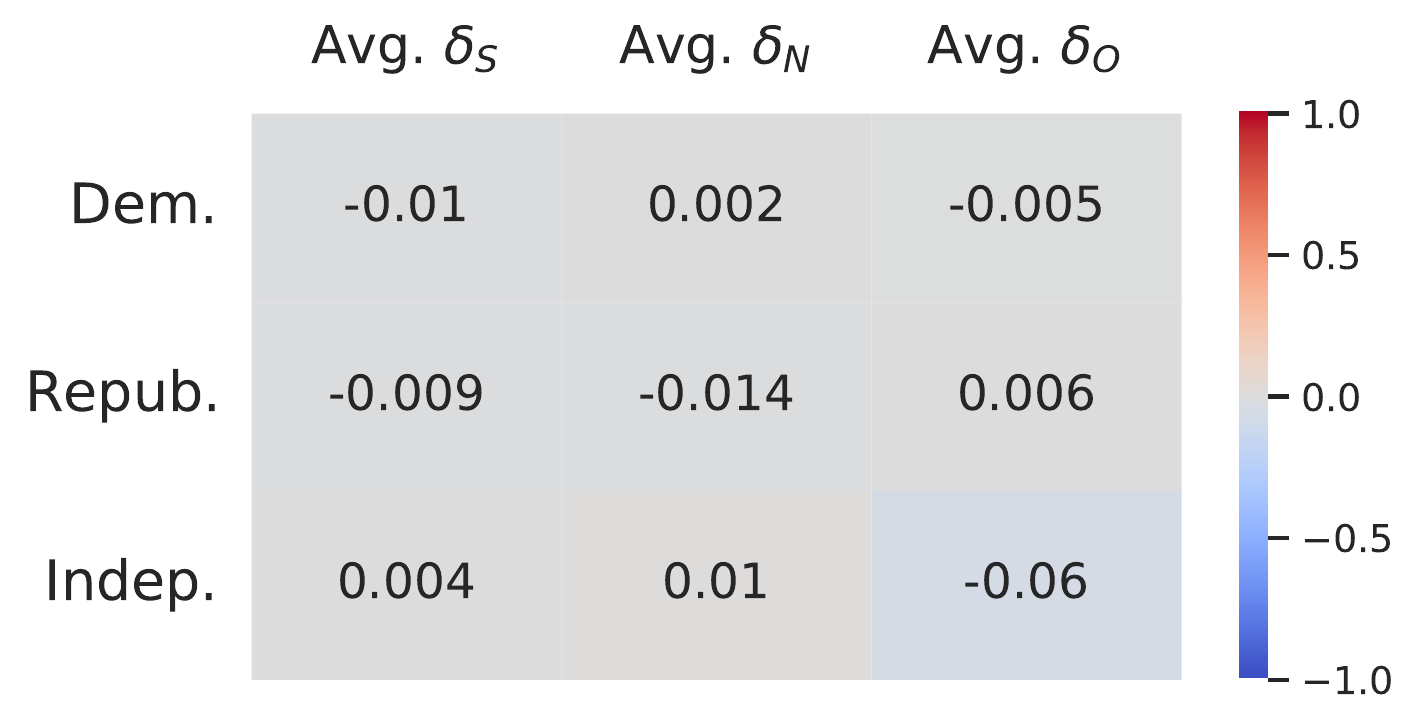}
    \caption{\textbf{RQ2c results:} Figure showing the average change in the amount of bias present in homepages because of watching a trail of up-next videos starting with either supporting, opposing, or neutral seed videos for democrats, republicans,  and independents.}
    \label{tab:delta}
    \Description{Figure showing the average change in the amount of bias present in homepages because of watching a trail of up-next videos starting with either supporting, opposing, or neutral seed videos for democrats, republicans,  and independents. All the values are equal to or less than 0.01.}
    \vspace{-0.5cm}
\end{figure}

\begin{figure*}[]
\centering
\begin{subfigure}[]{.4\textwidth}
  \centering
  \includegraphics[width=\linewidth,keepaspectratio]{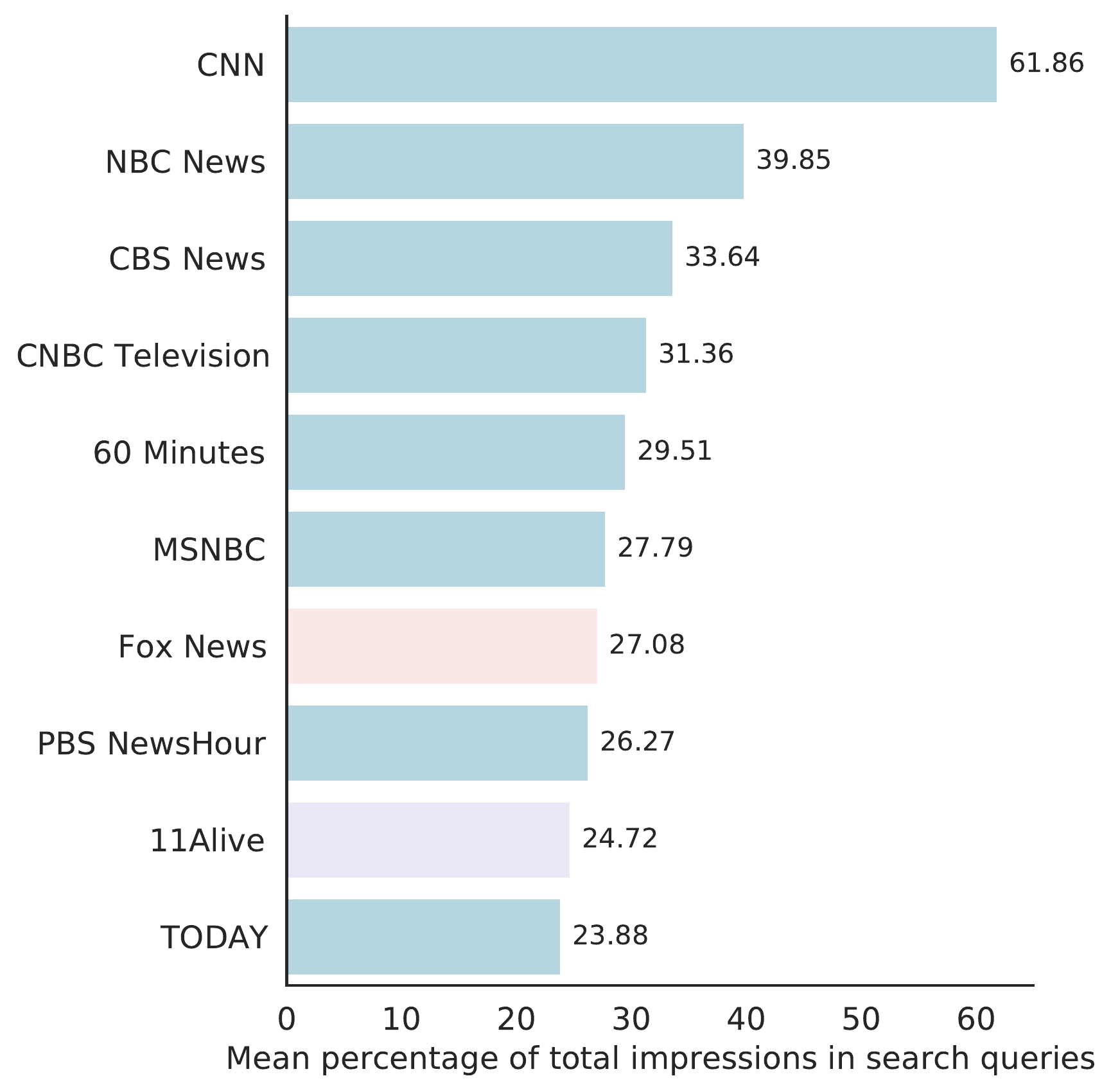}
  \caption{}
  \label{impr_search}
%   \hspace{1cm}
\end{subfigure}
% \hfill
\begin{subfigure}[]{.5\textwidth}
  \centering
  \includegraphics[width=\linewidth,keepaspectratio]{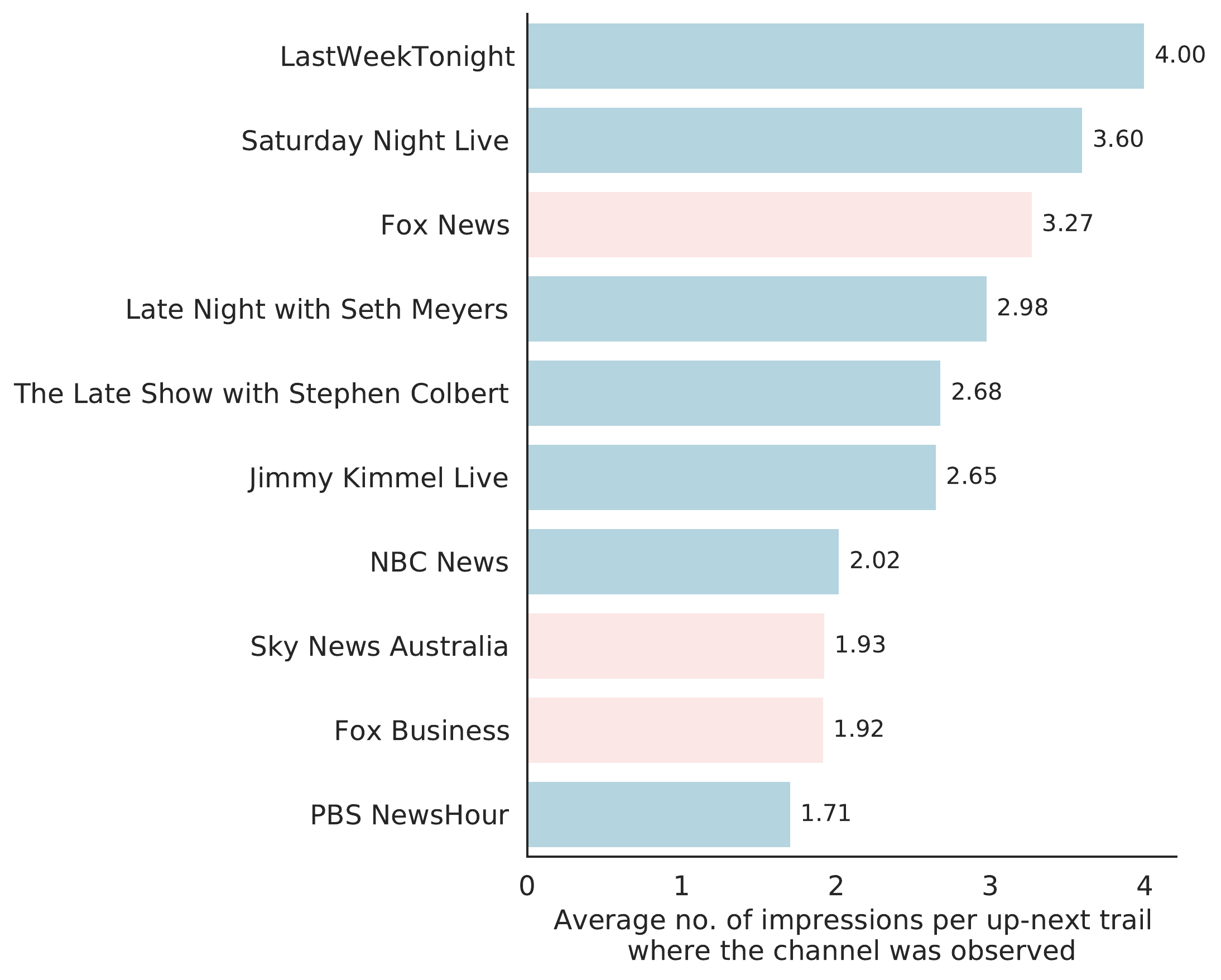}
  \caption{}
  \label{impr_trail}
\end{subfigure}
\caption{ \textbf{RQ3 results:} a) Figure showing Top-10 YouTube channels with impressions in the most number of search queries for all study participants. For example, on average CNN appears in 61.86\% of search queries for all our study participants. b) Figure showing the average number of impressions for Top-10 YouTube channels that appear in the most number of standard up-trails collected for users. For example, on average, videos from the Fox News channel appear 3.27 times in those up-next trails where videos from the channel are observed. \fcolorbox{labell}{labell}{\rule{0pt}{2pt}\rule{2pt}{0pt}} is a left-leaning channel, \fcolorbox{labelr}{labelr}{\rule{0pt}{2pt}\rule{2pt}{0pt}} is right-leaning and \fcolorbox{labelc}{labelc}{\rule{0pt}{2pt}\rule{2pt}{0pt}} is center-leaning.}
\label{fig:imp}
\Description{Figure (a) shows the Top-10 YouTube channels with impressions in the most number of search queries for all study participants. The channels are in order CNN (61.86\%), NBC News (39.85\%), CBS News (33.64\%), CNBC Television (31.36\%), 60 minutes (29.51\%), MSNBC (27.79\%), Fox news (27.08\%), PBS NewsHour (26.27\%), 11Alive (24.72\%), Today (23.88\%). Figure (b) shows the average number of impressions for Top-10 YouTube channels that appear in the most number of standard up-trails collected for users. The channels are in order LastWeekTonight, Saturday Night Live (4), Fox News (3.6), Late Night with Seth Meyers (3.27), The Late Show with Stephen Colbert (2.98), Jimmy Kimmel Live (2.65), NBC News (2.02), Sky News Australia (1.93), Fox Business (1.92), and PBS NewsHour (1.71)}
\end{figure*}

\subsection{RQ2c: Misinformation in homepages}
We collected participants' YouTube homepages to determine how the bias in the homepage changes ($\delta$) after watching a trail of videos starting with a seed video that is either supporting (${\delta}_{S}$), opposing (${\delta}_{O}$) or neutral (${\delta}_{N}$) in stance with respect to election misinformation. We calculated the impact of trails by using the following formula:-

${\delta}_{stance}$ = \textit{Misinformation}  $score_{Homepage\_before\_the\_trail}$- \textit{Misinformation} $score_{Homepage\_after\_the\_ trail}$

${\delta}_{S}$, ${\delta}_{N}$ and ${\delta}_{O}$ represent the change in the amount of bias present in homepages because of watching a trail of up-next videos starting with supporting, opposing and neutral seeds. A negative $\delta$ would indicate that the YouTube homepage collected after the trail  contained more opposing videos compared to the YouTube homepage before the trail. A positive $\delta$, on the other hand, indicates either presence of more videos supporting election misinformation  or a lesser number of opposing videos on the homepage collected after the trail as compared to the homepage collected before the trail. We consider the top ten recommendations present on the homepage for analysis. Figure \ref{tab:delta} shows $\delta$ values for all three kinds of trails for democrats, republicans,  and independents.  We discuss a few results. 
We observe that after following the up-next video trails starting from a neutral seed, the homepages of democrats and independents contain more supporting videos. However, recall that the average misinformation score of the up-next trails with neutral seeds for both democrats and independents was negative (Figure \ref{tab:misinfo scores}). This indicates that although the up-next trails with neutral seeds lead users to more opposing videos, the homepages, however, contain more misinformation or a lesser number of opposing videos after the trail. We also observe that after watching up-next trail videos with supporting seed, republicans' homepage  contain more opposing videos (Figure \ref{tab:delta}) while the trail itself contained more misinformation (Figure \ref{tab:misinfo scores}). However, note that the magnitude of the $\delta$ is low in all the conditions indicating that fewer videos  supporting or opposing election misinformation appear on the participants' homepages.

% \begin{table}[]
% \begin{tabular}{l|c|c|c}
% \cline{2-4}
%  & \multicolumn{1}{l|}{Avg., $\delta_S$} & \multicolumn{1}{l|}{Avg. $\delta_N$} & \multicolumn{1}{l}{Avg. $\delta_O$} \\ \hline
% Democrats & -0.01 & 0.002 & -0.005 \\ \hline
% Republicans & -0.009 & -0.014 & 0.006 \\ \hline
% Independents & 0.004 & 0.01 & -0.06
% \end{tabular}
% \caption{Table showing the average change in amount of bias present in homepages because of watching a trail of up-next videos starting with supporting, opposing and neutral seeds for democrats, republicans  and independents.}
% \label{tab:delta}
% \end{table}

% \begin{figure*}
%     \centering
%     \includegraphics[width=0.5\textwidth,keepaspectratio]{RQ3/impression_search.png}
%     \caption{search impressions per query.}
%     \label{fig:overlap}
% \end{figure*}

\section{RQ3: Composition and Diversity} \label{rq3}
In this research question, we want to characterize   source diversity in YouTube when users search for election misinformation on the platform. Source diversity in searches and recommendations is an important characterization of fairness \cite{ge2021towards}. Furthermore, given that the narratives about the election misinformation were closely intertwined with news sources and their leanings, it is important to determine what kinds of YouTube channels are users exposed to. News and media diversity can be characterized in multiple ways \cite{joris2020news}. One typology characterizes  media diversity with respect to \textit{source} (content providers), \textit{content} (perspectives) and \textit{exposure} (actual consumption of diverse content) \cite{napoli2011exposure,trielli2019search}. Our work analyzed the content diversity in RQ2  by analyzing the video's stance on election misinformation. We cannot study exposure diversity since it requires determining the actual content consumed (clicked, watched, etc) by our study participants in their naturalistic settings. For this study, we focus on source diversity in terms of the identity of top content providers (YouTube channels) and distribution and concentration of channels  in the standard SERPs and up-next trails. We  acknowledge that future studies should also examine the ideological position of news sources and study the filter bubbles of partisan content on the platform.

\begin{figure*}[]
  \begin{minipage}{\linewidth}
  \begin{subfigure}{0.32\textwidth}
    \centering
    \includegraphics[width=\textwidth]{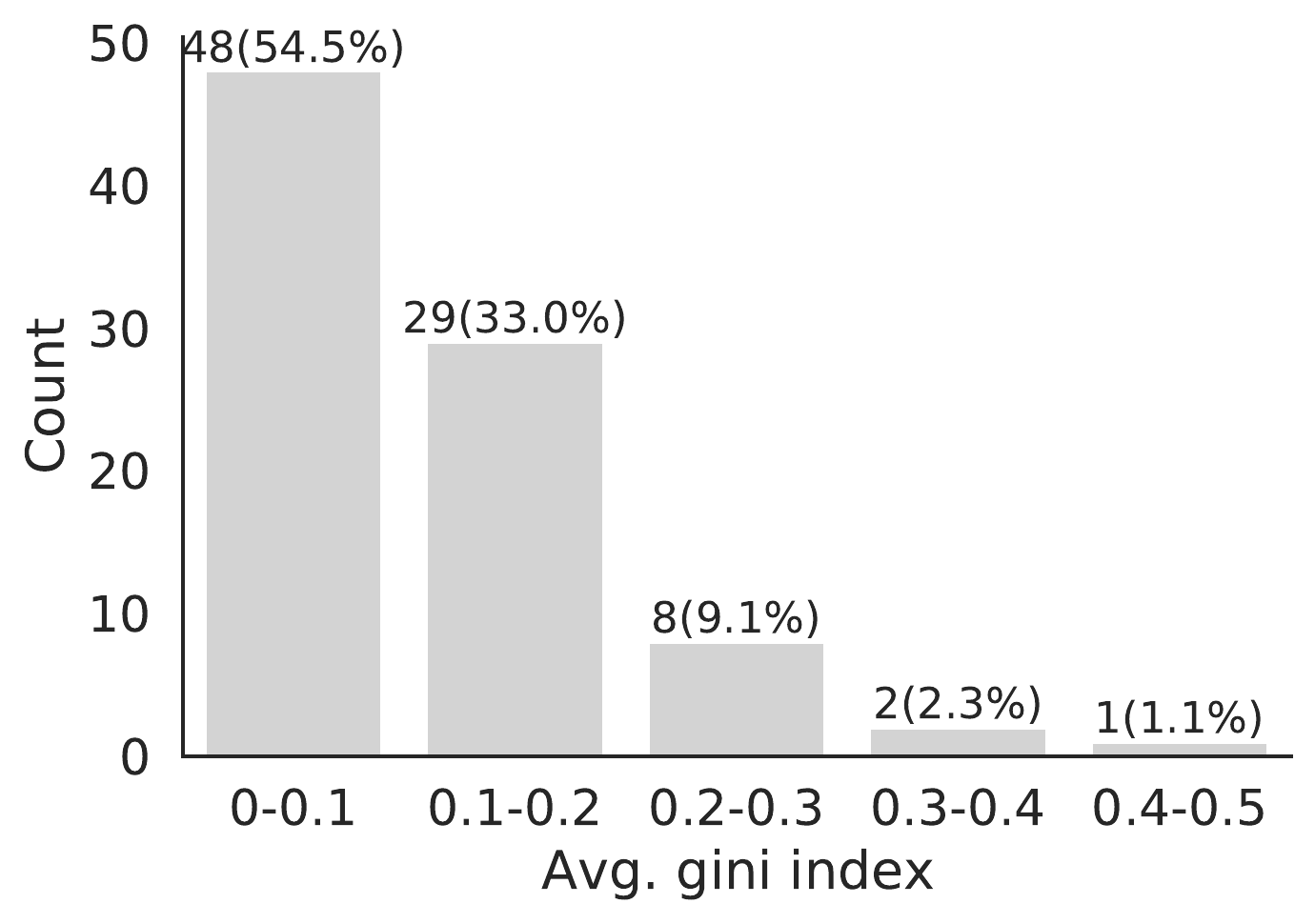}
    \caption{Democrats}\label{fig:1a}
  \end{subfigure}\hfill
  \begin{subfigure}{0.32\textwidth}
    \centering
    \includegraphics[width=\textwidth]{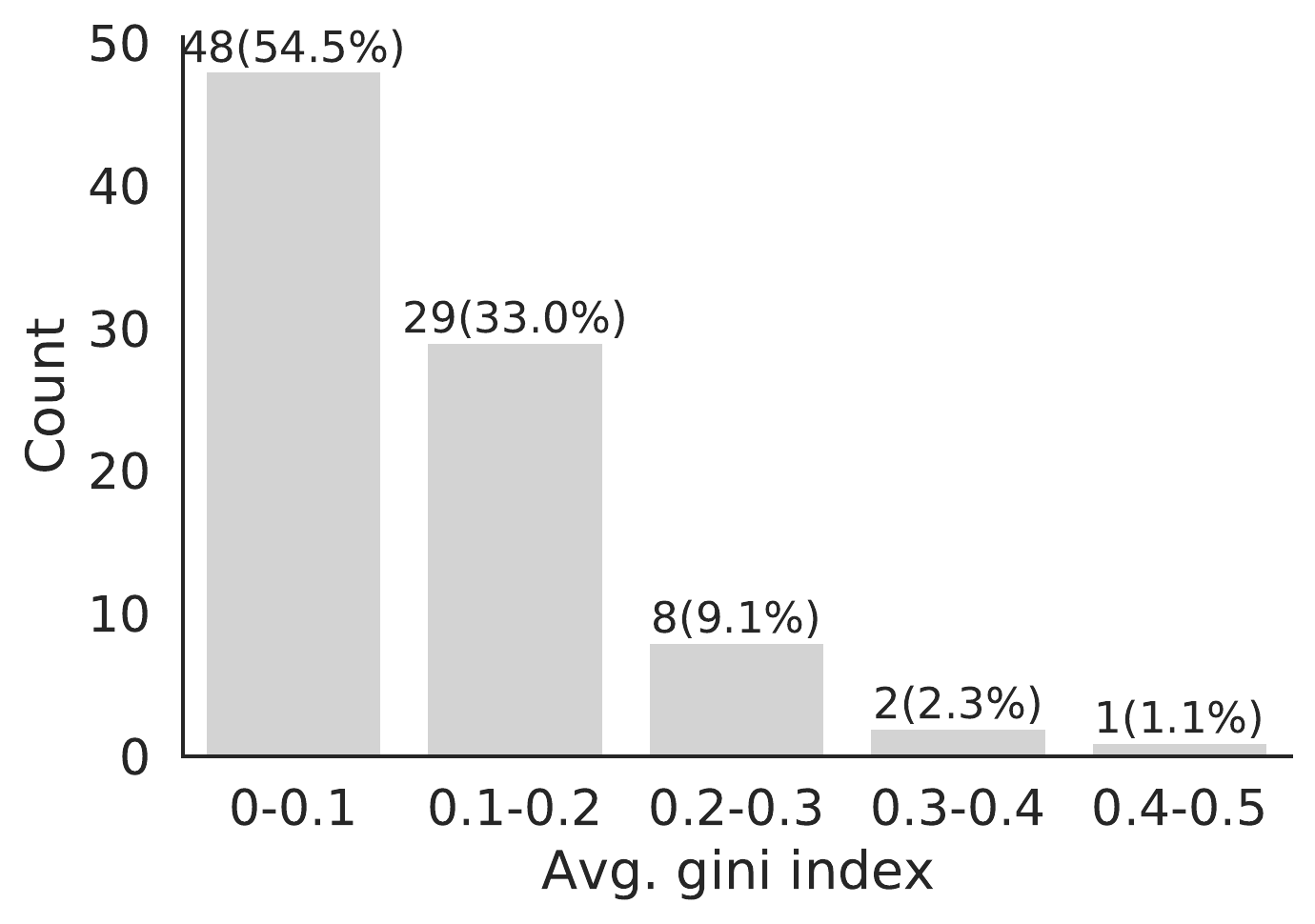}
    \caption{Republicans}\label{fig:1b}
  \end{subfigure}\hfill
  \begin{subfigure}{0.32\textwidth}
    \centering
    \includegraphics[width=\textwidth]{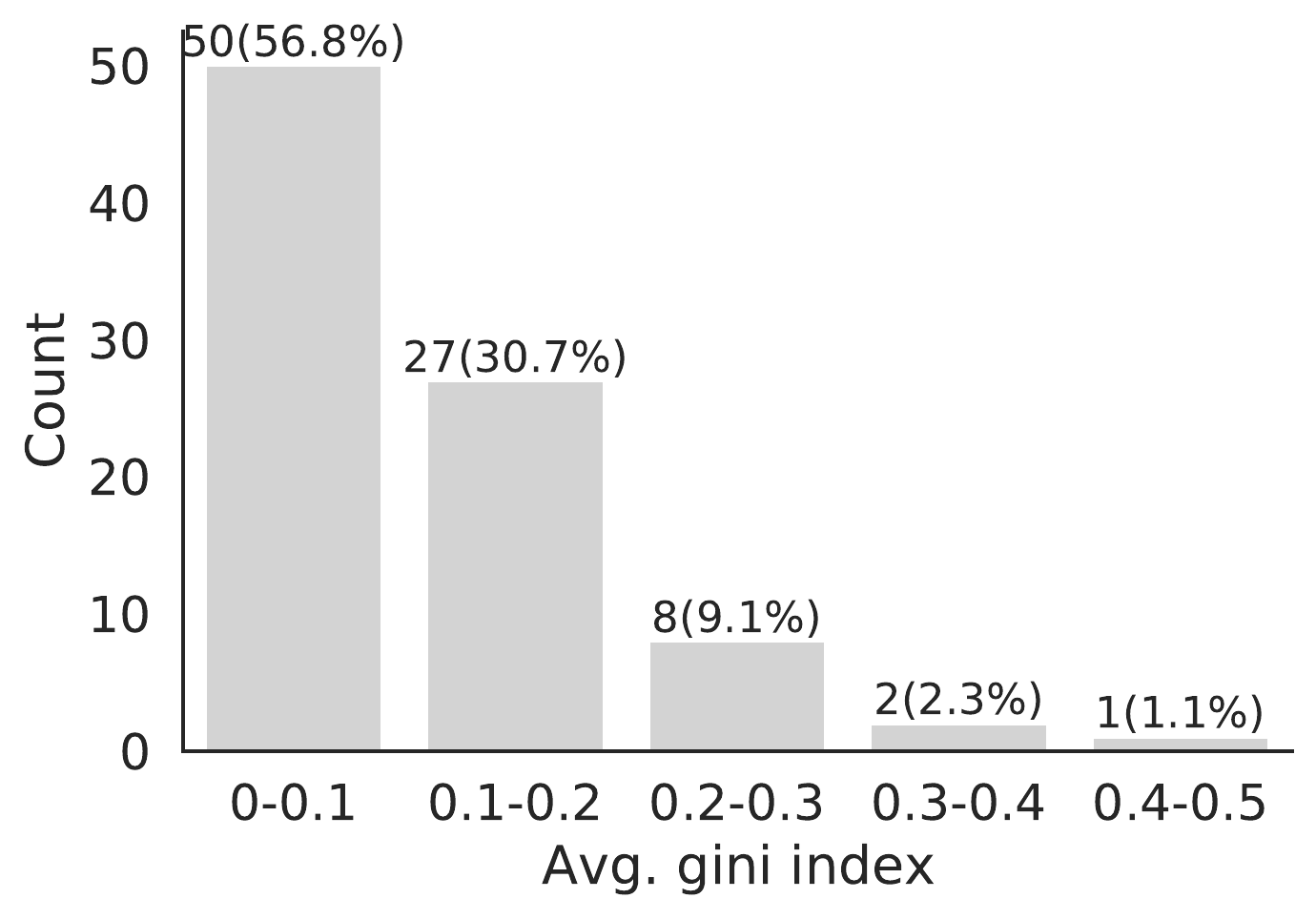}
    \caption{Independents}\label{fig:1b}
  \end{subfigure}
  \end{minipage}
 
  \caption{\textbf{RQ3a results:} Distribution of Gini coefficients for all
search queries (n=88) for a) Democrats, b) Republicans, and c) Independents, calculated based on the distribution of impressions of YouTube channels appearing in the search results. 
% The distribution of Gini coefficients is almost same for search results of users with different political leanings.
}
 \label{gini:search}
 \Description{Figure a shows the gini index distributions for search queries for democrats, 54.5\% search queries have gini between 0-0.1. 33\% have gini between 0.1-0.2. Figure b shows the gini index distributions for search queries for republicans, 54.5\% search queries have gini between 0-0.1. 33\% have gini between 0.1-0.2. Figure c shows the gini index distributions for search queries for democrats, 56.8\% search queries have gini between 0-0.1. 30.7\% have gini between 0.1-0.2.}
\end{figure*}

\begin{figure*}
  \begin{minipage}{\linewidth}
  \begin{subfigure}{0.32\textwidth}
    \centering
    \includegraphics[width=\textwidth]{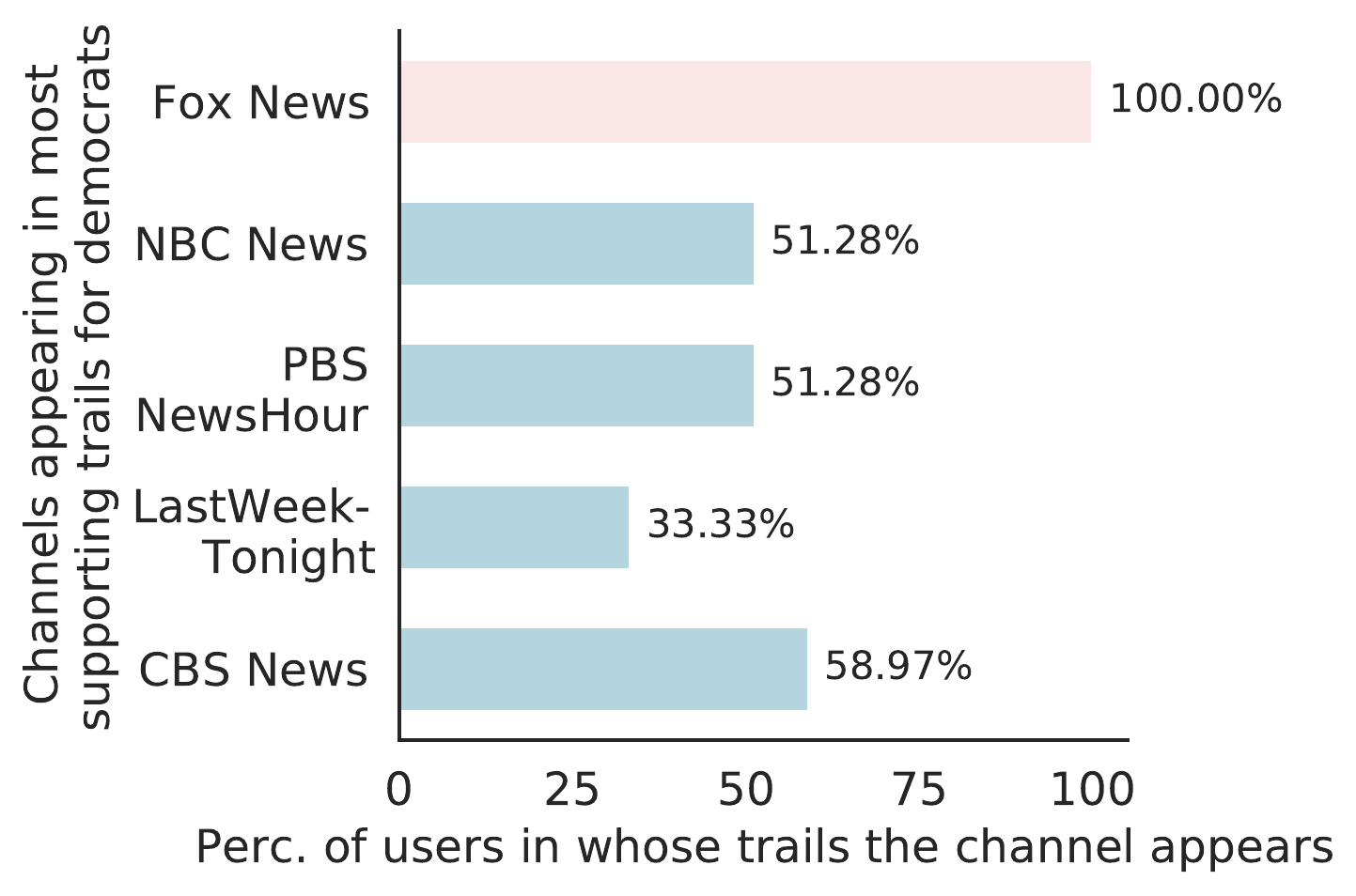}
    \caption{Democrats (supp. trails)}\label{dp}
  \end{subfigure}\hfill
  \begin{subfigure}{0.32\textwidth}
    \centering
    \includegraphics[width=\textwidth]{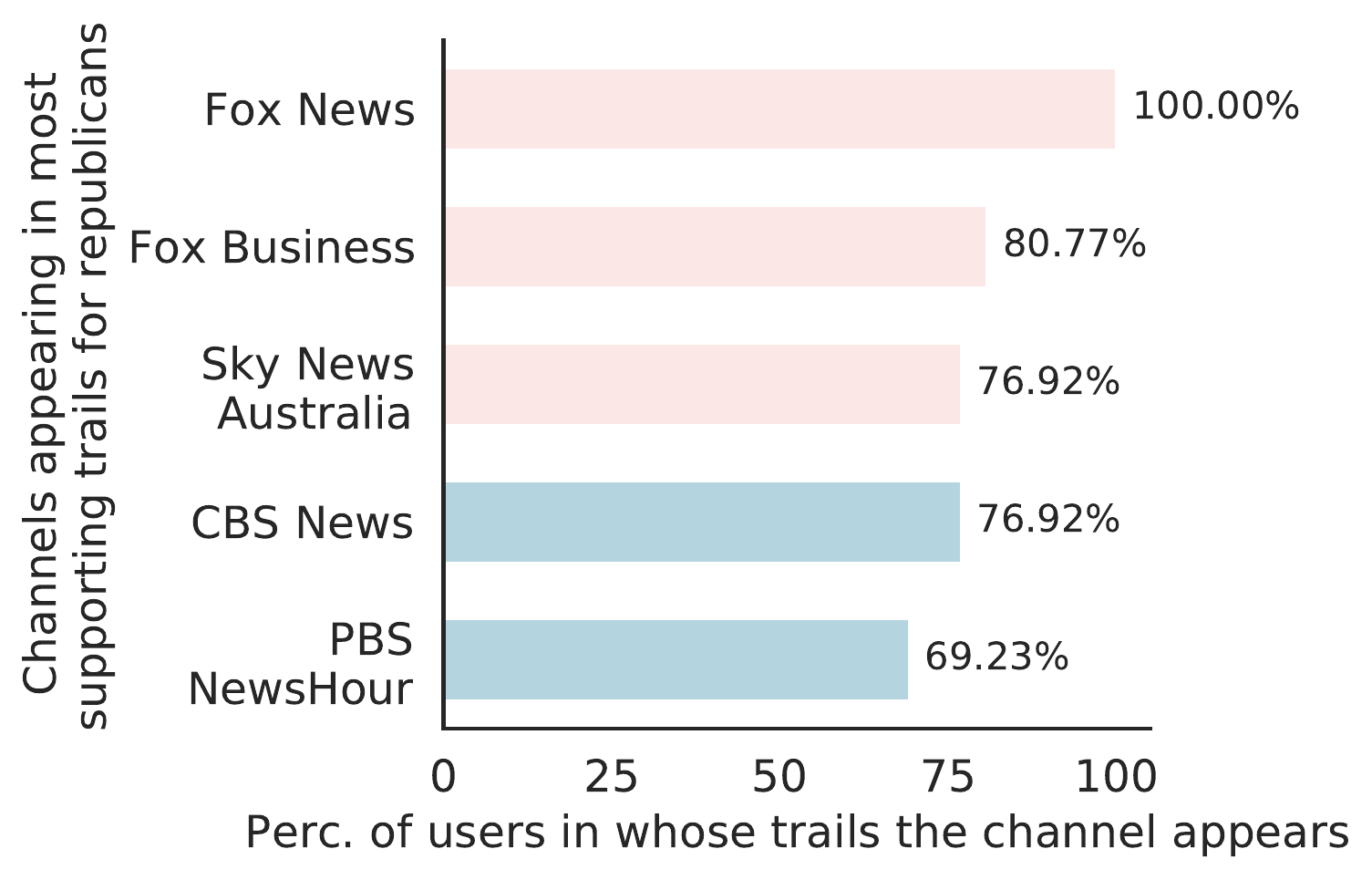}
    \caption{Republicans (supp. trails)}\label{rp}
  \end{subfigure}\hfill
  \begin{subfigure}{0.32\textwidth}
    \centering
    \includegraphics[width=\textwidth]{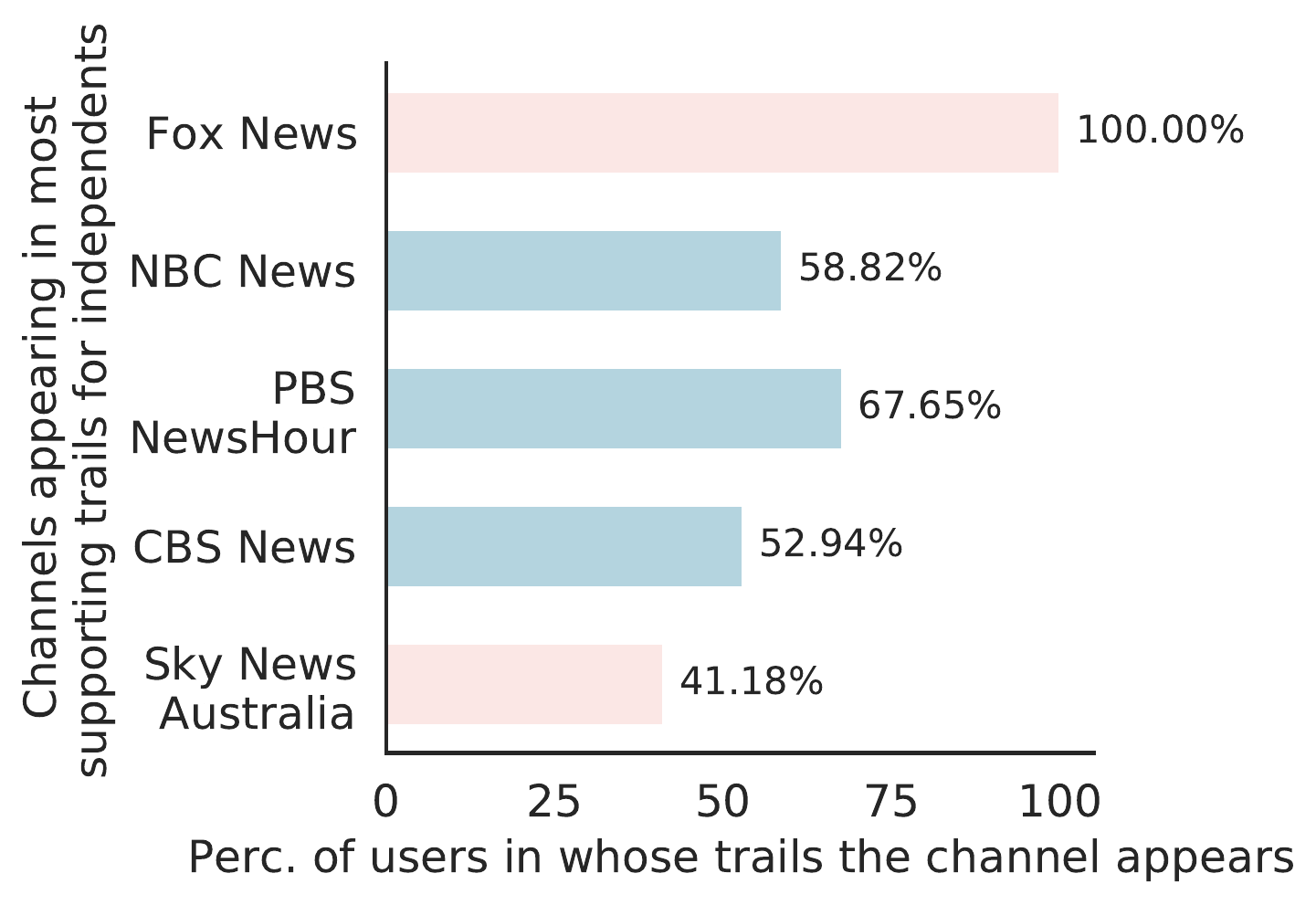}
    \caption{Independents (supp. trails)}\label{ip}
  \end{subfigure}
  \end{minipage}

   \begin{minipage}{\linewidth}
  \begin{subfigure}{0.33\textwidth}
    \centering
    \includegraphics[width=\textwidth]{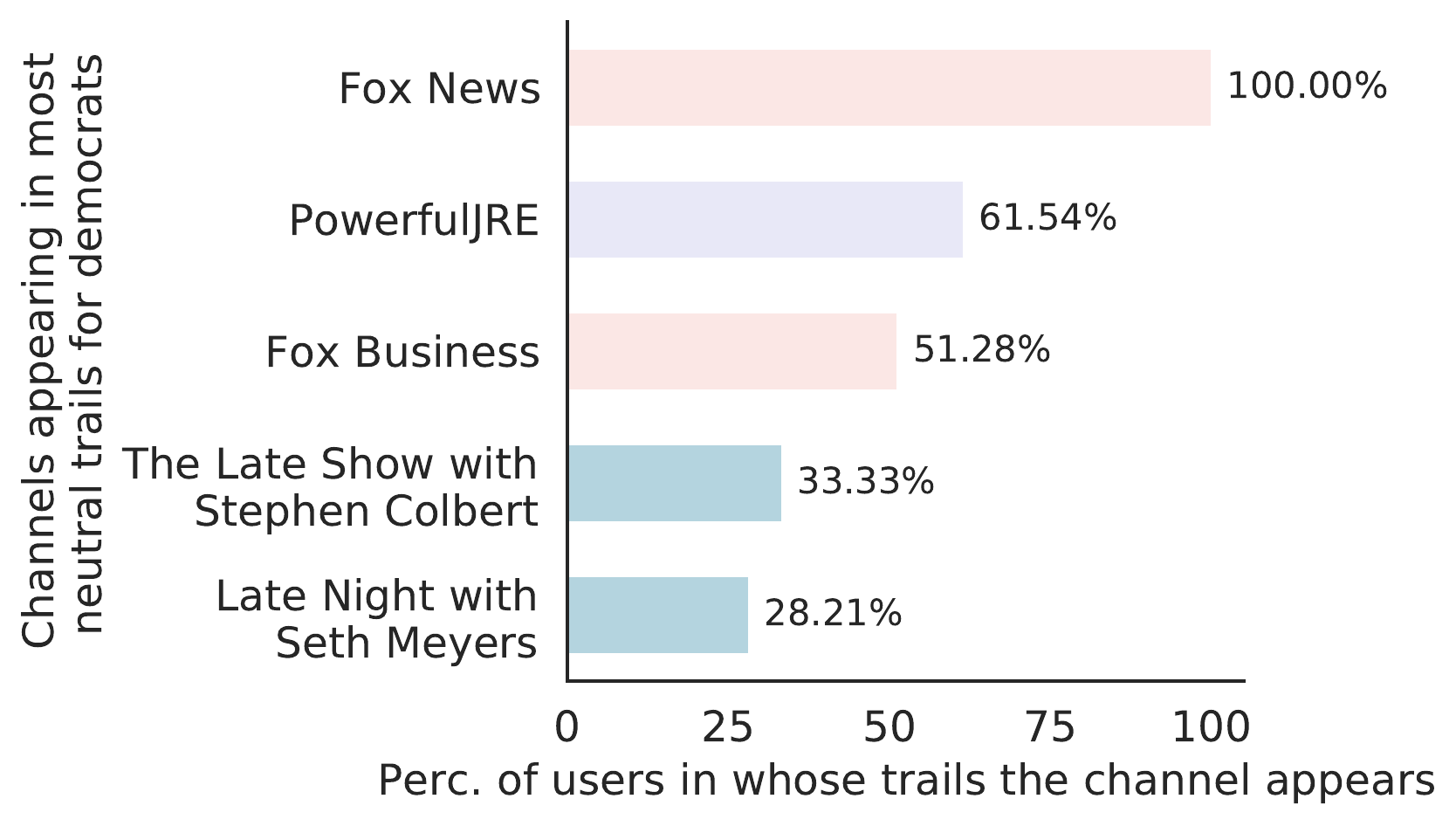}
    \caption{Democrats (neutral trails)}\label{dn}
  \end{subfigure}\hfill
  \begin{subfigure}{0.33\textwidth}
    \centering
    \includegraphics[width=\textwidth]{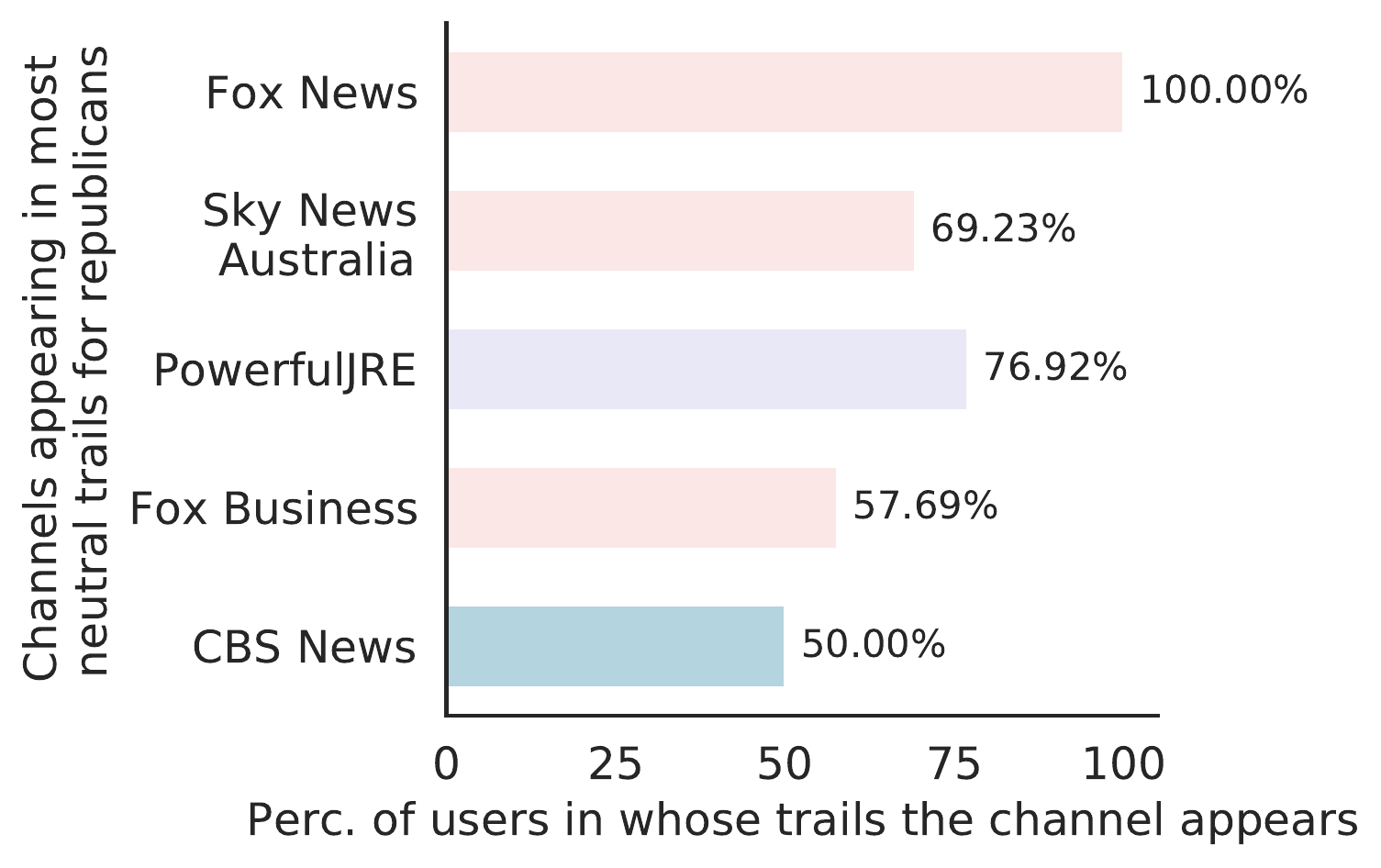}
    \caption{Republicans (neutral trails)}\label{rn}
  \end{subfigure}\hfill
  \begin{subfigure}{0.33\textwidth}
    \centering
    \includegraphics[width=\textwidth]{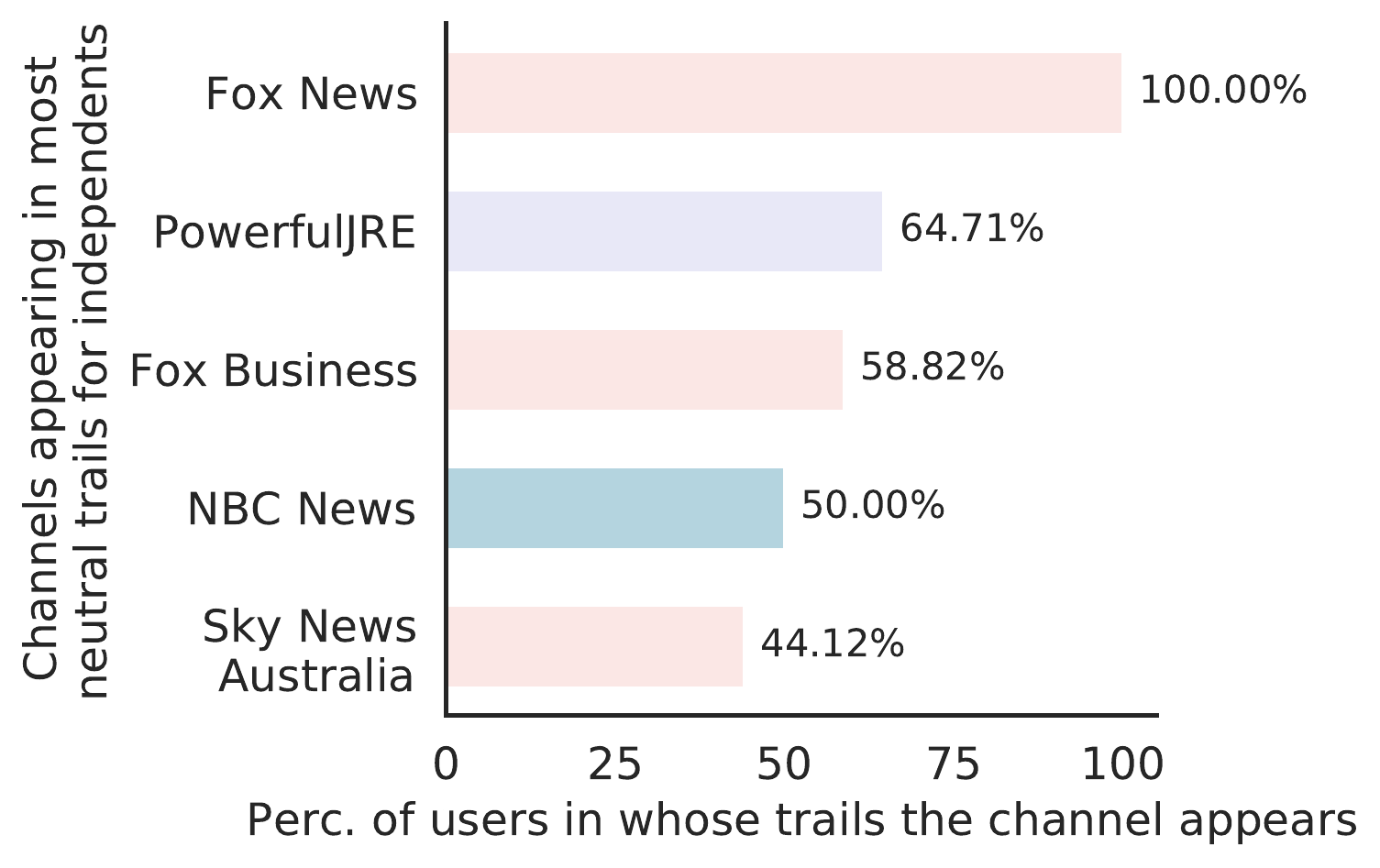}
    \caption{Independents (neutral trails)}\label{in}
  \end{subfigure}
  \end{minipage}

    \begin{minipage}{\linewidth}
  \begin{subfigure}{0.33\textwidth}
    \centering
    \includegraphics[width=\textwidth]{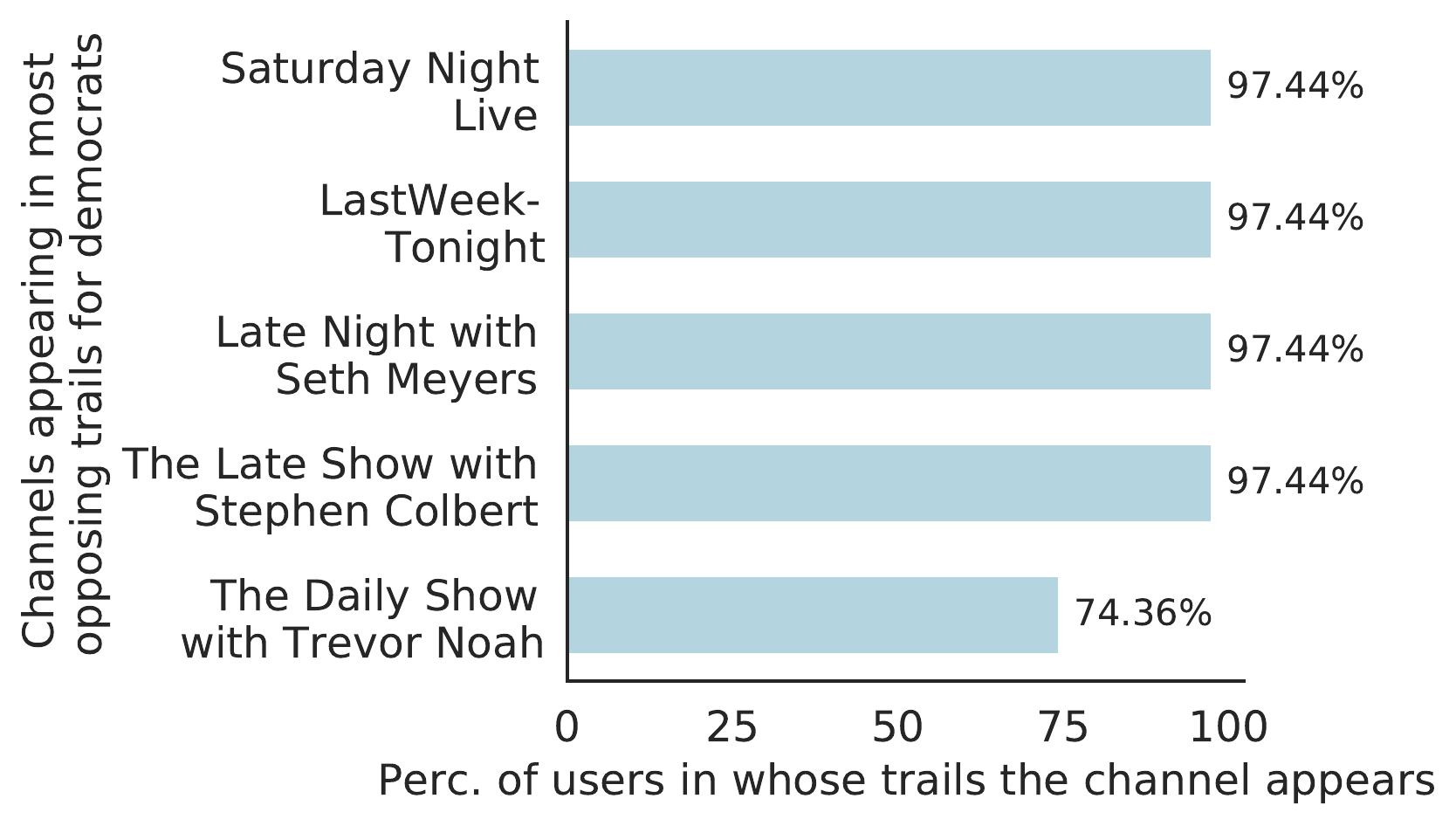}
    \caption{Democrats (oppos. trails)}\label{dd}
  \end{subfigure}\hfill
  \begin{subfigure}{0.33\textwidth}
    \centering
    \includegraphics[width=\textwidth]{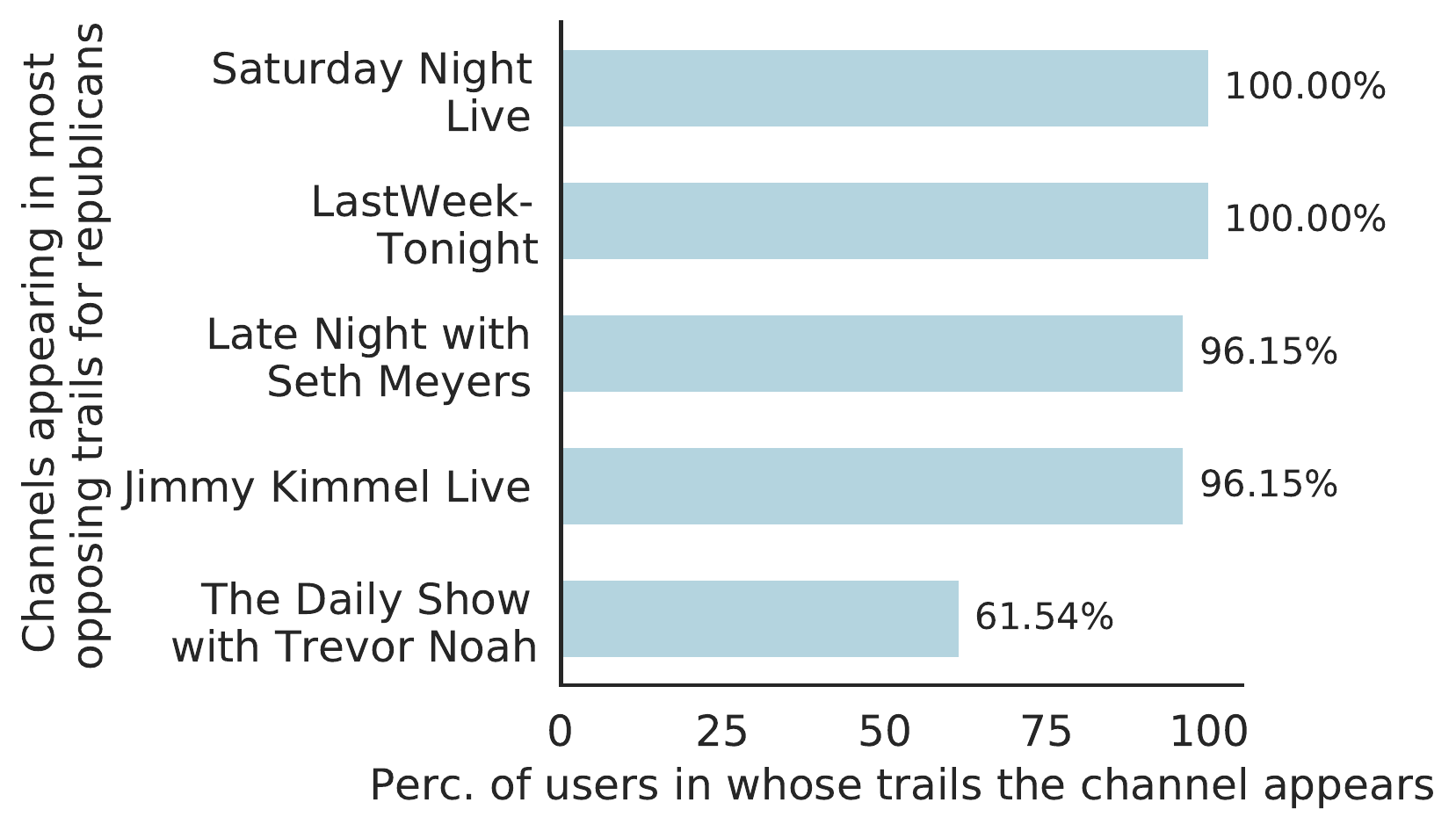}
    \caption{Republicans (oppos. trails)}\label{rd}
  \end{subfigure}\hfill
  \begin{subfigure}{0.33\textwidth}
    \centering
    \includegraphics[width=\textwidth]{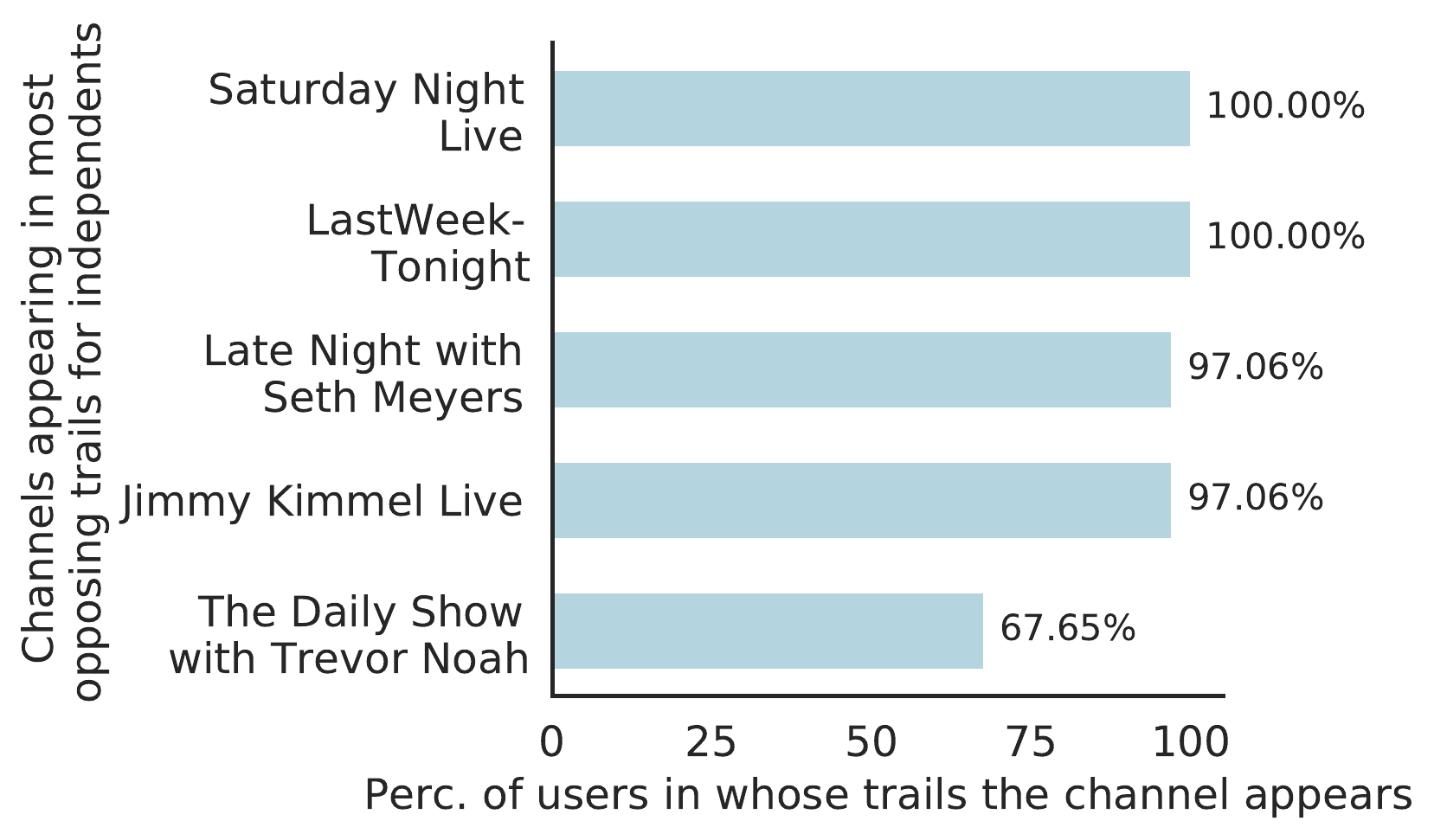}
    \caption{Independents (oppos. trails)}\label{id}
  \end{subfigure}
  \end{minipage}

  \caption{\textbf{RQ3b results:} Figure showing the top YouTube channels appearing in supporting, neutral, and opposing trails of democrats, republicans, and independents and the percentage of users in whose trails these channels appear. \fcolorbox{labell}{labell}{\rule{0pt}{2pt}\rule{2pt}{0pt}} is a left-leaning channel, \fcolorbox{labelr}{labelr}{\rule{0pt}{2pt}\rule{2pt}{0pt}} is right-leaning and \fcolorbox{labelc}{labelc}{\rule{0pt}{2pt}\rule{2pt}{0pt}} is center-leaning.}
 \label{div:trails}
 \Description{ Figure (a) shows the top YouTube channels appearing in supporting trails of Democrats (Fox news, NBC, News, CBS News). Figure (b) shows the top YouTube channels appearing in supporting trails of republicans (Fox news, Fox Business, Sky News Australia). Figure (c) shows the top YouTube channels appearing in supporting trails of independents (Fox news, NBC news, PBS Newshour). Figure (d) shows the top YouTube channels appearing in neutral trails of Democrats (Fox news, PowerfulJRE, Fox Business). Figure (e) shows the top YouTube channels appearing in neutral trails of republicans (Fox news, Sky News Australia, PowerfulJRE). Figure (f) shows the top YouTube channels appearing in neutral trails of independents ((Fox news, PowerfulJRE, Fox Business)). 
 Figure (g) shows the top YouTube channels appearing in opposing trails of Democrats (Saturday Night Live, Last Week Tonight, Late Night with Seth Meyers). Figure (h) shows the top YouTube channels appearing in opposing trails of republicans (Saturday Night Live, Last Week Tonight, Late night with Seth Meyers). Figure (i) shows the top YouTube channels appearing in opposing trails of independents (Saturday Night Live, Last Week Tonight, Late night with Seth Meyers).}
 \vspace{-0.5cm}
\end{figure*}

\subsection{RQ3a: Diversity in search results}
 For analysis, we consider the top ten search results in standard SERPs. Figure \ref{impr_search} shows the top 10 YouTube channels with impressions in the most number of search queries.\footnote{The top 10 YouTube channels and their mean percentage of total impressions were almost similar when calculated separately for democrats, republicans, and independents. Thus, we show the overall distribution for all users combined together.} Here, we define impression as  the occurrence of a channel's video in SERP. 
 We observe that the left-leaning channel CNN on average appears in  SERPs of more than half (61.86\%)  search queries. Additionally, except Fox news and 11Alive,  all other top channels  are left-leaning. We further analyzed which channels were responsible for the most relevant YouTube videos in our collected data. In our standard SERPs, we obtained a total of 4901 unique videos out of which  1940 (39.51\%) videos were relevant, i.e. related to elections (959 opposing, 865 neutral, and 103 supporting). Overall, in these relevant videos, most videos come from CNN and MSNBC. The most opposing videos come from channels MSNBC followed by CNN, most supporting videos come from Fox News followed by Daily Mail while most neutral videos come from NBC news followed by CNN. Given, CNN is one of the channels with the most opposing videos, it is encouraging to see that it has the most search query impressions.

% Counter({3: 2866, 0: 865, -1: 959, 1: 116, 4: 103})

Next, we determine the source diversity in the SERPs using gini coefficient
metric \cite{ge2021towards,xiao2019beyond,trielli2019search}. Gini coefficient determines  inequality in a frequency distribution. For our case, we use this metric to determine the inequality in the distribution of YouTube channel impressions. For a given SERP consisting of videos from \textit{n} unique channels, given a list of impressions for all YouTube channels [$g_1$, $g_2$,...$g_n$], then gini coefficient would be calculated as,

\textit{Gini coefficient} (G) = $\frac{1}{2\bar{g}n^2}$ $\Sigma^{|n|}_{i=1}$ $\Sigma^{|n|}_{j=1}$ |$g_i$ - $g_j$| where $\bar{g}$ is the mean of all impressions.

A fairer search engine would have lower values of gini coefficient indicating uniform distributions of YouTube channel impressions. Figure \ref{gini:search} shows the distribution of gini coefficients for all SERPs for democrats, republicans, and independents. The distributions are similar for users with different political leanings. Furthermore, for approximately 96\% of search queries, the gini coefficient of SERPs is less than 0.3 indicating that YouTube has mostly evenly distributed videos from different channels  in its search results.

% \begin{figure}[]
%  \centering
% %  \captionabove{Images}%
% %  \label{fig:images}
%  \subfloat[]{%
%       \includegraphics[width=0.45\textwidth,height=3.5cm]{type1/dp.png}}
%       \label{a}
%  \qquad
%  \subfloat[ ]{%
%       \includegraphics[width=0.45\textwidth,height=3.5cm]{type1/rp.png}}
%       \label{b}
% \\
%  \subfloat[]{%
%       \includegraphics[width=0.45\textwidth,height=3.5cm]{type1/ip.png}}
%       \label{c}
%     \qquad   
% \subfloat[]{%
%       \includegraphics[width=0.45\textwidth,height=3.5cm]{type1/dn.png}}
%       \label{d}
% \\
%  \subfloat[]{%
%       \includegraphics[width=0.45\textwidth,height=3.5cm]{type1/rn.png}}
%       \label{e}
% \qquad
%  \subfloat[]{%
%       \includegraphics[width=0.45\textwidth,height=3.5cm]{type1/in.png}}
%       \label{f}
% \\  
% \subfloat[]{%
%       \includegraphics[width=0.45\textwidth,height=3.5cm]{type1/dd.png}}
%       \label{g}
% \qquad 
%  \subfloat[]{%
%       \includegraphics[width=0.45\textwidth,height=3.5cm]{type1/rd.png}}
%       \label{h}
% \\
%  \subfloat[]{%
%       \includegraphics[width=0.45\textwidth,height=3.5cm]{type1/id.png}}
%       \label{i}     
% \caption{
% }
% \label{type1}
% \end{figure}

\subsection{RQ3b: Diversity in up-next trails}
Overall, we collected 6943 videos in standard trails out of which 1082 are  relevant, i.e. related to elections.  The most number of opposing videos in trails come from channels MSNBC and Late Night with Seth Meyers*, most supporting videos in trails come from Fox News* and Fox Business, and most neutral videos come from Fox News* and NBC News\footnote{* indicates that seed videos of our experiments also belonged to these channels.}. Next, we determine the top ten YouTube channels occurring  in the standard trails. Note, we do not consider the seed videos while analyzing the trails. Figure \ref{impr_trail} shows the average number of impressions of the top 10 channels appearing the most number of times  in the trails.
% in a trail in which a video belonging to that channel is present for the top 10 channels appearing most number of times  in the trails. 
Here, impression indicates the number of occurrences of a channel's videos in a trail, while considering trails containing videos from that channel. Note that  the top channels are also channels of some of the seed videos in our dataset. The figure reveals that on average, videos from LastWeekTonight, Saturday Night Live, and Fox News appear more than 3 times in a trail, when taking into account all the trails where the channel was observed. This finding indicates that videos from these channels lead to more videos from these channels in the up-next recommendations.

Next, to determine the diversity in trails, we determine the proportion of channels that are different than the channel of the seed video in the trails. We find that on average, in an up-next trail of length five, we find  2.07 YouTube channels other than the channel of the seed video. The number of non-seed channels in up-next trails is the least for trails with seed videos from Saturday Night Live (0.85), LastWeekTonight (0.86), and Late Night with Seth Meyers (1.07). Note, we did not calculate this metric for supporting, neutral, and opposing seeds separately since the channels of our supporting, opposing, and neutral  videos are not unique. For example, we have a supporting as well as a neutral seed from Fox news. Given this scenario, there is no way to determine whether the videos appearing in the trails are due to the channel lean of the seed video or because of other factors. We also refrain from determining the diversity in up-next trails using gini coefficient since several trails had just one or two unique channels (M=3.1, SD=1.46) in which case gini coefficient would not give a good representation of diversity.

To get a sense of what kinds of channels are presented to users in the up-next trails, we determine the channels appearing in the most number of trails of democrats, republicans, and independents for trails with supporting, neutral, and opposing seeds (Figure \ref{div:trails}). We observe that Fox news appears in up-next trails with  supporting and neutral seeds of all users. Fox Business and Sky News Australia appear in both the supporting and neutral   up-next trails of more than half of the republicans (Figure \ref{rp},  and \ref{rn}). None of the seed videos belonged to these channels and they still appear in the up-next trails. Similarly, Sky News Australia also appears in the neutral up-next trails of 44.12\% independents (Figure \ref{in}) despite no neutral seed belonging to the channel. Furthermore, PowerfulJRE (Joe Rogan's YouTube channel) did not appear in the neutral up-next trails of all the users even though two neutral seed videos belonged to the channel (Figure \ref{dn}, \ref{rn} and \ref{in}). On the other hand,  the top channels appearing in the up-next trails with opposing seeds of all users (Figure \ref{dd}, \ref{rd} and \ref{id}) are the channels of the opposing seed videos used in our experiment. Furthermore, three channels out of the top four appear in the trails of more than 96\% of the users. This indicates that watching a video belonging to these left-leaning channels will probably lead to one or more videos belonging to this channel in the up-next recommendation trail.

\section{Discussion} \label{disc}

In this paper, we  conduct a  crowd-sourced audit of the YouTube platform to determine 
 how effectively the platform removed election misinformation from its various components. We discuss the implications of our findings below.

%  Misinformation and disinformation
% can disenfranchise voters and diminish trust in the results of electoral contests,
% eroding public confidence in the integrity of democratic processes and leadership transitions overall.

\subsection{Standardization of search results}
We find little to no personalization in the search results.  We also did not find any effect of personalization
on the amount of misinformation returned in search results. Throughout the study period, the amount of personalization and misinformation  remained constant in the searches. On analyzing the standard SERPs, we find that YouTube returns more videos  opposing election misinformation in 95\% of the search queries that we tested. Interestingly, we see that misinformation scores of search queries having a misinformation lean  (e.g. dominion voter fraud) are more negative compared to misinformation scores of queries that are neutral in stance (e.g. presidential election 2020). This finding implies that YouTube has paid more attention to the queries with misinformation lean and ensured that users are exposed to more debunking information when they search about the fraudulent claims surrounding the elections. This selective attention is also in-line with results of past audits that showed YouTube improving the recommendations of  topics like vaccination over 9/11 conspiracies  \cite{hussein2020measuring}. 

% probably made an attempt to present debunking information to users who perhaps are seeking  information about the fraudulent claims surrounding the elections. 
% On qualitatively analyzing the misinformation scores of the SERPs, we see that 
% Interestingly, we see this trend for queries that are neutral in stance (e.g. presidential election 2020) and the ones that and have a misinformation lean (e.g. dominion voter fraud). On qualitatively analyzing the misinformation scores of the SERPs, we find that YouTube returns more opposing videos for search queries about election fraud as compared to neutral search queries about the elections. This finding implies that YouTube has probably made an attempt to present debunking information to users who perhaps are seeking  information about the fraudulent claims surrounding the elections. 
% We also find that YouTube has almost standardized the searches. 
Our  analysis  also indicates that gini index of 96\% of search queries is less than 0.3, with $\sim$54\% queries having a gini index of less than 0.1. Such low values of gini index imply that YouTube is ensuring source diversity in searches by evenly distributing  videos from different channels in its SERPs. Furthermore, the distribution of gini coefficients was similar for all users irrespective of their partisanship. This finding indicates  YouTube's attempt to expose users to videos from different channels rather than a select few based on participants' partisanship.   Interestingly, in line with a previous audit on Google search \citep{trielli2019search}, we find that CNN is one of the top channels whose videos appear in 61.8\% of search queries. Future studies can test whether the dominance is due to emergent bias or the strategies adopted by the channel to enhance algorithmic visibility \cite{trielli2019search}.  {Overall, our analysis reveals that YouTube's search results are largely unpersonalized and the platform has had varying levels of success in removing misinformation and presenting videos that debunk election-related falsehoods in different clusters of search queries.}

% standardized the search results and has mostly been successful inremoving election misinformation from its searches.}

% This seems to contradict the results in Figure 9 for RQ2a, which suggest that YouTube has been more successful in some query clusters and less successful in others. I think the paper should emphasize this variation when summarizing results, so as to avoid the impression that YouTube was uniformly successful.
% powerful algorithmic
% curators like Google have for mediating attention to news
% information

\subsection{Scope for improvement in up-next trail recommendations}
We find that up-next trails are highly personalized. However, for  50\% of the users, only up to 10\% videos in the up-next recommendations come from users' subscribed channels.  Future audit studies should further investigate the impact of users' channel subscriptions (both news and non-news channels) on the platform's recommendations.  We also find that there is no significant difference in the amount of misinformation that  users are exposed to in up-next recommendation trails in the signed-in standard window and unpersonalized incognito window.  On examining the standard up-next trails, we do find an echo-chamber effect. Users, irrespective of their partisanship, receive more misinformation in the up-next trails with supporting seeds as compared to the trails with neutral and opposing seeds (Figure \ref{tab:misinfo scores}). We also observe that the magnitude of misinformation scores of trails with opposing seeds is more than the magnitude of misinformation scores of trails with supporting seeds. This implies that users are  exposed to a small number of misinformative videos when they follow the up-next recommendations of a video supporting election misinformation. On the other hand, users are exposed to a larger number of opposing videos in the opposing up-next trails. This is a key finding  also supported by prior work that showed that echo chambers of misinformation can be burst by watching  debunking videos \cite{tomlein2021audit}. The platform can leverage this phenomenon by making its recommendation engine present more debunking videos to users which would then expose them to more credible videos in the recommendation trails. 

We also examine various transitions in the up-next trails to study how users get pushed towards misinformation. Overall, we observe that problematic transitions   where a supporting video is recommended in the up-next video recommendation of a supporting (S->S) or opposing video (O->S) are less than 2\%.   However, S->S transitions are more in trails with supporting seeds for independents compared to democrats and republicans. Furthermore, N->S transitions are also high in up-next trails with neutral seeds for independents. These findings are  problematic. Showing misinformative videos to  independents who might not have developed a strong opinion on the election fraud conspiracies could increase their chances of forming a pro-conspiracy belief. We also observe that N->S transitions are more for republicans in the up-next trails with neutral seeds (3.78\%) compared to trails with supporting seeds (1.61\%). This finding is again troublesome. Past studies have indicated that republicans are more susceptible to electoral fake news \cite{Republic19:online}. Thus, recommending  videos supporting election misinformation to republicans watching neutral videos would expose them to more misinformation which might reinforce or lead to forming conspiratorial beliefs.

On analyzing the up-next trails for channel diversity, we observe several interesting phenomena. First, the number of impressions for left-leaning late-night show channels on YouTube such as LastWeekTonight is very high. On average, approximately 3-4 videos from these channels appear in the up-next trails (of length five) when starting with opposing seed videos. Furthermore,  these channels appear in the video recommendations of almost all of our study participants. Similar to the late-night shows, we find that fox news also appears on average 3.27 times in the up-next trails of all participants. Future studies can look into the reasons behind the  strong ``algorithmic recognizability''  \cite{gillespie2017algorithmically} and high amplification of these channels in YouTube recommendations.  Overall, we conclude that while YouTube has reduced misinformative videos in its up-next recommendations, there is still scope for improving  the recommendation algorithm. 
\vspace{-0.3cm}

% %  The key finding is that watching misinformation debunking videos (e.g., credible news, scientific
% content) generally improves the situation (in terms of recommended items or search result personalization),
% albeit with varying effects and forms, mainly depending on particular misinformation topic.

% \subsection{Looking beyond the filter-bubble effect}
% {Our work shows that there is substantial overlap in the search results seen by users of different political interests }

\subsection{Participants' beliefs vs algorithmic reality}
The study survey conducted before our audit experiment provided us with an opportunity to map participants' beliefs about personalization and trust in YouTube's algorithms with the reality of the situation as determined by our audits.
The majority of participants believe that YouTube somewhat  personalizes search results. However, in reality, they are hardly personalized. On the other hand, only half of the participants believe up-next recommendations to be highly personalized which is in line with our findings. This mismatch in beliefs and reality indicates users' lack of algorithmic awareness. It also acts as a call to action for the platform  to make users aware  of the functioning of the algorithms. Users could be made aware of personalization or lack of it  by adding design features that promote algorithmic reflection, for example, seeing search results or recommendations of other users \cite{bhuiyan2022othertube}. 

Our survey also showed that, respectively, 19.2\% and 14.1\% users trust the credibility of information presented to them by YouTube in the search results and up-next recommendations to a great extent. This belief is problematic and indicates reliance on the platform's algorithms to show credible information. In reality, while we find the majority of YouTube's search results to be credible, up-next recommendations still contained misinformative videos. One way to make people spot misinformation on the platform and not blindly trust YouTube's recommendations could be by providing additional context about the content that the participant is searching for or viewing. While YouTube has started displaying Wikipedia links on the platforms \cite{YouTubea40:online}, additional cues in the form of credibility citations, existing fact-checks or knowledge panel\footnote{https://support.google.com/knowledgepanel/answer/9163198?hl=en} could also be helpful \cite{hughes2021introducing}.  

% Nevertheless, our audit showed that YouTube (similar to other platforms), despite their best efforts so far, can
% still promote misinformation seeking behavior to some extent. The results also motivate the need for independent
% continuous and automatic audits of YouTube and other social media platforms

% \subsection{}

\section{Limitations and future work}
Our work is not without limitations. Our audit study  is observational in nature, i.e  our experiment does not isolate user attributes that produce the differences in misinformation measurements. 
We only make observations on the differences in misinformation received in searches and recommendations of users  with different political affiliations. We recruited participants who used YouTube extensively to get information about the 2020 elections. However, for ethical reasons, we did not analyze participants' account histories to verify their self-reported data. Our participant sample was also not balanced with respect to demographic attributes and political affiliation. {We selected YouTube videos that had accumulated the most number of views as the seed videos for our audit experiments. One potential pitfall of such a sampling strategy is that it reduces the ecological validity of the experiment since the participants in our study might not have engaged with those videos in the past. 
 Another limitation is that YouTube might have specifically tailored the recommendations of popular misinformative videos. Future studies could consider  alternative strategies for sampling videos, such as  selecting videos that were more recently published on YouTube or sampling a combination of videos that have  accumulated the least and most amount of  engagement. The search queries used in our audit also might not be representative of  how our study participants formulate queries about the elections. }
% Additionally, the seed videos used in our study might not be representative of what our study participants witnessed in their own YouTube recommendations. 
{Future studies can survey the study participants to determine how they used YouTube searches in the context of political elections as well as their 
information needs about the elections.}

% \cite{mustafaraj2020case,kavnukova2019searching}. 
 % Additionally, we only analyzed  the top-occurring YouTube videos in SERPs and homepages. 
 {Our classifier developed to annotate the YouTube videos for election misinformation has an error rate of 9\% which could have affected the downstream analysis that we performed to quantify the amount of misinformation in various YouTube components. Additionally, we assign an annotation value of 0 to all videos that were removed from YouTube after our audit data collection. While the  number of such videos is very small (<1\%), it would  result in a conservative estimate of misinformation bias present in the search results and recommendations. We use the misinformation bias score adopted from Hussein and Juneja et al's study that captures the amount of misinformation along with the rank of the video \mbox{\cite{hussein2020measuring}}. However, this metric does not take into account the relevance of the videos. Future studies can use metrics that measure simultaneously the relevance and credibility in ranked lists such as Normalised Weighted Cumulative Score and Convex Aggregating Measure \mbox{\cite{lioma2017evaluation}}. In our audit experiment, after testing every condition (watching supporting, neutral, and opposing videos), we performed a step to delete users' YouTube history  created by our extension so that it  does not impact the other experimental condition.  The first author  tested out the effect of  deletion on users' search and watch history for a few sample queries and videos and found that the effect of such deletion is almost immediate. However, we did not test out this scenario for all search queries and videos used in our audit. Future studies can determine how soon the  deletion of history impacts users' recommendations and search results across various topics.} 
 % Despite these limitations, our work provides a broad overview of how YouTube's algorithm fares with respect to election misinformation in complex real-world settings. 
% \mbox{\cite{whittaker2021recommender}}

{Our study  focuses on users' beliefs about the personalization and credibility of content on YouTube as well as the  role of YouTube's algorithms in driving users to the filter bubbles of problematic content. Future studies can focus on the impact of algorithmic recommendations on  the radicalization of users.  There are several scholars who argue that algorithms are not centrally culpable for the polarization or the filter bubbles that users experience on online platforms \mbox{\cite{bruns2019filter,whittaker2021recommender,bruns2019filter2}}. Many times the users of social media have a more diverse
media diet than the non-users \mbox{\cite{bruns2019filter,bruns2019filter2}}.
Scholars posit that while algorithms can observe what a user consumes on social media, they cannot determine what the user actually prefers \mbox{\cite{dahlgren2021critical}}. In other words, a digital choice is not always a true reflection of an individual's preference \mbox{\cite{dahlgren2021critical}}.  Furthermore, users might use different online platforms for different types of content \mbox{\cite{dahlgren2021critical}}. Thus, to gain a holistic idea of the extent algorithms play a role in user polarization, future audit studies  can conduct multi-platform crowd-sourced audits for individuals. These audit studies can determine the impact of algorithmic recommendations on users' social/political viewpoints via surveys and  monitor users' patterns of content consumption  simultaneously on multiple search engines and social media platforms used by the users.  }

\section{Conclusion} \label{lim}

In this study, we conducted a crowd-sourced audit on YouTube to determine the effectiveness of its content regulation policies with respect to election misinformation.  We find that YouTube returns videos that debunk election misinformation in its searches. We also find that YouTube leads users to a small number of misinformative videos in up-next trails with  seed videos that support election misinformation. Overall, our study shows that while YouTube has been largely successful in removing election misinformation from its searches,  there is still scope to fix up-next recommendations.

% we considered such differences
% to be a part of an individual’s personalization experience
%  that our
% participant sample was not balanced in terms of their political preferences or demographics

% search queries used in our study 
% might not be representative of the real world homepage recommendations. Since the scenarios shown to users are 
% not representative of what they have witnessed in their own YouTube homepage, they  lack the
% significance and consequence of  real-world.

% Our analysis is observational in nature and does not isolate a mechanism that produces the relationships we observe. 
% \begin{acks}
% \end{acks}

\bibliographystyle{ACM-Reference-Format}
\bibliography{sample-base}


\appendix
 \section{Appendix}

 \subsection{Participants' characteristics} \label{charac}
 In our study survey, 
 we asked participants how often they used the YouTube platform for getting news about the 2020 US Elections. Out of the 99 users who participated in the study, 40.4\% reported using YouTube to access election-related news several times a day. 
% 'How often you do the following? - Perform searches on YouTube?'
% Counter({'Sometimes': 10, 'Always': 37, 'Often': 51, 'Rarely': 1})
When asked whether the 2020 election was stolen from Donald Trump,  59.59\% strongly disagreed, 14.14\% somewhat disagreed, 11.11\% somewhat agreed, and 11.11\% strongly agreed. 19.19\% participants believed that the US presidential elections were a result of illegal voting or election rigging while 70.70\%  believed it to be legitimate and accurate. 31.31\% participants believed fraud in the United States with respect to the presidential election 2020 – that is, votes being cast in the name of people who are not eligible to vote was a major problem. 41.41\% believed voter disenfranchisement in the United States with respect to the presidential election 2020, i.e, eligible voters being prevented from casting their ballots or not having their ballots counted was a major problem. 30.30\% participants believed fraud in voting by mail in the U.S. with respect to the presidential election 2020 to be a major problem. 57.57\% participants rated Donald Trump's conduct during the presidential elections as poor while 13.13\% rated it as excellent. On the other hand, 17.17\% of participants rated Joe Biden's conduct during presidential elections as poor while 16.16\% rated it as excellent.

 \subsection{Amazon Mechanical Turk Job} \label{amt}
 We used AMT  to get annotations for 545  videos for the ground truth dataset.
During our manual annotations, we realized that the majority of the videos were irrelevant. In order to get annotations for relevant videos---classes supporting, opposing, and neutral,  we curated a list of keywords (such as 'fraud', 'ballot', 'election', 'steal', etc.) and news channels (such as CNN, NTD, Fox news, etc.) on YouTube. Then, we filtered out videos that were published by the curated channels and had the keywords in the title or description. To get high-quality annotations for these videos, we trained and screened the AMT workers. 
% Using the criteria we got 50 videos annotated from the AMT platform.  We then iteratively updated the keyword and channel list to get more relevant videos. We repeated this process three times.   
Below we describe the screening process and our annotation task briefly.\\
% \\

\begin{figure*}[t]
    \centering
    \includegraphics[width=0.9\textwidth]{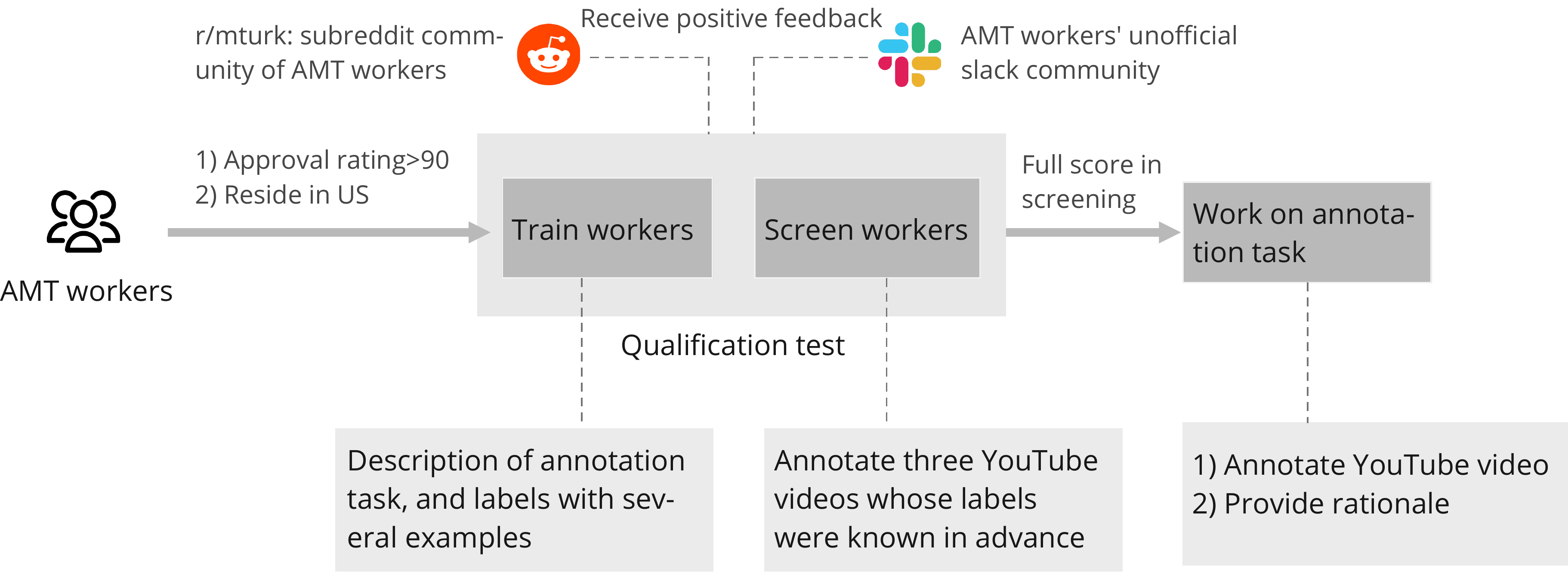}
    \caption{Figure illustrating the process of obtaining YouTube video annotations from AMT workers. The workers were screened via a qualification test where they were first trained by providing  detailed descriptions of the annotation labels. To test their understanding, they were  asked to  annotate three YouTube videos whose  labels were known in advance. Workers who correctly labeled the three videos proceeded to work on the annotation task. To ensure that our description of the annotation labels and the task was clear and comprehensive, we posted on r/mturk---a subreddit community of AMT workers and AMT workers' unofficial slack channel. We released our qualification test and annotation task after receiving positive feedback from the AMT community.} 
    \label{fig:amt}
    \Description{Figure illustrating the process of obtaining YouTube video annotations from AMT workers. The workers were screened via a qualification test where they were first trained by providing  detailed descriptions of the annotation labels. To test their understanding, they were  asked to  annotate three YouTube videos whose  labels were known in advance. Workers who correctly labeled the three videos proceeded to work on the annotation task. To ensure that our description of the annotation labels and the task was clear and comprehensive, we posted on r/mturk---a subreddit community of AMT workers and AMT workers' unofficial slack channel. We released our qualification test and annotation task after receiving positive feedback from the AMT community.}
\end{figure*}

\noindent\textit{Worker training and screening:} To train workers to do the annotation task and screen them on the basis of their understanding of the annotations, we created a qualification test. The test first described in detail the annotation labels, heuristics, and the annotation task. We provided several examples of YouTube videos for each annotation label and described the process and reason behind assigning a particular label to the video. To ensure that our description of the annotation labels and the task was clear and comprehensive, we posted on r/mturk--- a subreddit community of AMT workers as well as AMT workers' unofficial slack channel. After receiving positive feedback from the AMT community, we released
% finalized the training part of 
the qualification test.  We also included three questions in the qualifying test asking AMT workers to annotate YouTube videos whose annotation labels were known in advance. These  videos  were already annotated previously by the authors. AMT workers who correctly labeled all three videos (100\% score) qualified for the YouTube annotation task. In addition to getting a perfect score on the qualification test, we also required AMT workers to have at least a 90\% approval rating  on the AMT platform.  \\

\noindent\textit{YouTube annotation task:} The YouTube annotation task required AMT workers to assign a label to the video and also provide the rationale behind selecting the label. We enforced a minimum word limit of 10 characters for the rationale. 
We released YouTube videos in a batch sizes of 10 and 15 and obtained three annotations for each video. 
The majority response was selected as the annotation label for the video. In total, we obtained annotations for 545 videos, out of which consensus was reached on 516 videos (Supporting: 26, Opposing: 74, Neutral: 77, Irrelevant: 318, YouTube video in a language other than English: 6, URL not accessible: 15).

% To get high quality annotations for these videos, we trained and screened the AMT workers via a qualification test. Workers who 
% had 90\% approval rating  on the AMT platform and 
% obtained full score in the qualification test qualified to do the annotation task. Figure \ref{fig:amt} illustrates the entire process. We obtained three annotations for each video. 
% The majority response was selected as the annotation label for the video. In total, we obtained annotations for 545 videos, out of which consensus was reached on 516 videos (Supporting: 26, Opposing: 74, Neutral: 77, Irrelevant: 318, YouTube video in language other than English: 6, URL not accessible: 15).

 \subsection{Annotating YouTube channels for partisan bias} \label{partisanbias}
Our dataset of unique videos came from a large number of YouTube channels ($\sim$17.5K) devoted to  both news and  non-news content. We coded the leaning of the channel on a 5-point Likert scale (far-left, center-left, neutral, center-right, and far-right) using computational methods and several heuristics. First, to identify news-related channels, we used several  pattern-matching techniques (e.g., finding keyword \textit{news} in the channel's name, etc.) and discovered a total of 802 news channels. Then we  used existing datasets on media bias from \texttt{mediabiasfactcheck.com} and \texttt{allsides.com} for annotating the channels.
% to find related channels using string matching algorithms on channel title and description. 
% For the rest, we used several keyword and pattern matching algorithms (e.g., finding keyword \textit{news} in channel's name, etc.) to find news related YouTube channels. 
% For channels that fell under the news related YouTube channel category (n=802), 
For channels whose annotations were not available in the datasets, we manually went through their title, description, sample videos, and related information from their website, Wikipedia, and/or google search to identify their leaning or the leaning of their affiliations. Many local news channels such as KHOU\footnote{https://www.youtube.com/c/KHOU} or KPRC\footnote{https://www.youtube.com/c/KPRC2Click2Houston} are affiliated with national channels. If we did not find the bias ratings 
% from 
% existing datasets
% \texttt{mediabiasfactcheck.com} and \texttt{allsides.com} 
for such local channels, we assigned them the label of their affiliations. For example, KHOU is associated with center-left CBS and thus, was also assigned a center-left rating. We assigned channels that didn't fall under the news category the neutral label. We manually checked a random sample (n=50)  of  non-news channels and found only one channel that had content about the news. Therefore, this process produced channel bias annotations (to be used as a feature in our classifier) with reasonable accuracy for our study, given that channel bias detection is not the main focus of our work.

\end{document}